\documentclass[longauth]{aa}  
\usepackage{graphicx}
\usepackage{microtype}
\usepackage{txfonts}
\usepackage{ragged2e}
\usepackage{multirow}
\usepackage{booktabs}
\usepackage{tabularx,array}
\usepackage[colorlinks=true,linkcolor=blue,citecolor=blue, urlcolor=blue]{hyperref}
\usepackage{pgf}
\usepackage{aas_macros}
\usepackage{geometry}
\usepackage{layouts}
\usepackage{lscape}
\usepackage{setspace}
\usepackage{import}
\usepackage{color}
\usepackage{overpic}

\graphicspath{{fields/}}

\newcommand{\appendixlccaption}[1]{Light curves of #1.}

\begin{document} 

   \title{TDCOSMO. \textsc{XVII}. New time delays in 22 lensed quasars from optical monitoring with the ESO-VST 2.6m and MPG 2.2m telescopes}

   \titlerunning{Lensed QSO time delays from ESO VST and MPG 2.2m}

\author{
F. Dux \inst{\ref{eso}, \ref{epfl}} \and
M. Millon \inst{\ref{stanford}, \ref{ethz}} \and
A. Galan \inst{\ref{epfl}, \ref{tum}} \and
E. Paic \inst{\ref{epfl}} \and
C. Lemon \inst{\ref{epfl}} \and
F. Courbin \inst{\ref{barca}, \ref{icrea}, \ref{epfl}} \and
V. Bonvin \inst{\ref{epfl}} \and 
T. Anguita \inst{\ref{UNAB}, \ref{millenium}} \and
M. Auger \inst{\ref{camastro}, \ref{camkavli}} \and
S. Birrer \inst{\ref{stonybrook}} \and
E. Buckley-Geer \inst{\ref{fermilab}} \and
C.D. Fassnacht \inst{\ref{ucdavis}} \and
J. Frieman \inst{\ref{fermilab}, \ref{chicago}} \and
R.~G.~McMahon\inst{\ref{camastro}, \ref{camkavli}} \and 
P.J. Marshall \inst{\ref{stanford}} \and
A. Melo \inst{\ref{tum}} \and
V. Motta \inst{\ref{valpo}} \and
F. Neira \inst{\ref{epfl}} \and
D. Sluse \inst{\ref{liege}} \and
S.H. Suyu \inst{\ref{MPG}, \ref{tum}, \ref{ASIAA}} \and
T. Treu \inst{\ref{ucla}} \and
A. Agnello \inst{\ref{hartreecenter}, \ref{nielsbohrinst}} \and
F. Ávila \inst{\ref{valpo}} \and
J. Chan \inst{\ref{cuny}, \ref{MNU}} \and
M. Chijani \inst{\ref{UNAB}} \and
K. Rojas \inst{\ref{portsmouth}} \and
A. Hempel \inst{\ref{UNAB}, \ref{heidelberg}} \and 
M. Hempel \inst{\ref{UNAB}, \ref{heidelberg}} \and
S. Kim \inst{\ref{heidelberg}}\and
P. Eigenthaler \inst{\ref{heidelberg}, \ref{puc}}\and
R. Lachaume \inst{\ref{heidelberg}, \ref{puc}}\and 
M. Rabus \inst{\ref{uniconcepcion}} 
}

\authorrunning{Dux et al.}

\institute{
    European Southern Observatory, Alonso de Córdova 3107, Vitacura, Santiago, Chile \label{eso} 
\goodbreak \and
    Institute of Physics, Laboratory of Astrophysics, \'Ecole Polytechnique 
    F\'ed\'erale de Lausanne (EPFL), Observatoire de Sauverny, 1290 Versoix, 
    Switzerland \label{epfl}
\goodbreak \and
    Kavli Institute for Particle Astrophysics and Cosmology and Department of Physics, Stanford University, Stanford, CA, USA. \label{stanford} 
\goodbreak \and
    Institute for Particle Physics and Astrophysics, ETH Zurich, Wolfgang-Pauli-Strasse 27, CH-8093 Zurich, Switzerland \label{ethz} 
\goodbreak \and 
    Department of Physics, TUM School of Natural Sciences, Technical University of Munich, Garching, Germany. \label{tum} 
\goodbreak \and
    Institut de Ciències del Cosmos, Universitat de Barcelona, Martí i Franquès, 1, E-08028 Barcelona, Spain \label{barca}  
\goodbreak \and
    ICREA, Pg. Llu\'is Companys 23, Barcelona, E-08010, Spain \label{icrea}
\goodbreak \and
    Institute of Astrophysics, Universidad Andres Bello Fernandez 
    Concha 700, Las Condes, Santiago, Chile    \label{UNAB}\goodbreak \and
    Millennium Institute of Astrophysics, Chile \label{millenium}\goodbreak  \and
    Institute of Astronomy, University of Cambridge, Madingley Road, Cambridge CB3 0HA, UK {\label{camastro}}\goodbreak \and
    Kavli Institute for Cosmology, University of Cambridge, Madingley Road, 
    Cambridge CB3 0HA, UK {\label{camkavli}}\goodbreak \and
    Department of Physics and Astronomy, Stony Brook University, Stony Brook, NY 11794, USA \label{stonybrook} \goodbreak \and
    Fermi National Accelerator Laboratory, P.O. Box 500, Batavia, IL 60510, 
    USA \label{fermilab}\goodbreak \and
    Department of Physics and Astronomy, University of California, Davis, 1 Shields Ave., Davis, CA 95616, USA 
    \label{ucdavis}\goodbreak \and
    Kavli Institute for Cosmological Physics, University of Chicago, 
    Chicago, IL 60637, USA \label{chicago}\goodbreak \and
    Instituto de F\'isica y Astronom\'ia, Universidad de Valpara\'iso, Avda. 
    Gran Breta\~na 1111, Playa Ancha, Valpara\'iso 2360102, Chile 
    \label{valpo}\goodbreak \and
    STAR Institute, University of Li\`ege, Quartier Agora - All\'ee du six Ao\^ut, 19c B-4000 Li\`ege, Belgium \label{liege} \goodbreak \and
    Max Planck Institute for Astrophysics, Karl-Schwarzschild-Strasse
    1, D-85740 Garching, Germany \label{MPG}\goodbreak \and
    Institute of Astronomy and Astrophysics, Academia Sinica, P.O.~Box 
    23-141, Taipei 10617, Taiwan \label{ASIAA}\goodbreak \and  
    Department of Physics and Astronomy, University of California, Los 
    Angeles, CA 90095, USA \label{ucla}\goodbreak \and
    STFC Hartree Centre, Sci-Tech Daresbury, Keckwick Lane, Daresbury, Warrington (UK) WA4 4AD \label{hartreecenter} \goodbreak \and 
    DARK, Niels Bohr Institute, University of Copenhagen, Jagtvej 155A, DK-2200 Copenhagen N, Denmark \label{nielsbohrinst} \goodbreak \and
    Department of Physics and Astronomy, Lehman College of the CUNY, Bronx, NY 10468, USA \label{cuny} \goodbreak \and
    Department of Astrophysics, American Museum of Natural History, Central Park West and 79th Street, NY 10024, USA \label{MNU} \goodbreak \and
    Institute of Cosmology and Gravitation, University of Portsmouth, Burnaby Rd, Portsmouth PO1 3FX, UK \label{portsmouth} \goodbreak \and
    Max-Planck-Institut f\"ur Astronomie, K\"onigstuhl 17, 69117 Heidelberg, Germany \label{heidelberg}\goodbreak \and
    Centro de Astroingenier\'ia, Facultad de F\'isica, Pontificia Universidad 
    Cat\'olica de Chile, Av. Vicu\~na Mackenna 4860, Macul 7820436, 
    Santiago, Chile \label{puc}\goodbreak \and
    Departamento de Matem{\'a}tica y F{\'i}sica Aplicadas, Facultad de Ingenier{\'i}a, Universidad Cat{\'o}lica de la Sant{\'i}sima Concepci{\'o}n, Alonso de Rivera 2850, Concepci{\'o}n, Chile \label{uniconcepcion}
}

   \date{Received January 2025; accepted March 2025}

  \abstract{
  We present new time delays, the main ingredient of time delay cosmography, for 22 lensed quasars resulting from high-cadence \textit{r}-band monitoring on the 2.6$\,$m ESO VLT Survey Telescope and Max-Planck-Gesellschaft 2.2$\,$m telescope. 
  Each lensed quasar was typically monitored for one to four seasons, often shared between the two telescopes to mitigate the interruptions forced by the COVID-19 pandemic.
  The sample of targets consists of 19 quadruply and 3 doubly imaged quasars, which received a total of 1$\,$918 hours of on-sky time split into 21$\,$581 wide-field frames, each 320 seconds long.
  In a given field, the 5-$\sigma$ depth of the combined exposures typically reaches the 27$^\mathrm{th}$ magnitude, while that of single visits is 24.5 mag -- similar to the expected depth of the upcoming Vera-Rubin LSST.
  The fluxes of the different lensed images of the targets were reliably de-blended, providing not only light curves with photometric precision down to the photon noise limit, 
  but also high-resolution models of the targets whose features and astrometry were systematically confirmed in \textit{Hubble} Space Telescope imaging.
  This was made possible thanks to a new photometric pipeline, \texttt{lightcurver}, and the forward modelling method \texttt{STARRED}.
  Finally, the time delays between pairs of curves and their uncertainties were estimated, taking into account the degeneracy due to microlensing, and for the first time the full covariance matrices of the delay pairs are provided.
  Of note, this survey, with 13 square degrees, has applications beyond that of time delays, such as the study of the structure function of the multiple high-redshift quasars present in the footprint at a new high in terms of both depth and frequency. 
  The reduced images will be available through the European Southern Observatory Science Portal.
  }


\maketitle

\section{Introduction}

The precise measurement of the difference in the time of arrival between different lensed components of a gravitational lens is the first step in time-delay cosmography (TDC), proposed by \citet{Refsdal1964}. 
Time-delay cosmography is enabled by lensing, when the light of a single source experiences path differences through cosmological distances,
and opens the possibility of measuring the Hubble constant, $H_0$\footnote{
The $H_0$ measurement is most precise if assuming a cosmological model such as $\Lambda CDM$, however this assumption is not an absolute requirement: TDC fundamentally probes a ratio of angular diameter distances, which can be computed in arbitrary cosmologies.
}.
For the measurement of this difference in the time of arrival to be possible, the distant source must be very bright and variable: supernovae fit this requirement, and so do quasars.
In practice, the same photometric variations in the source can be observed at different times in the light curves of each lensed component, and the temporal shift required to align these light curves is called the time delay.
The total time delay, $\Delta t_{\rm tot}$, between the images depends on the geometric difference between the optical paths, $\Delta t_{\rm geom}$, and on the difference between the lens potential values at the positions of the lensed images, $\Delta t_{\rm grav}$. 
The total time delay is the sum of the geometric and gravitational delays and is directly proportional to the so-called time-delay distance, a ratio of the three angular-diameter distances, between the observer and the source, the observer and the lens, and the lens and the source \citep[e.g.][]{Refsdal1964, Suyu2010}. 

Time-delay cosmography does not rely on cosmic microwave background (CMB) measurements or on any intermediate standard ruler or standard candles, and it does not involve the difficult construction of a multi-step ladder.
The primary element in TDC is the time-delay measurement, but turning this delay into a time-delay distance, and finally into $H_0$, requires a mass model for the lens galaxy and a measurement of the contributions of nearby objects along the line of sight.
We refer the reader to \citet{ISSI-Birrer} and \citet{Treu2023} for reviews of the methodology of the modelling and line-of-sight mass contribution. 

For a given lensed quasar system, the uncertainty in the inferred $H_0$ value can be divided into systematic and statistical errors.
The systematic part of the error budget is typically dominated by the accuracy of the mass model and its inherent degeneracies, the most famous of which is the mass sheet degeneracy \citep[MSD; e.g.][]{Falco1985, Gorenstein1988, Kochanek2002, Schneider2013, Schneider2014, Blum2021}. 
These can be mitigated by using spatially resolved kinematics of the lensing galaxy~\citep[e.g.][]{Yildirim2020, Shajib2023}, which brings other but distinct degeneracies.
On average, the statistical uncertainty in $H_0$ is roughly divided among contributions from lens modelling, line-of-sight characterisation, and time delay measurements.

TDCOSMO (Time Delay COSMOgraphy) is a collaborative project resulting from the fusion of COSMOGRAIL~\citep{Courbin2018}, H0LiCOW~\citep{suyu2017}, STRIDES~\citep{treu2018}, and SHARP~\citep{chen2019}.
The current objective of TDCOSMO is reaching a 1\% precision on $H_0$, both by expanding the pool of ancillary-complete lensed quasars and understanding and controlling systematic errors \citep[see, e.g.,][]{tdcosmoV}.

This paper in the TDCOSMO series focuses on the primary ingredient needed for the method to work: time-delay measurements between multiple images of strongly lensed sources at cosmological distances, here quasars. 
Measuring lensed quasars time delays requires high-quality light curves with adequate temporal sampling over long periods of time, in an effort to avoid interruptions other than the inevitable seasonal gaps.
Typically, a one-year baseline is required, but most monitoring campaigns carried out so far have obtained several years per object, sometimes up to two decades. 

There are two main difficulties to overcome in order to achieve time-delay measurements at optical wavelengths. 
The first is that the lensed images are faint (20.2 \textit{r}-mag on average for the present sample), and most often blended in ground-based images: astronomical seeing is thereby a major limitation.
Given that the image separations in lensed quasars are typically not much larger than the average astronomical seeing, the data need to be processed with deblending techniques such as point spread function (PSF) fitting or more sophisticated forward modelling of the pixels \citep[e.g.][]{starredscience, starred, cantale2016, MCS}. 
High signal-to-noise ratio (S/N) imaging is mandatory to make the deblending reliable, requiring fairly large telescopes (typically 2$\,$m) and long exposure times (typically 30$\,$min on-target).
This achieves photometric precisions between a few milli-magnitudes to a few tens of milli-magnitudes for the brighter targets ($\sim 19$ $r$-mag).

The second difficulty is that microlensing by stars in the lensing galaxy introduces additional variations in the quasar images. 
These variations are uncorrelated in each quasar image and occur on timescales of the order of months to years. 
Notably, while this effect typically increases the uncertainty in time delays of lensed quasars, it also affects lensed type Ia supernovae and compromises their standard candle nature altogether~\citep{FoxleyMarrable2018, Weisenbach2024}.
For monitoring observations with temporal sampling of the order of one observation every 3-4 days and photometric precision of 10 milli-mag, it becomes necessary to monitor any given system for several years, averaging out microlensing over long periods of time. 
This has been the strategy adopted by most past monitoring campaigns 
\citep[e.g.][]{Burud2000, Burud2002, Burud2002b, Hjorth2002, Ullan2003, Kochanek2006b, Shalyapin2017, Shalyapin2019, Munoz2022, Shalyapin2023}, 
including the COSMOGRAIL program that has measured more than 30 time delays \citep[][]{Courbin2005, Eigenbrod2005, Bonvin2019, Millon2020}. 
The future Vera~C.~Rubin~Observatory \citep[Rubin-LSST]{lsst} observations of lensed quasars will fall in this regime, with the drawback that years will be necessary to measure time delays in new objects, but with the advantage that Rubin-LSST will monitor all targets in the southern hemisphere, even those not identified as lensed quasars yet.

In pre-Rubin-LSST times, when objects are still monitored one by one, it has been shown that time delays can be measured within 1-2 years, provided daily high-S/N observations are possible \citep[][]{Courbin2018, Millon2020b}. 
Such high-cadence and high-S/N data allow for the capture of low-amplitude and fast quasar variations occurring on shorter timescales than microlensing, and hence act as a natural frequency comb, discriminating between the two types of signals.

All known bright and well-separated quadruply lensed quasars in the southern hemisphere have been monitored for time delays in the past. 
This work tackles most of the known remaining such objects in the south, which are fainter and more narrowly separated (and thereby more difficult) than the ones monitored in earlier programs -- roughly doubling the sample of southern time-delay lenses. 
The determination of time delays was achieved thanks to consistent, high-quality optical observations at two telescopes, and refinements of the data processing.

It should be noted that while this program was originally designed to observe each object with a single facility to achieve homogeneous data quality and photometric calibration, there were  several interruptions due to the impact of the COVID-19 pandemic on the operations of the telescopes. 
As a consequence, we were affected by large gaps in our light curves and often had to observe objects with the two telescopes, as COVID-19 did not impact the two observatories in the same way. 

We describe the modalities, observations, and statistics in Sect.~\ref{sec:obs}.
The methodology employed for extracting the light curves from the 21$\,$582 exposures is detailed in Sect.~\ref{sec:lc_extraction}.
Section~\ref{sec:measure_tds} lays out the methodology of the estimation of the time delays, the results of which are presented in Sect.~\ref{sec:results}.
Next, in Sect.~\ref{sec:discussion} we briefly discuss what the future holds for time-delay determination in light of the upcoming Rubin-LSST observations and the findings of this program.
Finally, we draw conclusions in Sect.~\ref{sec:conclusions}. The magnitudes quoted herein are calibrated on the Vega scale.

\section{Observations and statistics}
\label{sec:obs}

\subsection{Targets}
We monitored 22 lensed quasars visible from the southern hemisphere, including 19 quadruply lensed ones.
This sample of 19 represents about half the known quadruply lensed quasars under 20$^\circ$ declination -- the other half being brighter and having mostly been monitored for time delays already.
We list co-ordinates, redshifts when available, and discovery papers in Table~\ref{tab:targets}. 

\begin{table*}
\caption{
Targets observed in this program, with J2000 co-ordinates, redshifts, and discovery papers. 
\label{tab:targets}}
\def\arraystretch{1.05}\vspace{-0.2cm}
\begin{tabular}{lrrrcl}
\toprule
 Name           &     R.A. &      Dec. &   $z_{\rm s}$ & $z_{\rm l}$               & Discovery                                        \\
\midrule
 DES$\,$J0029$-$3814 &   7.4208 & -38.2405  &         2.81  &                             & Schechter in prep.                                  \\
 PS$\,$J0030$-$1525    &   7.5636 & -15.4177  &         3.36  &                             & \citealt{lemon2018b}                                \\
 DES$\,$J0053$-$2012   &  13.4349 & -20.2091  &         3.8   &                             & \citealt{lemon2020}                                 \\
 WG$\,$J0214$-$2105    &  33.5681 & -21.0931  &         3.24  &                             & \citealt{spiniello2019}                             \\
 HE$\,$J0230$-$2130    &  38.1383 & -21.2905  &         2.162 & 0.523 \citep{Eigenbrod2006} & \citealt{wisotzki1999}                              \\
 WISE$\,$J0259$-$1635  &  44.9288 & -16.5953  &         2.16  &                             & \citealt{schechter2018}                             \\
 J0420$-$4037        &  65.1987 & -40.6184  &         2.4   &                             & \citealt{Ostrovski2017}                             \\
 DES$\,$J0602$-$4335 &  90.567  & -43.5945  &         2.92(5)   &                         & \citealt{dawes2023}                                 \\
 J0607$-$2152      &  91.7954 & -21.8715  &         1.302 &                                 & \citealt{stern2021, lemon2023}                          \\
 J0659+1629      & 104.7671  &  16.4858  &         3.09  & 0.766 \citep{stern2021}         & \citealt{Delchambre2019, lemon2023}                     \\
 SDSS$\,$J0832$+$0404  & 128.0711  &   4.06789 &         1.115 &                             & \citealt{oguri2008}                                 \\
 RX$\,$J0911$+$0551    & 137.8647  &   5.84834 &         2.763 & 0.769 \citep{kneib2000}     & \citealt{bade1997}                                  \\
 SDSS$\,$J0924$+$0219  & 141.2325  &   2.32358 &         1.685 & 0.393 \citep{ofek2006}      & \citealt{inada2003}                                 \\
 GRAL$\,$J1131$-$4419  & 172.7473  & -44.3359  &         1.09  &                             & \citealt{kronemartins2018}                          \\
 2M$\,$J1310$-$1714    & 197.5835  & -17.2494  &         1.975 & 0.293                       & \citealt{Lucey2018}                                 \\
 J1537$-$3010          & 234.3556  & -30.1713  &         1.721 &                             & \citealt{lemon2019a, Delchambre2019}                \\
 PS$\,$J1606$-$2333    & 241.5009  & -23.5561  &         1.69  &                             & \citealt{lemon2018b}                                \\
 WGD$\,$J2021$-$4115   & 305.414  & -41.266   &         1.39  & 0.335                       & \citealt{agnello2018}                               \\
 WFI$\,$J2026$-$4536   & 306.5434  & -45.6075  &         2.237 &                             & \citealt{morgan2004}                                \\
 WG$\,$J2038$-$4008   & 309.5113  & -40.1371  &         0.777 & 0.228 \citep{stern2021}     & \citealt{agnello2018}                               \\
 WG$\,$J2100$-$4452    & 315.0619  & -44.8685  &         0.92  & 0.203 \citep{spiniello2019} & \citealt{agnello2019}                               \\
 J2205$-$3727          & 331.4303   & -37.4531  &         1.848 &                             & \citealt{lemon2023}                                 \\
\bottomrule
\end{tabular}
\end{table*}

\subsection{Observing facilities}
The two instruments that observed our targets are multi-charged-coupled device imagers. For our purpose, a single charged-coupled device (CCD) provided a sufficient field of view, so we elected the one with the best read noise characteristics.

\subsubsection{OmegaCAM / VLT Survey Telescope}
The VLT Survey Telescope\footnote{
The ESO Program IDs associated to the observations of the present program at VST are \texttt{106.216P.001}, \texttt{106.216P.002}, \texttt{108.21Z4.001}, \texttt{1103.A-0801(A)}, \texttt{1103.A-0801(B)}, \texttt{1103.A-0801(C)}, \texttt{1103.A-0801(D)}.
} (VST) located at the ESO Paranal Observatory is a wide-field survey telescope with a primary mirror diameter of 2.65 meters. 
It is currently the largest telescope in the world solely dedicated to sky surveys in visible light, with a 1 square-degree field of view.
Its imager is the OmegaCAM, a large 268-megapixel camera, with 32 CCDs. 
Combining the 13 fields observed with this instrument yields a 13 square degrees survey.
We used CCD \#13 only for time delays -- a small fraction of the total area covered by OmegaCAM.

\subsubsection{WFI / MPG-ESO 2.2-meter telescope}
The MPG/ESO 2.2-meter telescope\footnote{
Those associated to the 2p2 are \texttt{098.A-9017(A)}, \texttt{099.A-9021(A)}, \texttt{0101.A-9011(A)}, \texttt{0106.A-9005(A)}, \texttt{0108.A-9005(A)}, \texttt{0109.A-9005(A)}
} (2p2), operational since 1984, is located at the La Silla Observatory. 
Its Wide Field Imager (WFI),
a focal reducer-type camera of 68 megapixels at the Cassegrain focus of the telescope, 
provides a field of view of roughly 0.25 square degree. Of the eight CCDs constituting the WFI, we used CCD \#4.

\subsection{Data acquisition and statistics}
\begin{figure}
    \centering
    \includegraphics[width=0.75\columnwidth]{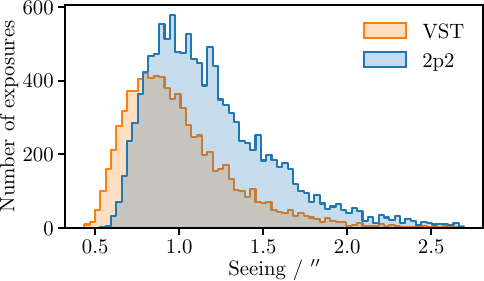}\vspace{-0.2cm}
    \caption{Astronomical seeing distribution in the images used to infer time delays in this data release.}
    \label{fig:seeing-stats}
\end{figure}

The photometric data were obtained from the VST and 2p2 telescopes, via daily monitoring sessions from September 2018 to December 2023. 
Each session consisted of four dithered 320-second exposures through the SDSS-\textit{r} (VST) or ESO-$Rc$ (2p2) filters.
Observations mostly occurred at an airmass below 1.5, with exceptions to extend the visibility window:
this is crucial for longer time delays, for which the shifted seasonal light curves overlap only briefly. 
The median cadence was 1 day for all the 22 monitored targets, which is precisely the objective of this program. 
However, some sessions were missed due to bad weather, technical problems, or maintenance. 
Furthermore, the exposures were filtered in the final step of the data processing based on the seeing, sky level, normalisation of the flux in the image, and shape of the PSF. 
Specifically, at the VST, 8$\,$821 exposures were kept out of a total of 8$\,$944 captured. 
At the 2p2, 12$\,$184 out of 12$\,$637 were kept. 
Thus, the average effective cadences are slightly higher than the medians -- ranging from 1.3 to 2.2 days at the VST, and from 1.3 to 1.8 day at the 2p2.
The median seeings (measured from the images) were $0\farcs91$ at the VST and $1\farcs06$ at the 2p2. Their distribution is shown in Fig.~\ref{fig:seeing-stats}. 
We provide a comprehensive overview of the key observation statistics of the monitored targets in Tables~\ref{tab:obsprops}~and~\ref{tab:obsprops2}.

\section{Light curve extraction}
\label{sec:lc_extraction}
We used the \texttt{lightcurver} Python package~\citep{Dux2024joss} to extract the photometry of our targets. \texttt{lightcurver} provides the infrastructure to apply \texttt{STARRED}~\citep{starred, starredscience} to time-series images.
The method implements a type of generalised PSF fitting method for light curve extraction, whereby we take advantage of the hundreds of dithered observations available for each target to accurately constrain the astrometry of the lensed images, as well as the non-point-source components of the targets.
The process is an extension of drizzling, where a high-resolution model represented on a grid of pixels is forward-fitted to all observations simultaneously -- but, unlike drizzling, the PSF and point-source fluxes here can vary between observations.
The method requires a precise zero-point calibration of the observed frames, which we detail in the paragraphs below and is implemented in \texttt{lightcurver}.
It should be noted that when a target was observed with the two telescopes, the light curve extraction process was carried out once per telescope-dataset due to different pixel scales.

\subsection{Pre-processing and bookkeeping}
Initial steps on the individual exposures included flat-fielding using sky flats, and bias subtraction. 
Flat-fielding is particularly important to achieve the target photometric precision: a constant zero point is needed across the field, because stars are used as references for calculating the relative flux normalisation between epochs.
Next, a conservative sky model was calculated and subtracted from each image using \texttt{sep}~\citep{Barbary2016, sextractor}. 
The subtraction of the sky was straightforward, probably due to our targets all being in extragalactic fields.
An astrometric solution was then found for each image using the \texttt{Astrometry.net} software~\citep{astrometrynet}.
This allowed us to check that the pixel scale (0$\farcs$213 and 0$\farcs$237 per pixel for OmegaCAM and WFI, respectively) was constant across epochs, and also permitted the easy identification and elimination of bad pointings. 
Unlike what was done in previous monitoring campaigns~\citep[e.g.,][]{Courbin2018,Bonvin2019,Millon2020,Millon2020b}, our frames were not interpolated onto a reference one,
because interpolation leads to the loss of sub-pixel information -- making the drizzling-like process mentioned above less effective.

\subsection{Selection of calibration stars}
The success of a light curve extraction depends enormously on the ability to precisely calculate the proper normalisation (zero point) of each image.
In turn, the normalisation requires an excellent PSF model.
Stars suitable for the construction of a PSF model near each lensed target were selected from the \textit{Gaia} catalogues~\citep{thegaiamission, gaiadr3}.
Specifically, we requested that each selected star has a $g$-mag between 16.8 and 19.5 (to avoid saturation given our set-up, while maintaining a high S/N), low variability (\texttt{phot\_g\_mean\_flux\_over\_error} above 100), and is well fitted by a point-source solution in \textit{Gaia} (\texttt{astrometric\_excess\_noise\_sig} below 3).
The (typically) ten closest such stars were elected both as PSF models and flux-normalisation references 
-- depending on the field, within one to three arcminutes away from the target.
Due to dithering and pointing irregularities, not all selected stars were always in the footprints of all frames, but this is not a problem.

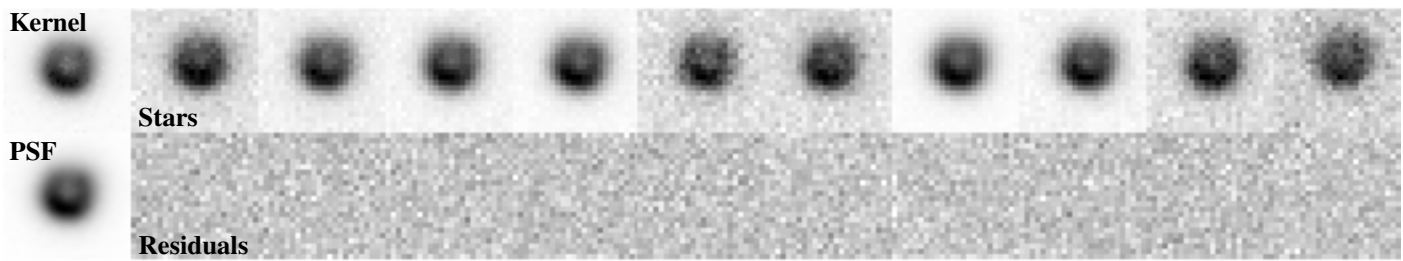
\begin{figure*}[ht]
    \centering
    \begin{overpic}[width=\textwidth,abs,unit=1mm]{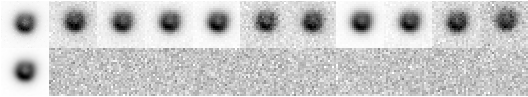}
        \put(1, 30.5){\textbf{Kernel}}
        \put(1, 13.5){\textbf{PSF}}
        \put(18, 18){\textbf{Stars}}
        \put(18, 1){\textbf{Residuals}}
    \end{overpic}
    \vspace{-0.5cm}
    \caption{
    Example simultaneous fit of a PSF model to ten stars, in a (rare) difficult case of unfocused observation.  
    The frame chosen for this figure could still be used for photometry, not leading to an outlier in the light curves.
    The kernel, at the top left, transforms a two-pixel FWHM Gaussian into the PSF at the bottom left. Note that for our lens photometry we used the kernels and not the PSFs (see Sect.~\ref{sec:lightcurveextract} and Fig.~\ref{fig:forwardmodellingmethod}).
    The bottom row shows the residuals of the joint fit.
    Similar residuals were obtained in all conditions of seeing, focusing, and tracking.
    }
    \label{fig:example_psf_fit}
\end{figure*}

\subsection{Empirical PSF modelling}
Cutouts of the stars and lens systems were extracted from each image.
The noise map of each cutout was calculated accounting for Poisson noise (square root of the data in electrons) and read plus sky
background (standard deviation of the noise in the background). 
Masks were implemented by strongly boosting the noise maps at the desired locations. Specifically, masks were needed for eliminating cosmic rays, identified using the \texttt{L.A.Cosmic} algorithm~\citep{lacosmic}, and bad detector rows.
Empirical PSF models were then constructed with \texttt{STARRED}:
\texttt{STARRED} first fits a single elliptical Moffat~\citep{moffat1969} profile to the stars cutouts, then optimises the pixels in a two-times supersampled pixelated grid to fit the remaining residuals. 
The natural positional shifts of the stars (with respect to the grid of pixels of the CCD) allow for the recovery of sub-pixel information, which is why the model is best described on a supersampled grid of pixels.
The overfitting of the noise is prevented by regularising with an isotropic family of wavelets (starlets).
We note that \texttt{STARRED} builds a PSF model, but also and more importantly a kernel that transforms a two-pixel full width at half maximum (FWHM) Gaussian into the PSF. 
We used these kernels to forward-model the lenses, which were also represented on a twice-supersampled pixelated grid with a two-pixel resolution, such that convolving with the kernel would reproduce the PSF of the data (see Sect.~\ref{sec:lightcurveextract} below).
\texttt{lightcurver} and \texttt{STARRED} provide infrastructure for dealing with spatial distortion of the PSF, but our datasets did not require it: there were no significant residuals when fitting the PSF on the stars.
Thus, we did not feel a need to account for colour differences between the stars either. 

Thanks to the precautions taken when preparing star cutouts, the building of a good initial guess based on measured seeing, and the robust optimisation algorithm leveraged by \texttt{STARRED}~\citep[\texttt{adabelief}, ][]{zhuang2020adabelief}, the PSF generation procedure was extremely reliable, even with the occasional out-of-focus or trailed exposures (see Fig.~\ref{fig:example_psf_fit}).
Only a few dozen images had to be discarded because of failed PSF modelling.

\subsection{Relative zero-point calibration}
\begin{figure*}
    \centering
    \scalebox{1.1}{
       \def\svgwidth{0.8\textwidth}
       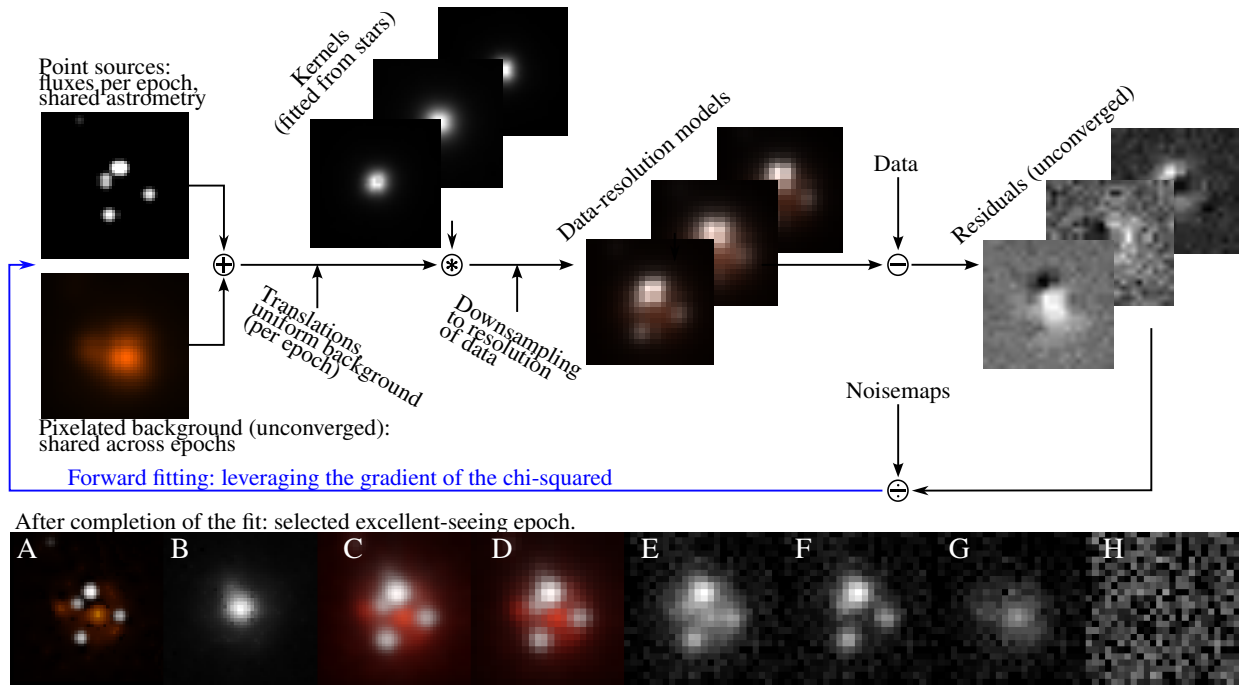
    }   
    \caption{ 
      Overview of the final step of light curve extraction: the forward-modelling of the lens cutouts. Note that the colour coding is only used to differentiate point-source and pixelated components of the models, and does not indicate observations in multiple filters.
      \textit{Top:} Components of the forward-modelling process of the zero-point-calibrated cutouts, starting with a high-resolution model containing a fully tunable grid of pixels and point sources of tunable flux (added to the grid of pixels as two-pixel FWHM Gaussians). The model is degraded to the resolution of the data, and the residuals are minimised through gradient-based optimisation of the high resolution models.
      \textit{Bottom:} Exploring the result of the forward modelling for a selected epoch. A is the fitted high-resolution model. B is the kernel, which transforms a point source in A into the PSF of the data. C is the convolution of A and B, which is then downsampled down to the pixel scale of the data, yielding D. 
      A, C, and D have the background component in red and the point-source component in white.
      E is the data cutout. F and G are again the data, but with the background and point-source components subtracted, respectively. 
      Finally, H shows the residuals upon subtraction of the two components from the data. 
      This method allows for the precise determination of the astrometry of the point sources, reveals the morphology of the non-point source components at high resolution, and provides precise photometry unbiased by flux leakage between point source and non-point source components.
    }
    \label{fig:forwardmodellingmethod}
\end{figure*}

The relative zero points between frames were calculated by comparing the fluxes of the stars that had also served as PSF references. 
To avoid systematic errors, we used the same infrastructure (see Fig.~\ref{fig:forwardmodellingmethod} and paragraph about lens photometry below) to extract both star fluxes and lens fluxes. 
For the stars, however, the pixelated background was set to zero, making the process equivalent to PSF photometry.
For each calibration star, we obtained a time series of flux measurements across all the frames in which it appeared. 
The raw flux time series for each star, $F_{i,j}$ (where $i$ indexes the stars and $j$ indexes the frames), was normalised by its median:
\begin{equation}
F_{i,j}^\text{norm} = \frac{F_{i,j}}{\text{median}(F_{i,:})}\,.
\label{eq:median_normalization}
\end{equation}
Next, we introduced a global scaling factor, $c_i$, for each star to minimise the intra-frame scatter among stars. 
These factors, $c_i$, were determined by minimising the scatter in the weighted mean flux for each frame, $j$, 
subject to the constraint that $\text{mean}(c_i) = 1$. 
The total cost function that the $c_i$ were requested to minimise was
\begin{equation}
\text{Cost} = \sum_j \frac{\sum_i w_{i,j} \left(c_i F_{i,j}^\text{norm} - \mu_j\right)^2}{\sum_i w_{i,j}},
\label{eq:cost_function}
\end{equation}
where the weighted mean for frame $j$ is
\begin{equation}
\mu_j = \frac{\sum_i w_{i,j} c_i F_{i,j}^\text{norm}}{\sum_i w_{i,j}},\quad
w_{i,j} = \frac{1}{\sigma_{i,j}^2}\, .
\label{eq:weighted_mean}
\end{equation}
After applying the optimised scaling factors, the adjusted normalised fluxes of all calibration stars within each frame were combined using a weighted average with $3$-$\sigma$ rejection, yielding the normalisation coefficients, $N_j$.
The frames were brought to the same zero point by dividing their pixels by the $N_j$ coefficients.
This single coefficient per frame neglects the inevitable spatial zero-point variation, 
often caused by flat-fielding (multiplicative) or sky subtraction (additive) errors. 
The hope is that this effect is mitigated by the calibration stars being spread in a small region around the target. 
Nonetheless, it might introduce systematic normalisation errors not fully accounted for by the scatter of the zero points derived from different stars\footnote{
This effect is typically very small (a few millimags), and thus matters for bright targets only. Our photometric precision is dominated by photon noise for most of our targets, see Fig.~\ref{fig:magscattermag}.}.
This scatter, $\sigma_\text{zp}$, was retained for later use in estimating the error bars on the fluxes of the lensed images.

\subsection{Absolute zero-point calibration}
The absolute magnitude calibration of each frame was made using \textit{Gaia} magnitudes of our chosen calibration stars: the \textit{Gaia} magnitudes were approximately converted to the Sloan Digital Sky Survey (SDSS) $r$ filter using the $G - r$ relation given by~\citet[Table A2]{gaiaphotomcalib}.
In the case of the 2p2 data, this introduces a further inaccuracy due to the slight mismatch between the SDSS-$r$ and ESO-$Rc$ filters, 
so, even though the relative night-to-night calibration is very precise, the absolute calibration is by nature approximate.
Overall, the average absolute Vega zero points were 31.44~mag and 31.54~mag for the VST and 2p2 telescopes, respectively, with global scatters (due to observing at different airmasses, with different sky transparencies) of 0.05~mag and 0.15~mag, respectively.

\begin{table}[h!]
\caption{Number of parameters optimised when forward-modelling the lenses. }
\label{tab:forward_model_param_number}
\begin{tabularx}{\columnwidth}{Xll}
\hline
\textbf{Parameter} & \textbf{Count} & \textbf{Case of Fig.~\ref{fig:forwardmodellingmethod}} \\
\hline
Astrometry & $2 N_s$ &  8 \\
Translations & $2 N_e$ &  1720 \\
Point Source Fluxes & $N_s \cdot N_e$ &  3440\\
Mean Background & $N_e$ & 860 \\
Pixelated Background & $(S \cdot N_x)^2$ &  5776\\
\hline
\end{tabularx}
\textbf{Notes:} $N_s$ is the number of point sources, $N_e$ the number of frames, $N_x$ the number of pixels in a side the (square) data cutouts, and $S$ the supersampling factor. In the example of Fig.~\ref{fig:forwardmodellingmethod}, we have $N_s=4$, $N_e=860$, $N_x=38$, and $S=2$.
\end{table}

\subsection{Photometry of lensed images}
\label{sec:lightcurveextract}

We performed the lens photometry using a forward modelling approach to accurately capture all the flux in the vicinity of the targets, without relying on high-resolution imaging for contaminant subtraction. 
Specifically, we constructed a high-resolution model that included point sources with adjustable flux (the quasar light curves) and a constant pixelated background component (absorbing extended photometric contaminants). 
This model was expressed on a twice-supersampled grid of pixels, which matched the supersampling of the PSF models. 
Each pixel in this supersampled grid was allowed to vary: we call this part the `pixelated background'.
On top of it, the point sources were injected as 2-pixel FWHM Gaussians. 
Together, the two components form our `supersampled model'.
The astrometry of the point sources and the pixelated background were common to all epochs\footnote{
This is the part that requires a constant zero point across epochs. Else, the common pixelated background would be washed away by the variation in normalisation.
}.
The point-source fluxes, translations (implemented by interpolation of the supersampled model), and constant background terms were allowed to vary per epoch. 
To compare the high-resolution models to the data, we convolved the models with the kernels, which convert the 2-pixel FWHM Gaussians to the epoch-specific PSFs. 
The resulting images (one per epoch) were then downsampled twice to match the resolution of the data. 
These steps of degrading the supersampled models to the resolution of the data are illustrated in the lower part of Fig.~\ref{fig:forwardmodellingmethod}.
The model parameters were tuned to minimise the chi-squared, computed by summing the squared noise-normalised residuals between data and model-images, across epochs and pixels. 
The number of fitted parameters was typically quite high ($\sim$10$\,$000, see Table~\ref{tab:forward_model_param_number}), 
but not an issue for gradient-based optimisation, which is enabled here by the automatic differentiability of \texttt{jax}, the framework in which \texttt{STARRED} is written.
The top of Fig.~\ref{fig:forwardmodellingmethod} provides a schematic depiction of the process. 
The modelling fully explained the data for all targets (reduced chi-squared$\sim$1) and yielded a flux for each point source, $p$, in a given exposure, $i$: $F_{p,i}$.

\begin{figure}[ht!]
    \centering
    \hspace{-0.3cm}
    \includegraphics[width=8.5cm]{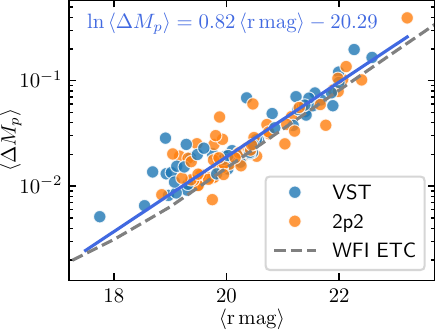}
    \vspace{-0.1cm}
    \caption{Average nightly scatter of the measured lensed-image magnitudes, plotted against the mean magnitude of the lensed image at hand.
    The scatters are empirical, and thereby include photon noise, read noise, deblending errors, and normalisation errors between frames.
    The dashed line is the noise estimate given by the ESO/WFI exposure time calculator (ETC), for a point source in the $r$ band in good seeing conditions (1$''$), average moon illumination, and 320 second exposures.
    The similarity between idealised and empirical errors indicates that our procedure has an excellent deblending performance. 
    For reference, the line fit in blue indicates the measured photometric performance given the magnitude of a given target point source.
    }
    \label{fig:magscattermag}
\end{figure}

\begin{figure*}
    \centering
    \includegraphics{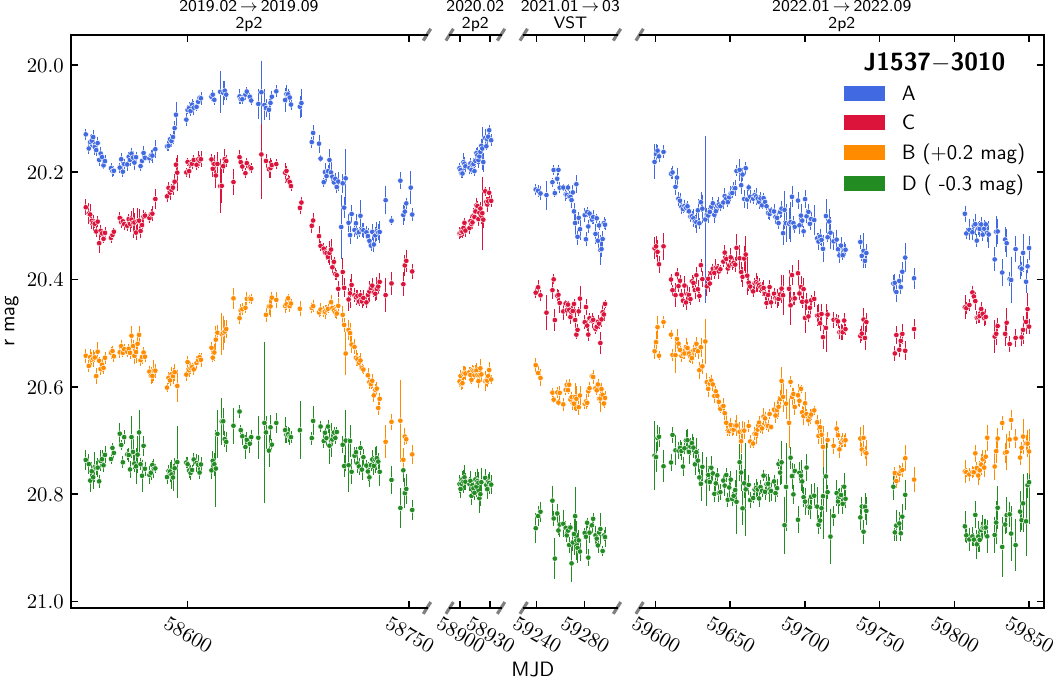}
    \caption{Light curves extracted from the imaging data of J1537$-$3010 (labelling of the lensed images in Fig.~\ref{fig:deconvs}), monitored at both observatories and spanning four seasons, with the 2020 and 2021 ones shortened by the COVID-19 pandemic.
    Note that the empty season gaps are cut from the plot.
    A magnitude offset was added to the individual curves as is indicated in the legend, for display purposes.}
    \label{fig:lcj1537}
\end{figure*}

\subsection{Flux uncertainty estimation}
Again thanks to the autodifferentiable nature of \texttt{STARRED}, we have access to an estimate of the uncertainties of each parameter through the Hessian matrix evaluated at the local minimum reached during optimisation. 
Specifically, we computed the diagonal of the Fisher information matrix, which,
in the absence of deblending issues, yields error bars that are representative of the photon noise error.
Thus, we denote this uncertainty as $\sigma^\mathrm{photon}_{p,i}$: the photon noise on the flux of a single point source. 
This is to be combined with the error on the relative zero point of the frame in question:
\begin{equation}
    \sigma_{p,i} = \sqrt{\left(\sigma_{p,i}^\mathrm{photon}\right)^2 + \sigma_{\rm zp}^2}.
\end{equation}
The final light curves were obtained by combining the measurements (usually four frames, $i=1,2,3,4$) within a single night using a weighted average:
\begin{equation}
    F_p = \frac{1}{Z} \sum_{i=1}^{4} \frac{F_{p,i}}{\sigma_{p,i}},
\end{equation}
with
\begin{align}
    Z &= \sum_{i=1}^{4} \frac{1}{\sigma_{p,i}}, \quad\; \sigma_p = 1 / Z.
\end{align}
$\sigma_p$ is the uncertainty on the flux of point source $p$ on a given night. 

We also kept track of the standard deviation within $F_{p,i}$, denoted hereafter by $\Delta F_p$, as it provides a more empirical estimation of the uncertainty. 
We denote the same quantity, but in units of magnitude, $\Delta M_p$.
The median $\sigma_p$ over the entire dataset is 0.017 mag, and 0.020 mag for $\Delta M_p$. 
This slightly higher observed scatter compared to photon noise and normalisation errors can be attributed to a few very narrowly separated objects, 
in which flux leakage between point sources occurs. 
These are discussed on a case-by-case basis in Appendix~\ref{subsec:discussion_individual}.
Figure~\ref{fig:magscattermag} shows what uncertainty can be expected in a night of monitoring, for a single lensed image of a given magnitude.  
For the purpose of time-delay determination, we took our final uncertainty on the magnitude of a point source on a given night as the average of $\Delta F_p$ and $\sigma_p$. 
This prescription qualitatively increased the robustness of the error bars compared to the local scatter in the extracted light curves.
As an example, we show the extracted light curves of J1537$-$3010 in Fig.~\ref{fig:lcj1537}, while the others are delegated to Appendix~\ref{appendix:lightcurves}.

\begin{figure*}
    \centering
    \vspace{-0.4cm}
    \includegraphics[width=\textwidth,height=\textheight,keepaspectratio]{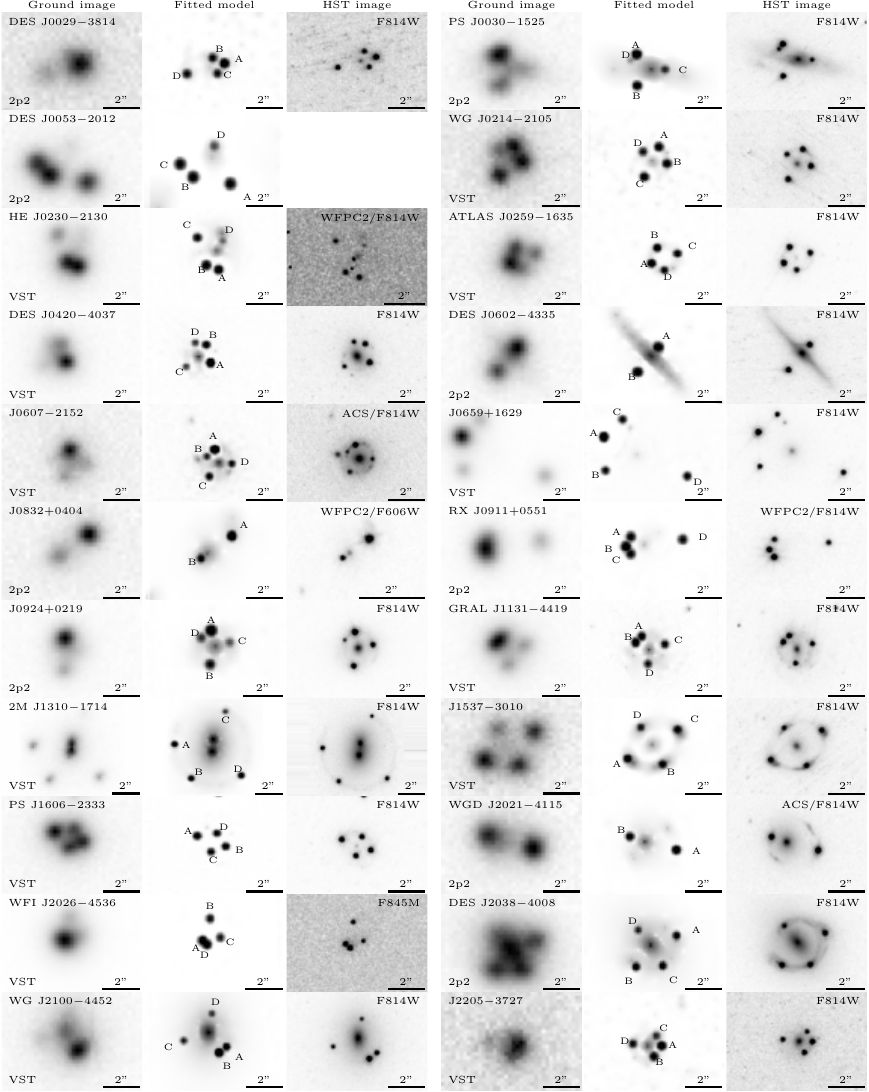}
    \vspace{-0.4cm}
    \caption{Imaging preview of all the monitored objects. For each lens, we include a representative exposure from the dataset with median seeing and low sky background, the high-resolution fitted model (see Sect.~\ref{sec:lightcurveextract}), and, when available, an HST/Wide Field Camera 3 (WFC3) image for comparison.
    The high-resolution models live on a grid twice-supersampled compared to the original images, with a two-pixel FWHM PSF. 
    The resolution (FWHM of the PSF) in the models is thereby 0$\farcs$213 for VST data, and 0$\farcs$237 for 2p2 data. 
    This is to be compared with the 0$\farcs$07 resolution of the HST/WFC3 images. 
    Note also that the F814W filter is much wider and redder than the $r$ band, such that more details can be revealed in the redshifted objects of our fields.
    All cutouts are oriented with north facing up and east facing left.
    }
    \label{fig:deconvs}
\end{figure*}

\begin{figure}[ht!!!!]
    \centering
    \includegraphics[width=0.95\columnwidth]{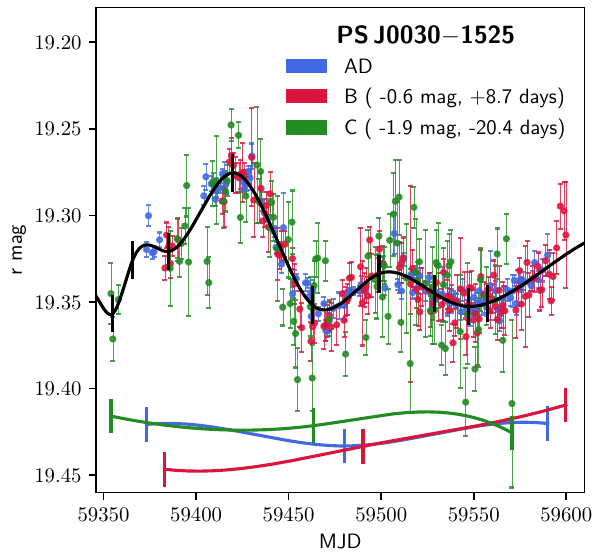}\vspace{-0.25cm}
    \caption{Curve matching example, in which a single spline (in black) matches all curves simultaneously.
    Due to microlensing in particular, which adds independent lower-frequency variations to each curve, the curves could not be matched in this way with time and magnitude shifts only. Thus, an extra modulation is applied to each separately.
    These modulations are represented by lower order splines, displayed at the bottom.
    Physically, the black spline models the intrinsic variations of the source, while the coloured splines undo the effects of microlensing.
    Linking with the notation of Sect.~\ref{sec:measure_tds}, $M_{\rm ext}$ is here a cubic spline with a single internal knot of fixed position, and $N_\text{int}=9$. 
    To determine a time delay and its uncertainty, this matching was repeated thousands of times with artificially further-modulated curves (mimicking microlensing), other realisations of the noise, and different freedoms given to the splines.
    What this matching is like for the other targets can be found in a Jupyter notebook
    \href{https://github.com/duxfrederic/TDCOSMO_XVII_time_delays/blob/main/initial_guess.ipynb}{\includegraphics[height=1em]{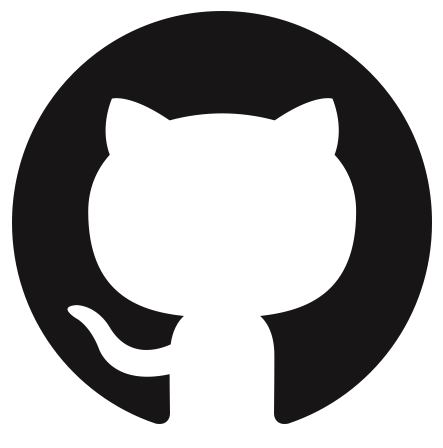}}. This notebook is part of the repository containing the photometry extracted herein and the code to estimate the time delays.
    \label{fig:example_matched_curves}}
\end{figure}

\subsection{The result: Light curves and models}
Besides the light curves themselves, the second products of our method are the high-resolution models fitted to all epochs simultaneously. 
These were fitted blindly together with the photometry, without prior knowledge of the astrometry or using existing \textit{Hubble} Space Telescope (HST) images as a prior.
Despite this, the fitted models accurately capture the morphology of Einstein rings, lensing galaxies, or nearby faint galaxies, while also recovering the HST astrometry with excellent precision (see Sect.~\ref{sec:astrometric_precision}).
The models are shown in Fig.~\ref{fig:deconvs}, in which we provide for each lens the labelling of the lensed images, a typical data frame from the telescope, the fitted high-resolution model, and an HST image when publicly available -- for many, HST imaging only became available after the photometric extraction was conducted.
The faithfulness of the high-resolution models, together with the reduced $\chi^2$ of the fits ($\sim 1$ in all cases), shows that the relative zero points between epochs were calibrated with high precision, and that the PSF models were well estimated in all epochs.

\section{Measuring time delays}
\label{sec:measure_tds}
Different methods have been proposed for the purpose of estimating time delays between light curves, such as autocorrelation~\citep{Press1992}, or minimising the dispersion of curve differences~\citep{Pelt1996}. 
These methods are fast, but they cannot properly account for complex extrinsic variations imprinted on the individual light curves. 
Extrinsic variations are always present in lensed quasar optical light curves, mainly due to the microlensing of the individual lensed images by stars of the lensing galaxy.
Thus, superimposing light curves requires accounting for both the intrinsic variation of the quasar, common to all the lensed images, and extrinsic variations individual to each lensed image.

\subsection{Matching curves with splines}
A purely data-driven method doing exactly this, which was proven to work well in the 2014 time-delay challenge~\citep{liao2014} as has been reported by~\citet{bonvin2016tdc}, is the simultaneous alignment of the curves using free-knot splines. 
In this method, each curve is given an extrinsic model of variation that modifies its magnitude in a smooth way. 
The modified curves are also shifted in time, until they all match a single free-knot spline representing the intrinsic variation in the quasar.
All parameters -- of the splines and the shifts -- are optimised simultaneously, until a match is found. 
An example is shown in Fig.~\ref{fig:example_matched_curves}, in which three curves were matched to a single spline with timeshifts and modulations.
This is the strategy we used herein, as is implemented in the PyCS3~\citep{millon2020pycs3} toolbox, which uses polynomials or a free-knot B-spline~\citep{MOLINARI2004} to represent curves.
The general methodology is given in the third figure of~\citet{Millon2020}, with the exception being that we did not include the regression difference method in the final estimates.
The splines were indeed found to be a more robust and accurate method than regression differences on simulated data~\cite[see][sec. 2.4.4]{martin_phd_thesis}.
The extrinsic variations were represented with polynomials or cubic splines, depending on the complexity of the light curves. 
In order of complexity, the extrinsic models, $M_\mathrm{ext}$, included polynomials of degree  one (a linear function) or three (a cubic polynomial), a cubic spline with one internal knot forced to the centre, or free-knot cubic spline with one or more internal knots.

However, not all extrinsic models were used on all sets of curves. 
Often, the simplest modulations would not have allowed for a proper matching of the curves: in such cases, strong systematic errors would have dominated the error budget due to the imperfect fit. 
Similarly, for curves with both little intrinsic and extrinsic variations, electing a complex extrinsic modulation would have resulted in a completely degenerate model, ultimately yielding overestimated uncertainties.
Hence, for each dataset we started with the simplest extrinsic model that permitted a good alignment of the curves, and included two more levels of complexity. 
For example, if a set of curves could be well aligned (that is, matched within their noise envelope over their entire length) with splines of one internal knot, splines with two and three internal knots were added to the pool of possible extrinsic variation models as well.

Together with the extrinsic models, the intrinsic variation of the quasar was modelled with a free-knot cubic spline, 
with $N_\mathrm{int}$ internal knots.\footnote{
Another meta-parameter of the intrinsic spline, besides the number of internal knots $N_\mathrm{int}$, is the minimum allowed spacing between two knots.
It was set to 10 days, allowing the capture of the finest features of all the curves presented herein within their noise.
}       
A first value of $N_\mathrm{int}$ was selected such that the finest features of the best curve could be well captured by the spline. 
Next, $N_\mathrm{int}$ was decreased until a certain smoothing of the finest features would be noticed. 
In the interval between the two values, an additional $N_\mathrm{int}$ value was added to the pool of possible models.
We call a pair of models $(N_\mathrm{int},\,M_\mathrm{ext})$ an estimator, and as such end up with a grid of nine estimators, with three different intrinsic splines of $N_\mathrm{int}$ internal knots, and three models of extrinsic variations, $M_\mathrm{ext}$.

\subsection{Microlensing bias and error estimation}
To estimate the reliability and uncertainty of a time-delay estimation, the inference was performed repeatedly with each estimator on mock curves with distinct realisations of the observed noise.
Importantly, red noise mimicking the effects of microlensing is injected into the mock curves: this essentially transforms a systematic error -- arising from fitting microlensing, which distorts the curves in a way that can bias the time-delay estimates -- into a statistical error.
The main estimate and its uncertainty can then be read from the histogram of optimised time delays.
The separate pools of delays measured from mocks (one per estimator) are then combined as is described in Sect.~3.3~of~\cite{Millon2020}, 
with the tension parameter, $\tau$~\citep{Bonvin2018}, set to $0.5$.
This provides a way to combine estimates from different microlensing and intrinsic variation models, 
halfway between selecting the estimate with the best precision and marginalising over all estimators, and effectively eliminates outliers.
This process can also lead to asymmetrical error bars, but on its own cannot account for the covariance between delay pairs.

\subsection{Covariance between delays}
A quadruply lensed quasar provides six independent pairs of curves from which a delay can be measured. 
However, the strategy of PyCS3 is maximising the available signal in each temporal bin by shifting all four curves together: implying only three independent time-shifts. 
As such, a covariance is bound to exist between different delay pairs, 
and can slightly change the overall precision in the application of TDC.

To compute our covariance matrices, we assume that the delays obtained from shifting mock light curves are normally distributed: this is a good approximation, but the mocks further than 4$\sigma$ from the mean were discarded as to not artificially increase the error bars because of failed fits.
Then, the covariance is read from its definition:
\begin{equation}
    \label{eq:covariance}
    \mathrm{cov}(p_1, p_2) = \mathbb{E}\left[\left(p_1 - \mathbb{E}[p_1]\right)\left(p_2 - \mathbb{E}[p_2]\right)\right].
\end{equation}
This was computed on the pools of mocks that passed the $\tau=0.5$ selection mentioned above.
We also made sure that the square root of the diagonal -- the standard deviations, $\sigma_d$, on individual delays, neglecting covariance -- were always within 20\% of the confidence interval read directly from the histograms of mocks; that is,
\begin{equation}
    \sigma_d \sim \frac{P_{84} - P_{16}}{2},
\end{equation}
where $P_n$ are the $n^\mathrm{th}$ percentiles read from the histograms of mocks.
This ensured that the standard deviations given by the diagonal of the covariance matrix are indeed representative of the actual widths of the distributions of the delays optimised on mock curves.
Moreover, any bias in the median value (recovered versus injected) of the mock-optimised time-delays was added in quadrature to the diagonal of the covariance matrix, but this contribution was always negligible compared to the statistical uncertainties.
We note that the resulting covariance matrix is ill-conditioned: for a quadruply lensed quasar,
it is a $6\times 6$ matrix of rank 3 (due to dealing with only three independent shifts).
Nonetheless, for convenience, we provide all six delays together with their singular $6\times6$ covariance matrix. 
In practice, one should choose a reference lensed image and keep only the three delays relative to it, with the corresponding $3\times3$ covariance sub-matrix.

\section{Results}
\label{sec:results}
Overall, this set of TDCOSMO lenses has proven more challenging than lenses of previous COSMOGRAIL publications, in particular due to the fainter (20.2 average magnitude versus 19.2 in \citeauthor{Millon2020b}~\citeyear{Millon2020b}) and more compact lensing configurations, all the brighter, well-separated targets having been monitored for time delays in the past.
Nevertheless, at least one reliable time delay could be inferred for each monitored lens.
Still, this shows that obtaining reliable photometry of even fainter and more compact lensed quasars will become more difficult, at least with seeing-limited observations.
Not all the targets presented in this work have been monitored for the first time: four were already discussed in COSMOGRAIL publications; namely, 
HE$\,$J0230$-$2130,
J0832+0404,
WFI$\,$J2026$-$4536 by~\citet{Millon2020},
and WG$\,$J0214$-$2105 as well as PS$\,$J1606$-$2333 by the previous data release of the present program~\citep{Millon2020b}.

We delay the discussion of the challenges encountered in each target, as well as delay values and covariance matrices, to Appendix~\ref{subsec:discussion_individual}. 
Here, we provide a summary of the relative uncertainties in each target in Table~\ref{tab:delays}, 
and the labelling of the lensed images can be found in Fig.~\ref{fig:deconvs}.

\begin{table*}
\caption{Systems and their most precise delay, as well as precision estimates resulting from the combination of independent delays in a given system. \label{tab:delays}}
\begin{tabularx}{\textwidth}{>{\raggedleft\arraybackslash}X c r r r}
\toprule
System & Most precise delay pair & Most precise delay value (days) & Worst case precision & Best case precision \\
\midrule
DES$\,$J0029$-$3814 & C$\,$D & $49.9 \pm 2.7$ & 5\% & 2\% \\
PS$\,$J0030$-$1525 & B$\,$C & $-28.5 \pm 3.5$ & 12\% & 7\% \\
DES$\,$J0053$-$2012 & A$\,$D & $-90.2 \pm 6.7$ & 7\% & 3\% \\
WG$\,$J0214$-$2105 & B$\,$C & $15.7 \pm 0.7$ & 5\% & 2\% \\
HE$\,$J0230$-$2130 & C$\,$D & $-45.0 \pm 4.5$ & 10\% & 4\% \\
ATLAS$\,$J0259$-$1635 & B$\,$C & $17.0 \pm 1.5$ & 9\% & 4\% \\
DES$\,$J0420$-$4037 & A$\,$C & $7.9 \pm 2.6$ & 33\% & 11\% \\
DES$\,$J0602$-$4335 & A$\,$B & $23.6 \pm 2.1$ & 9\% & 9\% \\
\textit{(1) J0659+1629} & \textit{B$\,$D} & $\mathit{277.0 \pm 11.4}$ & \textit{4\%} & \textit{1\%} \\
\textit{(2) J0659+1629} & \textit{B$\,$D} & $\mathit{317.3 \pm 13.4}$ & \textit{4\%} & \textit{2\%} \\
J0607$-$2152 & C$\,$D & $-29.7 \pm 3.4$ & 11\% & 7\% \\
SDSS$\,$J0832+0404 & A$\,$B & $-128.0 \pm 7.8$ & 6\% & 6\% \\
RX$\,$J0911+0551 & C$\,$D & $160.0 \pm 8.9$ & 6\% & 2\% \\
SDSS$\,$J0924+0219 & B$\,$C & $-21.5 \pm 3.5$ & 16\% & 7\% \\
GRAL$\,$J1131$-$4419 & C$\,$D & $-13.6 \pm 0.8$ & 6\% & 3\% \\
2M$\,$J1310$-$1714 & A$\,$B & $-55.9 \pm 1.5$ & 3\% & 1\% \\
J1537$-$3010 & B$\,$C & $37.7 \pm 0.8$ & 2\% & 1\% \\
PS$\,$J1606$-$2333 & A$\,$C & $-28.9 \pm 1.4$ & 5\% & 2\% \\
WFI$\,$J2026$-$4535 & AD$\,$B & $16.3 \pm 3.2$ & 19\% & 12\% \\
WGD$\,$J2021$-$4115 & A$\,$B & $-90.8 \pm 9.7$ & 11\% & 11\% \\
WG$\,$J2038$-$4008 & A$\,$D & $-33.3 \pm 6.3$ & 19\% & 8\% \\
WG$\,$J2100$-$4452 & C$\,$D & $-12.3 \pm 1.5$ & 12\% & 7\% \\
J2205$-$3727 & A$\,$D & $12.3 \pm 1.8$ & 15\% & 7\% \\
\bottomrule
\end{tabularx}
\textbf{Note:} The J0659+1629 rows are emphasised to highlight the multimodality suffered by the delays of this system. The best and worst case precisions are given by Eq.~(\ref{eq:best_uncertainty}) and (\ref{eq:worst_uncertainty}) respectively. 
\end{table*}

\subsection{Uncertainty propagating to the Hubble constant}
\label{sec:uncertainty_propagation}
The time-delay part of the $H_0$ error budget is roughly directly given by the uncertainty of the time delay itself. 
This is a valid working assumption, but the situation is complicated when several delays are available. 
In the absence of correlation, we could argue that a quadruply lensed quasar provides three fully independent delays.
These can then be combined as independent Gaussian estimates, with the resulting uncertainty scaling as
\begin{equation}
    \label{eq:best_uncertainty}
    \mathrm{relative\;uncertainty} \sim \frac{1}{\sum_i {\frac{\tau_i}{\Delta\tau_i}}},
\end{equation}
where the $\tau_i$ and $\Delta\tau_i$ are three individual delay and uncertainty pairs.
This is a benchmark of the best possible achievable precision, and works in the absence of correlation between delays.
On the other hand, if the delays are closely correlated, then the relative uncertainty that maps to $H_0$ is roughly that of the most precise delay:
\begin{equation}
    \label{eq:worst_uncertainty}
    \mathrm{relative\;uncertainty} \sim \min_j \left(\frac{\Delta\tau_j}{\tau_j}\right) \,,
\end{equation}
where $j$ denotes all possible delay pairs.
It is likely that the actual uncertainty mapping to $H_0$ in a TDC analysis will fall somewhere between the two above estimates.
We give both estimates and the most precise delay for each system in Table~\ref{tab:delays}.

\section{Discussion}
\label{sec:discussion}
\subsection{A look back on intrinsic and extrinsic variability}
\begin{figure}[h]
    \centering
    \includegraphics[width=0.85\columnwidth]{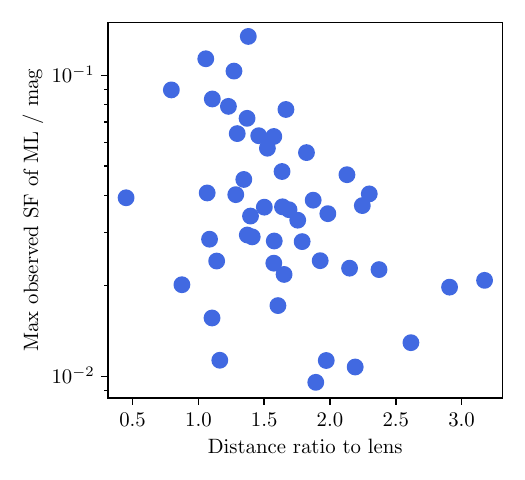}\vspace{-0.3cm}
    \caption{Maximum observed SF of the microlensing curves, plotted against the distance to the lens of the lensed image divided by the effective radius of the lensing galaxy.}
    \label{fig:ML_vs_distance}
\end{figure}

This program aimed for a very high cadence of monitoring and high photometric precision per epoch so that extrinsic variations could be better disentangled from the faster intrinsic variation of the quasar. 
To verify whether this was realised, we provide an estimate of the intrinsic variability of each source quasar, and a comparison to that of the extrinsic variability of their lensed images.
We turned to the structure function (SF) to measure the variability of a light curve:
\begin{equation}
    \mathrm{SF}(\Delta\tau)=\sqrt{\left<\big(m_{j}-m_{i}\big)^{2}-\sigma_{i}^{2}-\sigma_{j}^{2}\right>_{t_{i}-t_{j}\in[0,\Delta\tau)}} \;\;,
    \label{eq:SF}
\end{equation}
where $i$ and $j$ denote epochs.
So, for each bin of lag time $\Delta\tau$, the SF is the average squared differences of points of magnitude, $m$. 
We also subtracted the squared uncertainties in each point to avoid an artificial contribution to the SF from noise.
However, given that all lensed images will show at least some extrinsic variability, one cannot access the true intrinsic variability of the quasar: 
we were instead forced to elect a reference lensed image, in which we assumed no extrinsic modulation.
An initial working hypothesis is that the further a lensed image is from the halo of stars of the lensing galaxy, the less microlensing it will experience on average.
Hence, assuming that the estimated delays are correct, microlensing curves were calculated by subtracting the furthest-image curve from shifted and interpolated versions of the other light curves.
If the initial working hypothesis is correct, then we should observe a further dependence of the amount of variability in the microlensing curve as a function of distance to the lens.
This was realised in our light curves, as is shown in Fig.~\ref{fig:ML_vs_distance}: the maximum observed SF of extrinsic variations tends to be lower when the lensed image is further from the lensing galaxy.

\begin{figure*}
    \centering
    \includegraphics[width=\textwidth]{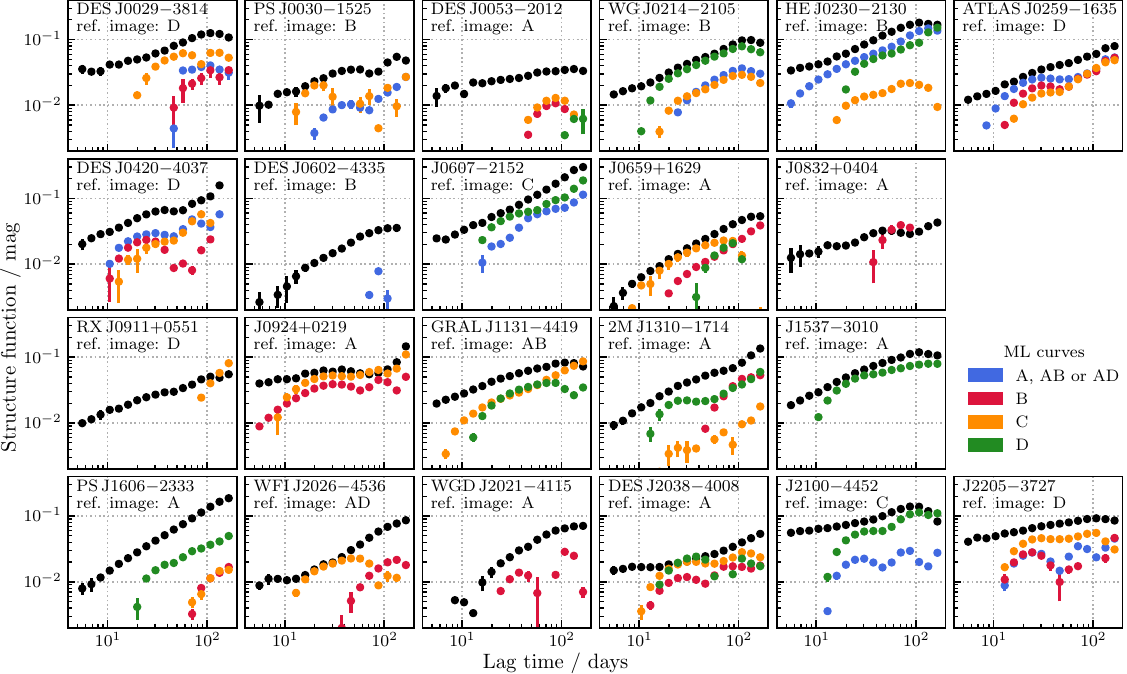}
    \caption{
       Structure functions in the observed frame, of the best estimate of intrinsic variability of the source quasars (in black), as well as that of the extrinsic variations mainly due to microlensing.
       The absence of markers at certain lag times indicate that any potential variation was below the noise level. 
       The error bars were estimated by bootstrapping. 
       The SF of the extrinsic variations only catches up with the intrinsic SF at later time lags, 
       which this program of high frequency monitoring takes advantage of by being able to probe lag times below 10 days.
       The SF is cut at 200 days, the characteristic duration of our monitoring seasons.
     }
    \label{fig:structure_functions}
\end{figure*}

Next, the quasars' `intrinsic' curves were built by aligning all the curves with their respective delay, and modulating them with the smoothest extrinsic variability model used in the time-delay estimation so that they fitted the  curve of the reference image. 
The SFs were then calculated after Eq.~(\ref{eq:SF}); they are displayed in Fig.~\ref{fig:structure_functions}. 
We see that the intrinsic variations in the source quasar (in black) are mostly dominant at short lag times, a regime that can only be probed with high-frequency and -precision monitoring.
However, extrinsic variation can also operate on short timescales. 
An example is HE$\,$J0230$-$2130, in which a sudden rise was observed in image A at around MJD 60$\,$275 (Fig.~\ref{fig:lcsHE0230-2130}). 
This event is visible in the SFs as well, with that of A rising at much lower lag times than that of the other lensed images. 
Fast events observed in a single light curve only are common, and while they could enable source science by combining light curves in multiple colours and a mass model of the lens, they also complicate the time-delay estimation, especially in regimes with lower sampling frequencies.

\subsection{Comparison with LSST}
In terms of depth, this program has remarkable similarities to the upcoming LSST. 
The 5-$\sigma$ depth of a single LSST visit (two back-to-back 15 seconds exposures) is expected to be about 24.5 mag in the $r$ band~\citep{lsst_sim_forecast}, similar to the empirical depth measured on single visits of this program (24.5 mag, in 4$\times$320 seconds exposures). 
Overall, this indicates that the per-visit photometric precision and performance in deblending can be expected to be identical.
Next, the total number of visits including a given target is expected to be about 180 in the $r$ band for LSST over the 10-years operation period. 
The mean total number of visits per target of this program is 227, ranging from 85 (WFI$\,$J2026$-$4536) to 435 (GRAL$\,$J1131$-$4419).
These are again comparable figures, with LSST falling slightly short but making up for it by providing this number of visits multiple times in different filters.
The main difference, however, is cadence, with visits in a given filter spaced by about 20 days on average for LSST, compared to the much faster one-day cadence of this program.
Looking at the SFs of Fig.~\ref{fig:structure_functions}, we see that even with 20-day sampling, the intrinsic variability of our pool of lensed quasars is still much higher than that of microlensing in about half the sample, likely permitting the capture of large and fast variations in multiple images.

The other half will fall into a regime similar to that of the original COSMOGRAIL strategy, in which the contributions of extrinsic and intrinsic variation can only be disentangled thanks to the longer baseline.
LSST will still have the advantage of observing in multiple filters: with the prior knowledge of how intrinsic variation transforms from band to band (shift and distortion, see, e.g., \citeauthor{chan2020twistedlcs}~\citeyear{chan2020twistedlcs}), one might be able to disentangle the features due to actual intrinsic variation (to be matched when shifting curves), and those due to extrinsic variation (to be ignored for the purpose of time-delay estimation).
However, this would introduce a dependence on our understanding of the geometry of the source quasar to hand.
It would also require additional work on the side of time-delay estimation and modelling and would not be as straightforward as the shifts permitted by the high-frequency monitoring of this program.

\subsection{Astrometric precision and lens models}
\label{sec:astrometric_precision}
Currently, high-quality time series ground imaging is a lot more expensive than high-resolution imaging. This will not hold true any longer once LSST starts operations -- in the southern sky at least.
Thus, being able to vet lensed quasars from ground-based time series imaging -- that is, evaluate their suitability for cosmography or other applications -- would be useful.
For this, we need (i) to detect lensing features that can be used to constrain a lens model, which we showed in Fig.~\ref{fig:deconvs}, and (ii) accurate astrometry for preliminary lens modelling.
We compare the astrometry of the lensed images derived from our high-resolution model fitting with measurements from HST imaging data reported in Table~3~of~\citet{schmidt_automatic_modelling}. 
We find a root-mean-squared deviation of 11 milliarcseconds (mas), which roughly matches the uncertainties reported by~\citet{schmidt_automatic_modelling} (6$\,$mas per axis).
In this sample of lenses, only one problematic case was found: PS$\,$J0030$-$1525, for which the position of image D was off by 100$\,$mas. 
This was caused by the narrow separation between the bright image A (r $\sim$ 19.3) and faint image D (r $\sim$23.2), as well as contamination by the bright lensing galaxy.
Thus, our method achieves a level of precision comparable to HST imaging in non-pathological cases -- the pathological ones being easily identifiable. 

We also suggest the possibility of fitting lens models directly on the ground imaging time series data. 
This is achievable by connecting, for example, \texttt{STARRED} and Herculens~\citep{Galan2022_herculens}: given a lensing model, Herculens would predict the lens-plane image, while \texttt{STARRED} would provide a chi-squared to minimise by comparison with the dithered time series of imaging, given the PSF of each frame.
Because both code bases are autodifferentiable, the gradient of the lensing parameters can be back-propagated from the chi-squared to the lensing parameters. This strategy will be the subject of future work.

\subsection{Identifying lensed quasars through matched variability}
The infrastructure of light curve extraction presented in this work is largely automated and can scale up to a large number of blended targets.
It automatically selects calibration stars from \textit{Gaia}, adapts to those available in each epoch, and accounts for frame rotation and PSF distortion, making it ready for the future LSST. 
Thus, for lensed quasar searches where spectroscopic follow-up is the limiting factor, the present infrastructure can leverage imaging time series to create pure samples of lensed quasars by selecting candidates exhibiting identical (modulo time delay) variations in all their point-like components.

\section{Conclusions}
\label{sec:conclusions}
We presented new time delays in 22 lensed quasars, a sample constituting the faintest and most narrowly separated targets for which time delays could be measured to date.
Of the 22 systems, 8 are TDC-ready with a time-delay error budget below 5\%, and 9 will likely contribute errors comparable to  those of the other components of a TDC analysis. 
Regarding infrastructure, the improved photometric pipeline enables an extraction of light curves limited by photon noise only, for separations between lensed images down to 0$\farcs$5, which is an improvement over the work presented in~\cite{Millon2020} in which deblending errors still play a role for lower-separation objects.
Previously measured delays from shorter light curves have also been confirmed, with us finding compatible results even in unlucky cases in which the first season was strongly affected by strong microlensing. This demonstrates the robustness of our time-delay measurements and error estimation techniques.
We have also shown how the careful matching of the zero points of the wide-field frames and forward-fitting  the pixels can yield high-resolution models of the targets, compatible with HST observations in morphology and astrometry.
In the times of LSST, the capacity for high-resolution follow-up imaging will be dwarfed by the number of unveiled targets, but each will have been imaged from the ground hundreds of times.
Thus, the ability to reliably convert a deep, seeing-limited survey to a high-resolution image will become very valuable.

\begin{acknowledgements}
The access to the 2.2m telescope was made possible through an agreement with the Max Planck Institute for Astronomy (MPIA):
we would like to thank Thomas Henning, Hans-Walter Rix, and Roland Gredel for their help in making the agreement possible.
We thank the European Southern Observatory (ESO) for accommodating the use of the VLT Survey Telescope (VST) amid the COVID-19 pandemic constraints.
COSMOGRAIL is supported by the Swiss National Science Foundation (SNSF) and by the European Research
Council (ERC) under the European Unions Horizon 2020 research and innovation programme 
(COSMICLENS: grant agreement No 787886). This project has received funding from the European Union’s Horizon Europe research and innovation programme under the Marie Sklodovska-Curie grant agreement No 101105725.
Paul Schechter generously allowed us to add his weekly-spaced points of J$\,0924$+$0219$ captured at the 2p2 to our time-delay estimation.
The data analysis was made accessible by the Python ecosystem, relying heavily on libraries such as 
Numpy~\citep{harris2020array}, 
Scipy~\citep{2020SciPy-NMeth},
Astropy~\citep{astropy:2013, astropy:2018},
and
Matplotlib~\citep{Hunter:2007}.
\textbf{Gaia:}$\;$
This work has made use of data from the European Space Agency (ESA) mission
{\it Gaia} (\url{https://www.cosmos.esa.int/gaia}), processed by the {\it Gaia}
Data Processing and Analysis Consortium (DPAC,
\url{https://www.cosmos.esa.int/web/gaia/dpac/consortium}). Funding for the DPAC
has been provided by national institutions, in particular the institutions
participating in the {\it Gaia} Multilateral Agreement.
TA acknowledges support from ANID-FONDECYT Regular Project 1240105, ANID Millennium Science Initiative AIM23-0001 and ANID BASAL project FB210003.
Support for MR is provided by the Direcci{\'o}n de Investigaci{\'o}n of the Universidad Cat{\'o}lica de la Sant{\'i}sima Concepci{\'o}n with the project DIREG 10/2023. MM acknowledges support by the SNSF (Swiss National Science Foundation) through mobility grant P500PT\_203114 and return CH grant P5R5PT\_225598.
CDF acknowledges support for this work from the National Science Foundation under Grant No. AST-2407278 and the UC Davis College of Letters and Sciences.
V.M. acknowledges support from ANID FONDECYT Regular grant number 1231418, Millennium Science Initiative, AIM23-0001, and Centro de Astrof\'{\i}sica de Valpara\'{\i}so CIDI N21.
\end{acknowledgements}

\bibliographystyle{aa} 
\bibliography{aa} 
---------------------------------------------------------

\appendix

\section{Notes about individual objects and time delays}
\label{subsec:discussion_individual}
Below we comment on the data, difficulties and results regarding each target. 
Note the convention of our time delays: a positive delay implies the second image arrives first. For example, $\mathrm{A}\,\mathrm{B}=10$~days means that a given feature will first be seen in B, and will arrive 10 days later in A. 
We also estimate how the uncertainties in the obtained delays will map into $H_0$ by providing a benchmark precision: that is, we take all the delays involving one lensed image, and combine their relative precision assuming Gaussian uncertainties.
The ideal scenario is reaching a below 5\% combined precision, as this is where the time-delay errors become subdominant compared to that of the modelling and line-of-sight components of TDC~\citep{suyu2014, suyu2017, wong2020}.
Note that this only is a benchmark estimation of the precision that will map to $H_0$, 
while in reality the covariance between delays will also play a role in the cosmographic inference (see Sect.~\ref{sec:uncertainty_propagation}).

\begin{figure}
    \centering
    \includegraphics[width=0.49\textwidth]{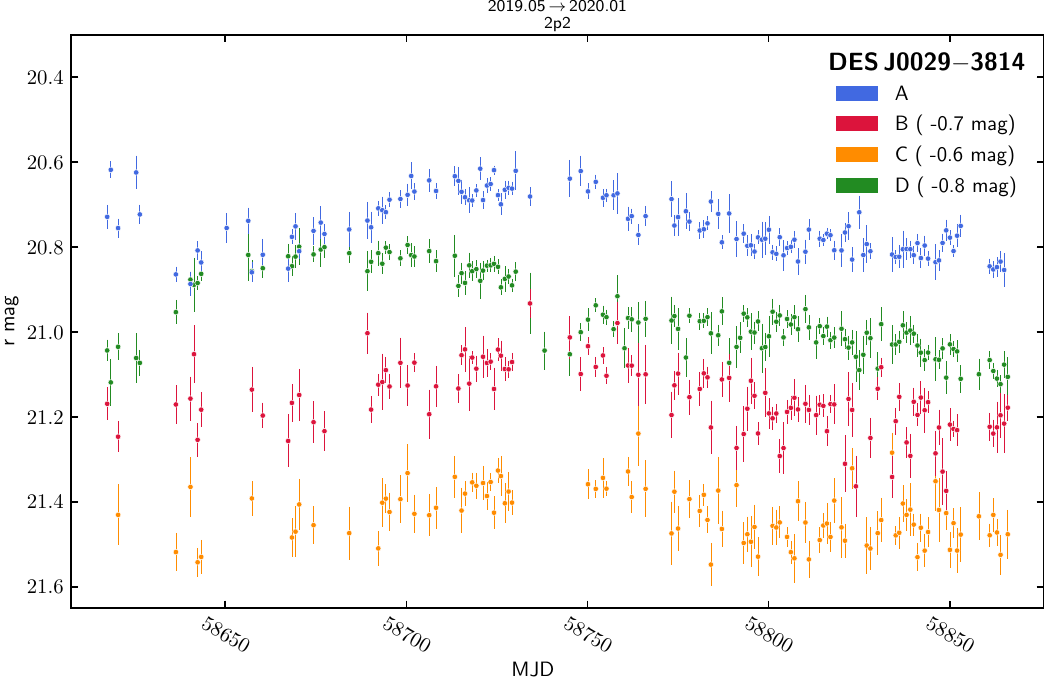}
    \caption{\appendixlccaption{DES$\,$J0029$-$3814}}
    \label{fig:lcsDESJ0029-3814}
\end{figure}

\paragraph{DES$\,$J0029$-$3814}
This quadruply lensed quasar~(Schechter et al. in prep.) 
has source redshift $z_{\rm s}\sim2.81$.
It was monitored at the 2p2 telescope from May 2019 to January 2020. 
The resulting one-season-long light curves, visible in Fig.~\ref{fig:lcsDESJ0029-3814}, have well defined long term variations which allow a relatively precise alignment. 
The low frequency nature of the observed variation makes it degenerate with potential microlensing, such that 
different freedoms afforded to the extrinsic modulations yield slightly different solutions.
Some higher frequency variations are also observed, but their amplitude close to the noise level makes them unable to significantly 
break the microlensing degeneracy.  
This is reflected in the uncertainties we claim on the different delays.
Overall, as expected because of the symmetry axis of the system, we observe a very small delay between B and C, and a larger delay between D and the trio A, B and C.
The best precision on a single delay is 7\% for B$\,$D, and the combined precision estimate that will map to $H_0$ is 3\%, combining the delays relative to D.
We provide the delay values and covariance matrix in Table~\ref{tab:0029_vals}, below.

\begin{table}[h!]
\tabcolsep=0.17cm
\centering
\caption{Delay values in days, and covariance matrix elements in squared days, for DES$\,$J0029$-$3814.}
\vspace{-0.1cm}\label{tab:0029_vals}
\begin{tabular}{r|r|rrrrrr}
\toprule
 & \multirow{2}{*}{\textbf{Delays}} & \multicolumn{6}{c}{\textbf{Covariance Matrix}} \\
  &  & A$\,$B & A$\,$C & A$\,$D & B$\,$C & B$\,$D & C$\,$D \\
\midrule
A$\,$B & -6.5 & 9.1 &  &  &  &  &  \\
A$\,$C & -6.6 & 2.0 & 7.1 &  &  &  &  \\
A$\,$D & 43.1 & 2.4 & 2.4 & 5.7 &  &  &  \\
B$\,$C & -2.0 & -5.7 & 4.5 & 0.1 & 12.2 &  &  \\
B$\,$D & 46.7 & -5.0 & 0.3 & 2.3 & 5.5 & 8.4 &  \\
C$\,$D & 49.9 & 0.3 & -4.1 & 2.5 & -4.8 & 2.0 & 7.5 \\
\bottomrule
\end{tabular}
\end{table}

\begin{figure}
    \centering
    \includegraphics[width=0.49\textwidth]{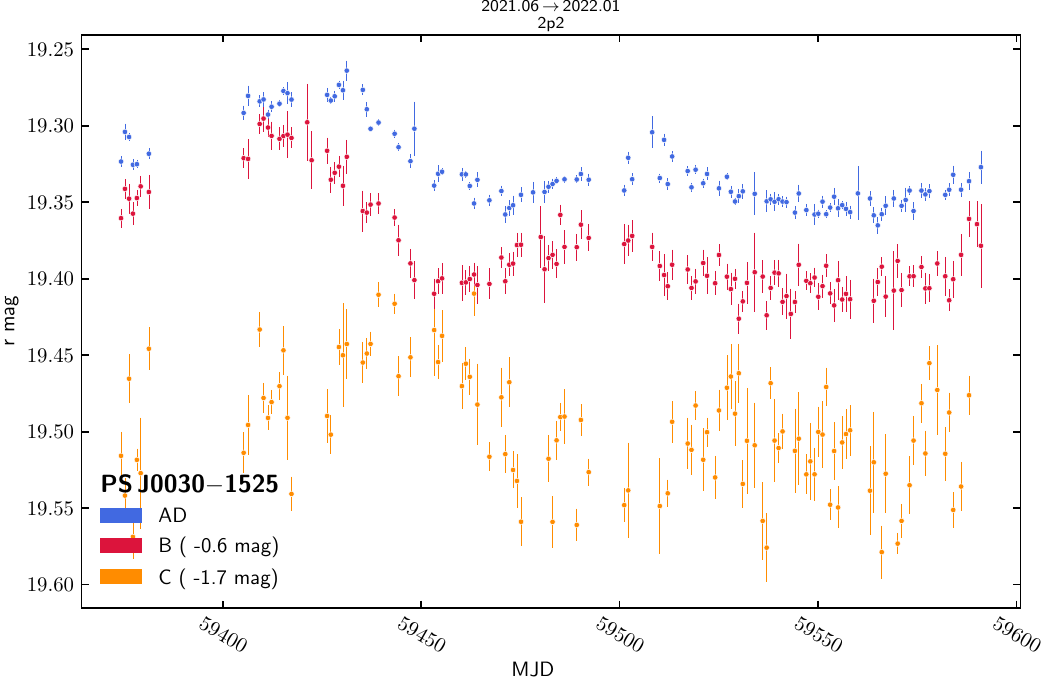}
    \caption{\appendixlccaption{PS$\,$J0030$-$1525}}
    \label{fig:lcsPSJ0030-1525}
\end{figure}

\paragraph{PS$\,$J0030$-$1525}
This $z_{\rm s}=3.36$ quadruply lensed quasar was discovered by \citet{lemon2018b}. 
We collected one season worth of imaging data from June 2021 to January 2022 with the 2p2 telescope, and even though it looks like a doubly imaged quasar only in the ground-based imaging, the forward modelling allows for reliable deblending of the lensing galaxy and the four images, revealing even the faint galactic satellite to the East of the lens. 
Due to the low separation in the folding pair however, we can only perform reasonable photometry of the sum of their fluxes.
This is not a problem for TDC as their difference in time of arrival is expected to be very small.
The data results in high-quality light curves with sharp features (Fig.~\ref{fig:lcsPSJ0030-1525}),
even though the quality of the C curve is degraded by the photon noise of the lensing galaxy.
A robust determination of the time delay between AD, the folding pair, and B and C could be conducted thanks to the two relatively high frequency peaks observed in all the curves. 
With a 50 days timescale of the intrinsic variations, the time delays are insensitive to the freedom given to the extrinsic variation models of each curve.
Overall, the time-delay uncertainties are dominated by the photon noise of the curves.
The best precision on a single delay is 12\% for B$\,$C, and the combined precision estimate that will map to $H_0$ is 7\%, combining the delays relative to C.

\begin{table}[h!]
\tabcolsep=0.17cm
\centering
\caption{Same as Table \ref{tab:0029_vals} but for PS$\,$J0030$-$1525.}
\vspace{-0.1cm}\label{tab:j0030_cov}
\begin{tabular}{r|r|rrr}
\toprule
 & \multirow{2}{*}{\textbf{Delays}} & \multicolumn{3}{c}{\textbf{Covariance Matrix}} \\
  &  & AD$\,$B & AD$\,$C & B$\,$C \\
\midrule
AD$\,$B & 9.3 & 2.6 &  &  \\
AD$\,$C & -19.3 & 0.4 & 10.7 &  \\
B$\,$C & -28.5 & -2.0 & 9.9 & 12.4 \\
\bottomrule
\end{tabular}
\end{table}

\begin{figure}
    \centering
    \includegraphics[width=0.49\textwidth]{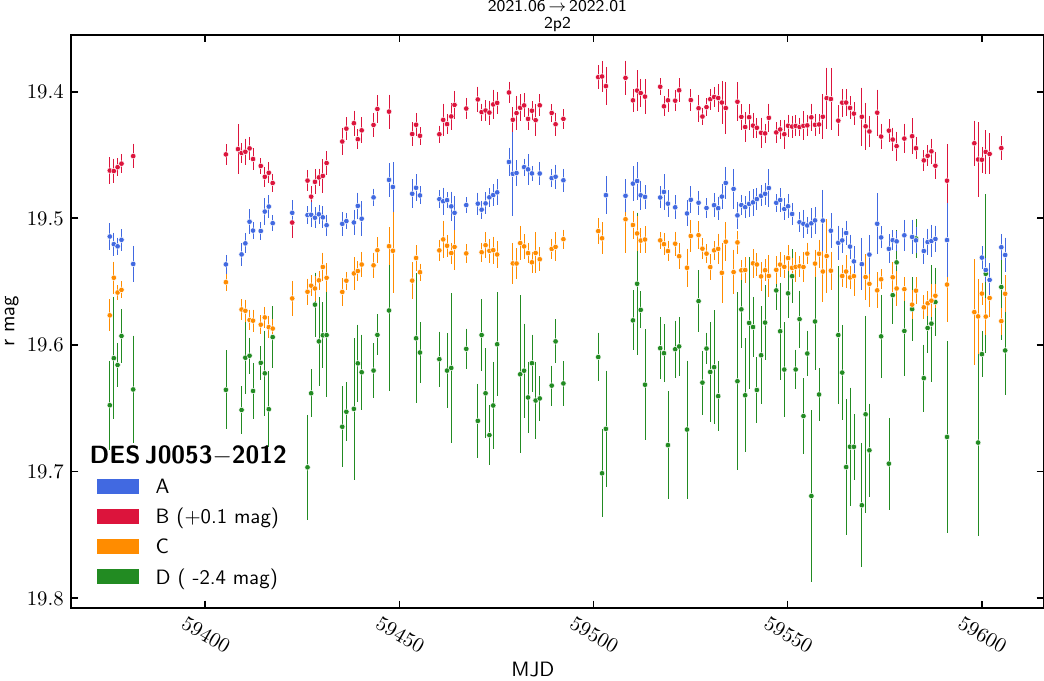}
    \caption{\appendixlccaption{DES$\,$J0053$-$2012}}
    \label{fig:lcsDESJ0053-2012}
\end{figure}

\paragraph{DES$\,$J0053$-$2012}
Discovered by~\citet{lemon2020}, this is a quadruply lensed quasar with source redshift $z_{\rm s}\sim3.8$.
Four light curves were extracted from a full season at the 2p2, starting in June 2021 and ending in January 2022, resulting in the light curves shown in Fig.~\ref{fig:lcsDESJ0053-2012}.
The three brightest images, A, B and C, show enough high-frequency variation to reliably constrain time delays, 
all insensitive to how free the extrinsic modulation is.
The D curve, however, is extremely noisy due to the faintness of image D, whose $r$-magnitude lies beyond 21.9. 
From prior knowledge that the A$\,$D delay should be quite large and negative, it appears that the best match is around $\mathrm{A}\,\mathrm{D}\sim -90\,$days.
No strong microlensing is observed in any of the images as the curves can be matched within their noise with only time and magnitude shifts, 
but D is both the noisiest and the closest lensed image to the lensing galaxy: the most prone to undetectable extrinsic variation.
Thus, even though more flexible microlensing models were included in the estimation of the error bars of the delays, those involving D should still be taken with care.
So, excluding D, the best precision on a single delay is 10\% for A$\,$B, and the combined precision estimate that will map to $H_0$ is 6\%, combining the delays relative to A.

\begin{table}[h!]
\tabcolsep=0.17cm
\centering
\caption{Same as Table \ref{tab:0029_vals} but for DES$\,$J0053$-$2012.}
\vspace{-0.1cm}\label{tab:j0053_cov}
\begin{tabular}{r|r|rrrrrr}
\toprule
 & \multirow{2}{*}{\textbf{Delays}} & \multicolumn{6}{c}{\textbf{Covariance Matrix}} \\
  &  & A$\,$B & A$\,$C & A$\,$D & B$\,$C & B$\,$D & C$\,$D \\
\midrule
A$\,$B & -26.7 & 6.9 &  &  &  &  &  \\
A$\,$C & -20.2 & 4.1 & 7.0 &  &  &  &  \\
A$\,$D & -90.2 & 4.5 & 4.3 & 44.9 &  &  &  \\
B$\,$C & 6.3 & -1.8 & 1.8 & -0.5 & 3.9 &  &  \\
B$\,$D & -63.5 & -1.1 & 0.1 & 31.4 & 1.4 & 34.8 &  \\
C$\,$D & -70.2 & 0.5 & -1.7 & 32.6 & -2.3 & 31.1 & 36.9 \\
\bottomrule
\end{tabular}
\end{table}

\begin{figure}
    \centering
    \includegraphics[width=0.49\textwidth]{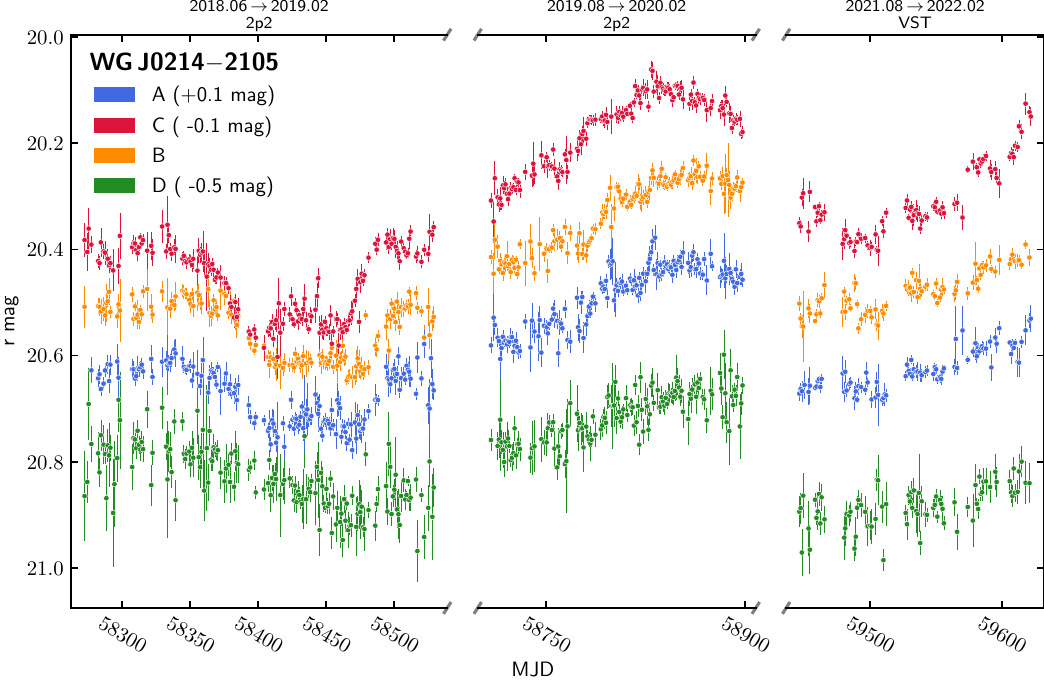}
    \caption{\appendixlccaption{WG$\,$J0214$-$2105}}
    \label{fig:lcsWG0214-2105}
\end{figure}

\paragraph{WG$\,$J0214$-$2105}
Discovered by~\citet{spiniello2019}, this quadruply imaged quasar has source redshift $z_{\rm s}\sim3.24$.
This object was monitored at the 2p2 for two seasons (June 2018 -- February 2019, August 2019 -- February 2020), and then for one more season at the VST (August 2021 -- February 2022). The two latter seasons were shorter than the first due to the COVID-19 pandemic. 
The 2p2 data were already presented in~\cite{Millon2020b}, but the curves were extracted again with the new method.
These and the new ones are shown in Fig.~\ref{fig:lcsWG0214-2105}.
We provide a re-analysis together with the new VST curves, improving the precision and changing the value of the A$\,$B delay slightly. The A$\,$D delay is also changed (although by less than 2-$\sigma$ upon quadratically combining error bars), possibly due to the more flexible microlensing models allowed for in the present estimation.
Due to combining three seasons worth of data, the microlensing effects become very noticeable,
such that splines with several internal knots (up to 10 for curve D, whose shape is qualitatively different from the other curves) modulating each curve were needed to properly align the light curves. 
Nevertheless, the many high-frequency and S/N features observed in the light curves permitted a precise alignment.
The best precision on a single delay is 5\% for B$\,$C, and the combined precision estimate that will map to $H_0$ is 2\%, combining the delays relative to C.

\begin{table}[h!]
\tabcolsep=0.17cm
\centering
\caption{Same as Table \ref{tab:0029_vals} but for WG$\,$J0214$-$2105.}
\vspace{-0.1cm}\label{tab:j0214_cov}
\begin{tabular}{r|r|rrrrrr}
\toprule
 & \multirow{2}{*}{\textbf{Delays}} & \multicolumn{6}{c}{\textbf{Covariance Matrix}} \\
  &  & A$\,$B & A$\,$C & A$\,$D & B$\,$C & B$\,$D & C$\,$D \\
\midrule
A$\,$B & -5.0 & 0.4 &  &  &  &  &  \\
A$\,$C & 10.7 & 0.2 & 0.5 &  &  &  &  \\
A$\,$D & -2.9 & 0.2 & 0.3 & 1.9 &  &  &  \\
B$\,$C & 15.7 & -0.2 & 0.2 & 0.0 & 0.5 &  &  \\
B$\,$D & 2.1 & -0.2 & 0.0 & 1.6 & 0.3 & 1.9 &  \\
C$\,$D & -13.6 & 0.0 & -0.2 & 1.5 & -0.2 & 1.5 & 1.7 \\
\bottomrule
\end{tabular}
\end{table}

\begin{figure}
    \centering
    \includegraphics[width=0.49\textwidth]{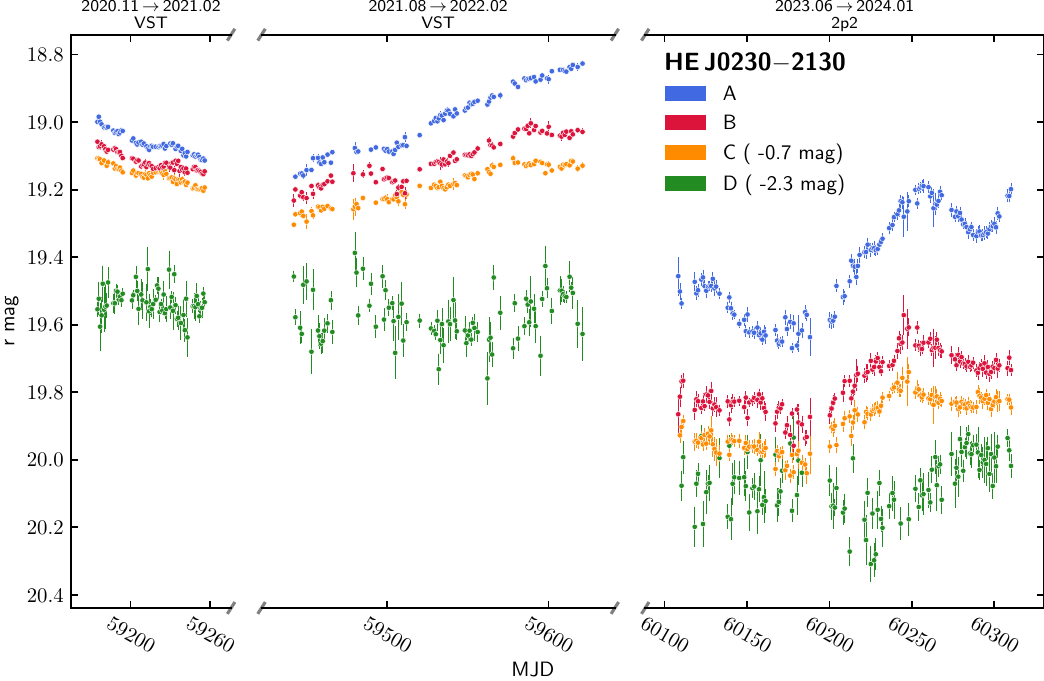}
    \caption{\appendixlccaption{HE$\,$J0230$-$2130}}
    \label{fig:lcsHE0230-2130}
\end{figure}

\paragraph{HE$\,$0230$-$2130}
Discovered by~\citet{wisotzki1999}, this is a quadruply lensed quasar with source redshift $z_{\rm s}\sim2.162$ and lens redshift $z_{\rm l}\sim0.523$~\citep{Eigenbrod2006}. 
It was monitored for a short season (started late due to the pandemic), from December 2020 to February 2021. Next, another season was attempted, starting in August 2021 but quickly interrupted the next month, again due to the pandemic. 
The resulting curves are visible in Fig.~\ref{fig:lcsHE0230-2130}.
The first short season could constrain the time delays quite well, but it was clear that additional data would help: thus, it was scheduled again at the 2p2 starting in June 2023 and ending in January 2024.
During this last 2p2 season, a clear microlensing event affected the A image, making the estimation of the time delays involving this image more difficult:
the alignment of the curves required very flexible extrinsic modulations of A. This is reflected in the uncertainties of the A$\,$B, A$\,$C and A$\,$D delays.
This object has also been the subject of a two- and a three-years monitoring campaigns at the Euler Swiss telescope, the results of which were presented in~\cite{Millon2020}.
Thanks to the higher S/N provided by the 2p2 telescope compared to the 1.2$\,$m Euler Swiss Telescope, we were this time able to reliably deblend the A and B images, and to also provide the delays involving the fourth image, D.
Our time-delay estimations are in slight tension with the measurement presented in~\cite{Millon2020}: 1.2$\sigma$ and 1.5$\sigma$ for the A$\,$C and B$\,$C delays respectively (both referred to as A$\,$C in~\citet{Millon2020}, as their A is the sum of our A and B).
A re-analysis of the light curves presented in~\citet{Millon2020} by~\citet{donnan2021} provides much tighter error bars on the delays, and is in even more tension with the measurements we present herein from the new, deeper data.
In particular, we find a tension of 2- and 3-$\sigma$ respectively when comparing our A$\,$C and B$\,$C delays to their $(17.7\pm2.1)\;$days A$\,$C value. Their A$\,$D and C$\,$D values are, also, roughly 2- and 3-$\sigma$ away from ours.
We explain this difference by the higher quality data used herein, with uncertainties that account for the possible degeneracies due to extrinsic modulation.
Overall, with the delays of this work, the best precision on a single delay is 10\% for C$\,$D, and the combined precision estimate that will map to $H_0$ is 4\%, combining the delays relative to D.

\begin{table}[h!]
\tabcolsep=0.17cm
\centering
\caption{Same as Table \ref{tab:0029_vals} but for HE$\,$J0230$-$2130.}
\vspace{-0.1cm}\label{tab:j0230_cov}
\begin{tabular}{r|r|rrrrrr}
\toprule
 & \multirow{2}{*}{\textbf{Delays}} & \multicolumn{6}{c}{\textbf{Covariance Matrix}} \\
  &  & A$\,$B & A$\,$C & A$\,$D & B$\,$C & B$\,$D & C$\,$D \\
\midrule
A$\,$B & -1.2 & 7.3 &  &  &  &  &  \\
A$\,$C & 9.5 & 5.6 & 6.5 &  &  &  &  \\
A$\,$D & -35.5 & 5.4 & 5.3 & 28.1 &  &  &  \\
B$\,$C & 10.7 & -1.2 & 0.5 & -0.1 & 1.8 &  &  \\
B$\,$D & -34.2 & -1.1 & 0.0 & 14.2 & 1.1 & 20.2 &  \\
C$\,$D & -45.0 & 0.1 & -0.5 & 14.4 & -0.6 & 15.1 & 19.9 \\
\bottomrule
\end{tabular}
\end{table}

\paragraph{SDSS$\,$J0248$+$1913}
This quadruply imaged quasar has source redshift $z_{\rm s}\sim2.424$, and was discovered both with the technique presented in ~\citet{Ostrovski2017} and with HST follow-up presented in \citet{shajib2019}. The system was also discovered independently by ~\citet{Delchambre2019}. 
A monitoring season was started at the VST at the end of September 2021, but the pandemic effects quickly thwarted the attempt at the end of December. No time delay could be extracted from the resulting light curves, which we do not present due to the poor sampling and short duration. 
Nevertheless, the existence of 8 hours of OmegaCAM exposure time on this field is worth mentioning.

\begin{figure}
    \centering
    \includegraphics[width=0.49\textwidth]{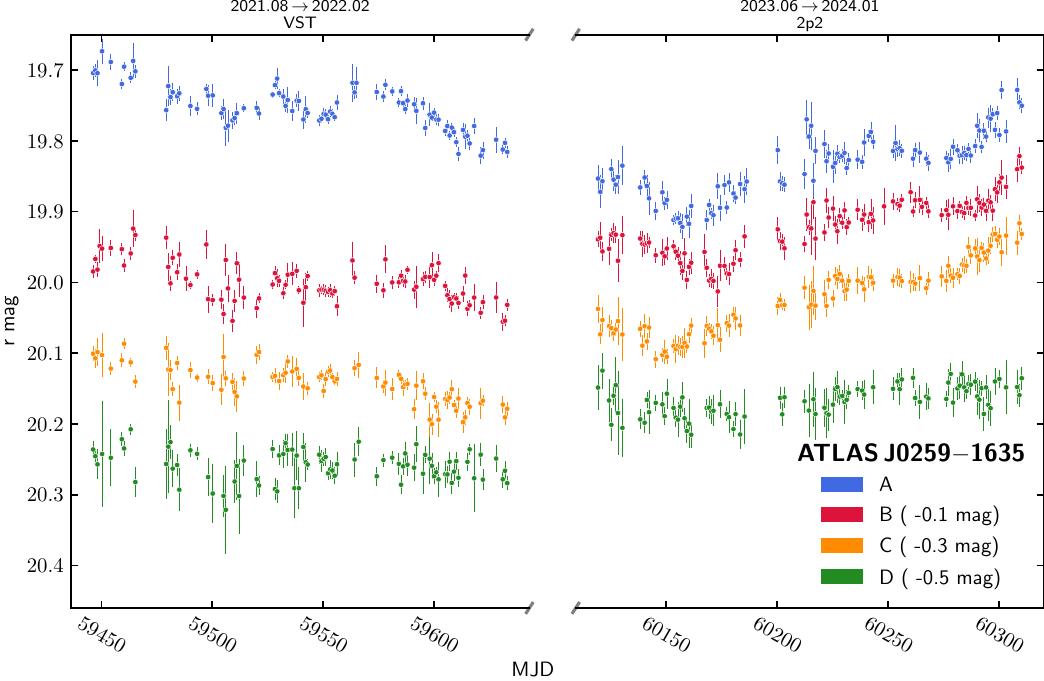}
    \caption{\appendixlccaption{ATLAS$\,$J0259$-$1635}}
    \label{fig:lcsATLAS0259-1635}
\end{figure}

\paragraph{WISE$\,$J0259$-$1635}
Discovered by~\citet{schechter2018}, this quadruply imaged quasar has source redshift $z_{\rm s}\sim2.16$.
A first monitoring season was captured at the VST, starting in August 2021 and ending in March 2022.
A second season was then taken at the 2p2, starting in June 2023 and ending in January 2024.
High-precision time delays could be obtained thanks to the high S/Ns of the extracted curves, shown in Fig.~\ref{fig:lcsATLAS0259-1635}.
The recorded features have timescales of 5-15 days, which is well separated from the $\sim100$~days observed extrinsic modulations attributable to microlensing.
Thus, the time delays are insensitive to how flexible the extrinsic modulation models are, such that the uncertainty budget is mostly due to photon noise.
The best precision on a single delay is 9\% for B$\,$C, and the combined precision estimate that will map to $H_0$ is 4\%, combining the delays relative to C.

\begin{table}[h!]
\tabcolsep=0.17cm
\centering
\caption{Same as Table \ref{tab:0029_vals} but for ATLAS$\,$J0259$-$1635.}
\vspace{-0.1cm}\label{tab:j0259_cov}
\begin{tabular}{r|r|rrrrrr}
\toprule
 & \multirow{2}{*}{\textbf{Delays}} & \multicolumn{6}{c}{\textbf{Covariance Matrix}} \\
  &  & A$\,$B & A$\,$C & A$\,$D & B$\,$C & B$\,$D & C$\,$D \\
\midrule
A$\,$B & -9.0 & 2.8 &  &  &  &  &  \\
A$\,$C & 8.3 & 1.1 & 2.1 &  &  &  &  \\
A$\,$D & -17.5 & 1.3 & 1.1 & 11.9 &  &  &  \\
B$\,$C & 17.0 & -1.6 & 0.8 & -0.2 & 2.4 &  &  \\
B$\,$D & -8.5 & -1.1 & 0.0 & 9.4 & 1.3 & 11.5 &  \\
C$\,$D & -25.3 & 0.3 & -0.7 & 9.6 & -1.0 & 9.4 & 11.4 \\
\bottomrule
\end{tabular}
\end{table}

\begin{figure}
    \centering
    \includegraphics[width=0.49\textwidth]{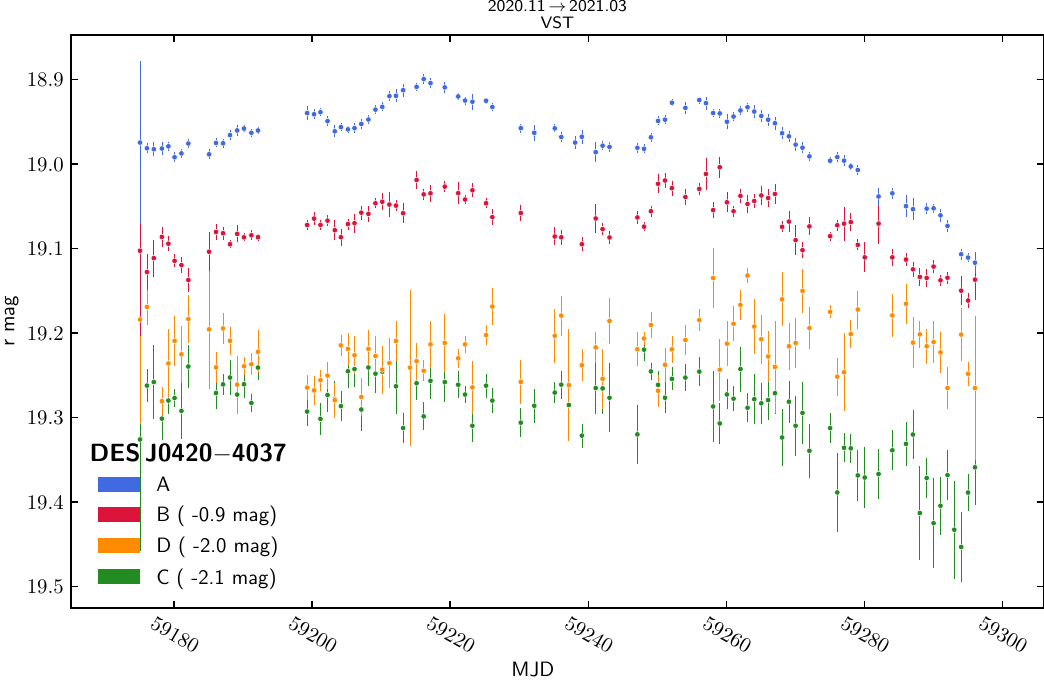}
    \caption{\appendixlccaption{DES$\,$J0420$-$4037}}
    \label{fig:lcsDESJ0420-4037}
\end{figure}

\paragraph{J0420$-$4037}
This quadruply lensed quasar has source redshift $z_{\rm s}\sim2.4$. It was discovered~(Ostrovski et al. in prep.) with the method described in~\cite{Ostrovski2017}.
It was monitored at the VST for one season from October 2020 to March 2021, resulting in the four curves provided in Fig.~\ref{fig:lcsDESJ0420-4037}.
Two of these show high S/N variations (A and B), while the other two are noisier, albeit with visible features still (C and D). 
The short delays in this lens, however, prevent us from reaching a high relative precision in any of the pairs of images: the only definitely incompatible with zero delays are those involving the D lensed image.
The best precision on a single delay is 33\% for A$\,$C, and the combined precision estimate that will map to $H_0$ is 11\%, combining the delays relative to C.

\begin{table}[h!]
\tabcolsep=0.17cm
\centering
\caption{Same as Table \ref{tab:0029_vals} but for DES$\,$J0420$-$4037.}
\vspace{-0.1cm}\label{tab:j0420_cov}
\begin{tabular}{r|r|rrrrrr}
\toprule
 & \multirow{2}{*}{\textbf{Delays}} & \multicolumn{6}{c}{\textbf{Covariance Matrix}} \\
  &  & A$\,$B & A$\,$C & A$\,$D & B$\,$C & B$\,$D & C$\,$D \\
\midrule
A$\,$B & 1.7 & 5.8 &  &  &  &  &  \\
A$\,$C & 7.9 & 2.7 & 6.6 &  &  &  &  \\
A$\,$D & -7.2 & 3.0 & 3.2 & 25.6 &  &  &  \\
B$\,$C & 6.2 & -2.0 & 2.7 & 0.3 & 5.3 &  &  \\
B$\,$D & -8.9 & -2.4 & 0.4 & 19.8 & 2.4 & 24.4 &  \\
C$\,$D & -14.9 & 0.2 & -2.9 & 19.2 & -2.6 & 19.1 & 24.3 \\
\bottomrule
\end{tabular}
\end{table}

\begin{figure}
    \centering
    \includegraphics[width=0.49\textwidth]{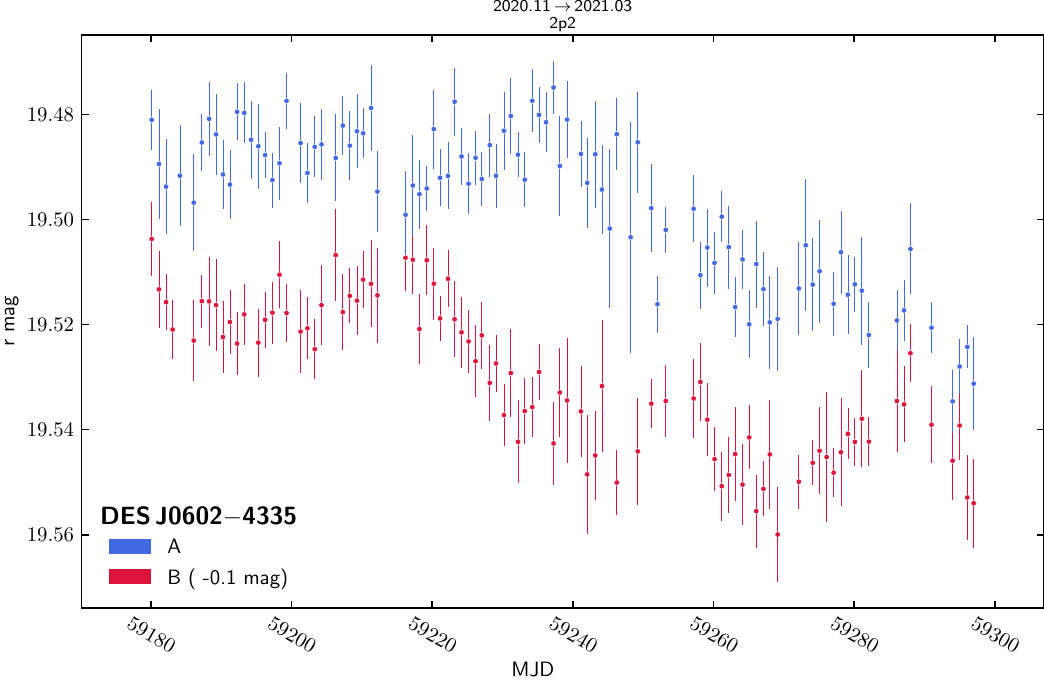}
    \caption{\appendixlccaption{DES$\,$J0602$-$4335}}
    \label{fig:lcsDESJ0602-4335}
\end{figure}

\paragraph{DES$\,$J0602$-$4335}
This doubly imaged quasar, discovered by \citet{dawes2023}, was monitored at the 2p2 from the end of November 2020 to the end of March 2021. 
The redshift of the source is $z_{\rm s}\sim2.92(1)$, determined from archival ESO EFOSC2 data.\footnote{Proposal ESO 0100.A-0297(B), PI: Timo Anguita.}
The lens is an edge-on galaxy, which the forward-modelling could readily deblend from image A.
Even though the extracted light curves (Fig.~\ref{fig:lcsDESJ0602-4335}) are only 120~days long, two rises with timescale $\sim$20~days were captured in both curves.
These can hardly be degenerate with extrinsic variations given their short timescale, and as such, the measured delays are not sensitive to the freedom given to the extrinsic modulation. 
The resulting time delay is therefore very precise, its uncertainty mostly being due to the photon noise of the curves.
The precision is here 6\% on the single delay, A$\,$B.

\begin{table}[h!]
\tabcolsep=0.17cm
\centering
\caption{Same as Table \ref{tab:0029_vals} but for DES$\,$J0602$-$4335. }
\vspace{-0.1cm}\label{tab:j0602_cov}
\begin{tabular}{r|r|r}
\toprule
 & \multirow{2}{*}{\textbf{Delay}} & \multicolumn{1}{c}{\textbf{Variance}} \\
  &  & A$\,$B \\
\midrule
A$\,$B & 23.6 & 4.5 \\
\bottomrule
\end{tabular}
\tablefoot{Note that the provided value is a variance, the associated uncertainty is its square root.} 
\end{table}

\begin{figure}
    \centering
    \includegraphics[width=0.49\textwidth]{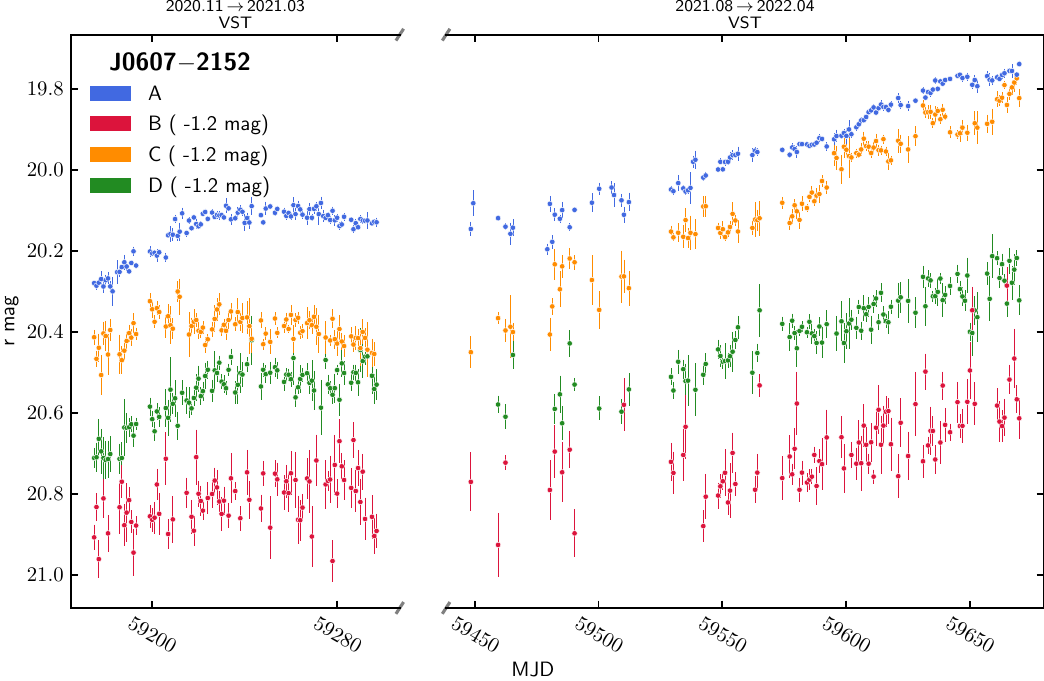}
    \caption{\appendixlccaption{J0607$-$2152}}
    \label{fig:lcsJ0607-2152}
\end{figure}

\paragraph{J$0607$-$2152$}
Discovered by~\citet{stern2021} and~\citet{lemon2023}, this quadruply imaged quasar has source redshift $z_{\rm s}\sim1.302$. 
This object was first monitored for a short season at the VST, starting in October 2020 and ending in March 2021. Monitoring was then resumed for an additional season, from August 2021 to April 2022, resulting in the light curves displayed in Fig.~\ref{fig:lcsJ0607-2152}.
The light curve of the B image was unfortunately too noisy to produce a reliable estimation of a time delay, due to its faintness (22.1 \textit{r}-mag), but also to the presence of a very bright star in the field of view (YY Leporis), adding to the photon noise and degrading the quality of the background subtraction. A, C and D however, all show sharp features that could be aligned without degeneracies with potential extrinsic modulations.
Overall, the best precision on a single delay is 12\% for C$\,$D, and the combined precision estimate that will map to $H_0$ is 7\%, combining the delays relative to D.
Additionally, the high-resolution model shows a bright Einstein ring with visible structure, making this target a good candidate for deep, sharp follow-up imaging.\footnote{
HST/ACS imaging was obtained for this lens short before submission of this paper, confirming our high resolution model. PI Lemon, HST gap program SNAP 17308.
} 
Moreover, \citet{lemon2023} suggest that a single galaxy model is not sufficient to explain the lensing configuration -- indeed our fitted model shows a potential second galaxy very close to image B that supports their suggestion. We can also confirm the image configuration suggested by \citet{lemon2023} over that of \citet{stern2021} who suggested image B was between A and D.

\begin{table}[h!]
\tabcolsep=0.17cm
\centering
\caption{Same as Table \ref{tab:0029_vals} but for J0607$-$2152.}
\vspace{-0.1cm}\label{tab:j0607_cov}
\begin{tabular}{r|r|rrr}
\toprule
 & \multirow{2}{*}{\textbf{Delays}} & \multicolumn{3}{c}{\textbf{Covariance Matrix}} \\
  &  & A$\,$C & A$\,$D & C$\,$D \\
\midrule
A$\,$C & 16.2 & 7.0 &  &  \\
A$\,$D & -13.4 & 0.8 & 5.6 &  \\
C$\,$D & -29.7 & -6.1 & 4.5 & 11.4 \\
\bottomrule
\end{tabular}
\end{table}

\begin{figure}
    \centering
    \includegraphics[width=0.49\textwidth]{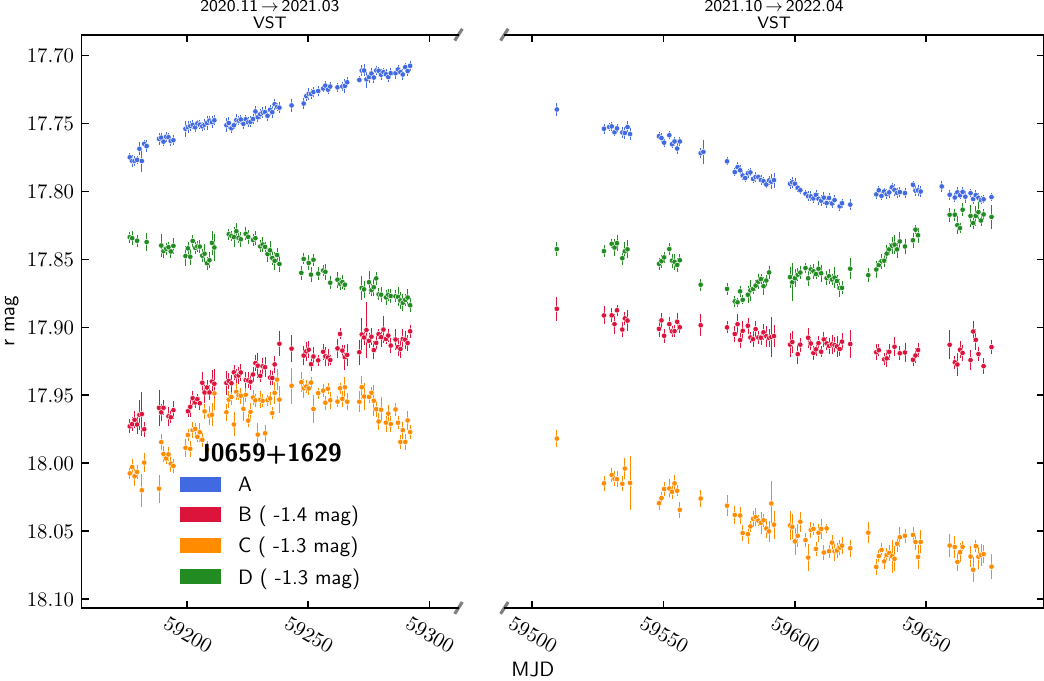}
    \caption{\appendixlccaption{J0659+1629}}
    \label{fig:lcsJ0659+1629}
\end{figure}

\paragraph{J0659$+$1629}
Discovered by~\citet{Delchambre2019} and \citet{lemon2023}, this is a quadruply lensed quasar with source redshift $z_{\rm s}\sim3.09$ and lens redshift $z_{\rm l}\sim0.766$~\citep{stern2021}.
This object was observed for two seasons at the VST: a short one from November 2020 to March 2021, and another from October 2021 to April 2022.
We also include archival data from the Las Cumbres Observatory\footnote{Proposal NOAO2020A-007.}, with a mean cadence of 2 days, ranging from December 2019 to May 2020. 
The resulting light curves (Fig.~\ref{fig:lcsJ0659+1629}) can be aligned without strong extrinsic modulations, except for that of image C: this is not surprising considering its proximity to a second galaxy, which both the HST imaging and high-resolution model fitted on VST data reveal.
Once the required freedom is given to the C curve, the curves can be aligned within their noise envelope with a preferred solution involving D ahead by about 270 days.
However, it is clear that a 330 days shift of D could fit the curves almost as well, albeit involving stronger extrinsic modulations. 
Moreover, our data cannot constrain potential delays beyond 400 days.
We provide in Table~\ref{tab:j0659_cov} delays and uncertainties around the best solution, but one should keep in mind the second possible solution (Table~\ref{tab:j0659_cov_2}), and the possibility that the delays are much bigger.
To use this lens in a TDC study, one should carefully confront the mass model predictions of the time delays with the different solutions we find herein, and only move forward if one of the possibilities is favoured beyond all doubts by the mass model.
Because J0659$+$1629 will land in the LSST footprint, there is hope for additional constraints on the delay in the long run.

\begin{table}[h!]
\tabcolsep=0.17cm
\centering
\caption{Same as Table \ref{tab:0029_vals} but for our preferred time-delay solution of J0659+1629.}
\vspace{-0.1cm}\label{tab:j0659_cov}
\begin{tabular}{r|r|rrrrrr}
\toprule
 & \multirow{2}{*}{\textbf{Delays}} & \multicolumn{6}{c}{\textbf{Covariance Matrix}} \\
  &  & A$\,$B & A$\,$C & A$\,$D & B$\,$C & B$\,$D & C$\,$D \\
\midrule
A$\,$B & -16.1 & 4.7 &  &  &  &  &  \\
A$\,$C & -14.2 & 2.7 & 165 &  &  &  &  \\
A$\,$D & 262.2 & 5.6 & 45.5 & 123 &  &  &  \\
B$\,$C & 1.1 & -2.0 & 145 & 20.7 & 188 &  &  \\
B$\,$D & 277.0 & 0.8 & 25.9 & 97.6 & 57.2 & 129 &  \\
C$\,$D & 277.7 & 2.9 & -120 & 77.6 & -125 & 71.7 & 197 \\
\bottomrule
\end{tabular}

\end{table}

\begin{table}[h!]
\tabcolsep=0.17cm
\centering
\caption{Same as Table \ref{tab:0029_vals} but for the second possible solution of J0659+1629.}
\vspace{-0.1cm}\label{tab:j0659_cov_2}
\begin{tabular}{r|r|rrrrrr}
\toprule
 & \multirow{2}{*}{\textbf{Delays}} & \multicolumn{6}{c}{\textbf{Covariance Matrix}} \\
  &  & A$\,$B & A$\,$C & A$\,$D & B$\,$C & B$\,$D & C$\,$D \\
\midrule
A$\,$B & -12.5 & 9.5 &  &  &  &  &  \\
A$\,$C & -12.2 & 11.6 & 446 &  &  &  &  \\
A$\,$D & 309.6 & 9.6 & 134 & 243 &  &  &  \\
B$\,$C & -0.5 & 2.8 & 300 & 23.0 & 436 &  &  \\
B$\,$D & 317.3 & 1.0 & 16.6 & 84.6 & 90.9 & 180 &  \\
C$\,$D & 318.1 & -1.5 & -266 & 85.5 & -264 & 63.4 & 406 \\
\bottomrule
\end{tabular}
\end{table}

\begin{figure}
    \centering
    \includegraphics[width=0.49\textwidth]{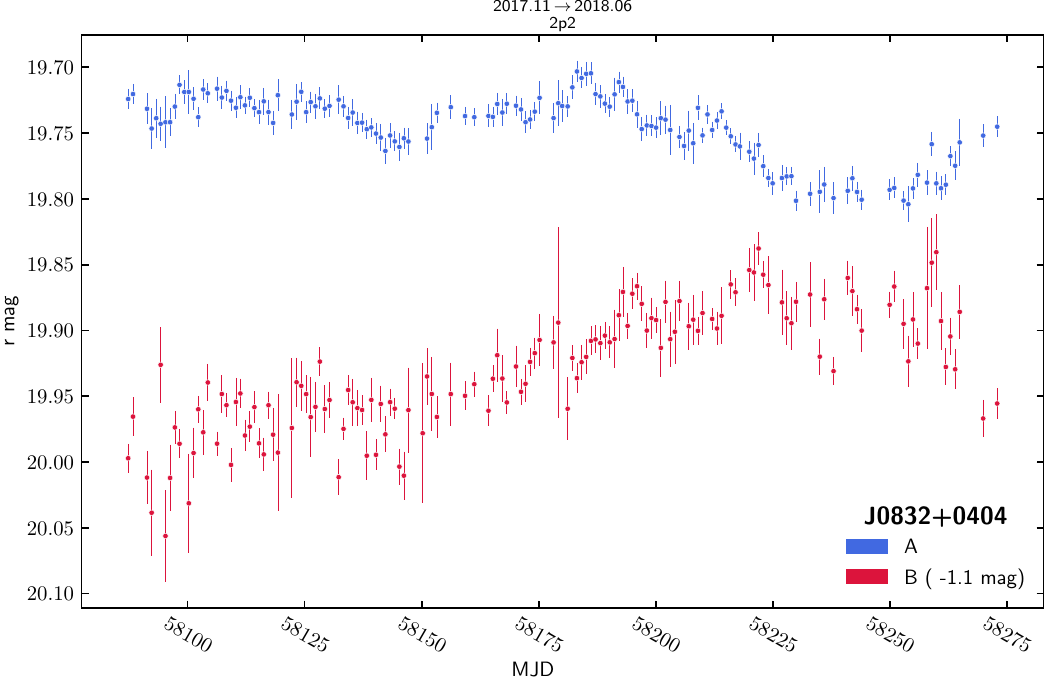}
    \caption{\appendixlccaption{J0832+0404}}
    \label{fig:lcsJ0832+0404}
\end{figure}

\paragraph{SDSS$\,$J0832$+$0404}
This doubly imaged quasar was discovered by~\citet{oguri2008}. It has source redshift $z_{\rm s}\sim1.115$.
It was monitored at the 2p2 from the end of November 2017 to the beginning of June 2018.
The extracted curves (Fig.~\ref{fig:lcsJ0832+0404}) show plenty of high-frequency features. 
However, due to the long delay, the overlap is only $\sim$70 days.
This makes the time delay sensitive to the local slope of the extrinsic variation in each lensed image, which is reflected in the uncertainty on the time delay.
The obtained value is, however, perfectly compatible with that reported by~\citet{Millon2020} at the Euler Swiss telescope. 
It is also compatible with the re-analysis of the Swiss telescope curves by~\citet{donnan2021}.
Combining the estimation of \citet{Millon2020} with ours as independent measurements, assuming Gaussian uncertainties, we obtain a precision of 6\%.

\begin{table}[h!]
\tabcolsep=0.17cm
\centering
\caption{Same as Table \ref{tab:0029_vals} but for J0832+0404. }
\vspace{-0.1cm}\label{tab:j0832_cov}
\begin{tabular}{r|r|r}
\toprule
 & \multirow{2}{*}{\textbf{Delay}} & \multicolumn{1}{c}{\textbf{Variance}} \\
  &  & A$\,$B \\
\midrule
A$\,$B (ours) & -129.4 & 141 \\
A$\,$B (combined with  \citeauthor{Millon2020}) & -128.0 & 61 \\
\bottomrule
\end{tabular}
\tablefoot{Note that the provided values are variances, the associated uncertainties are their square root. } 
\end{table}

\begin{figure}
    \centering
    \includegraphics[width=0.49\textwidth]{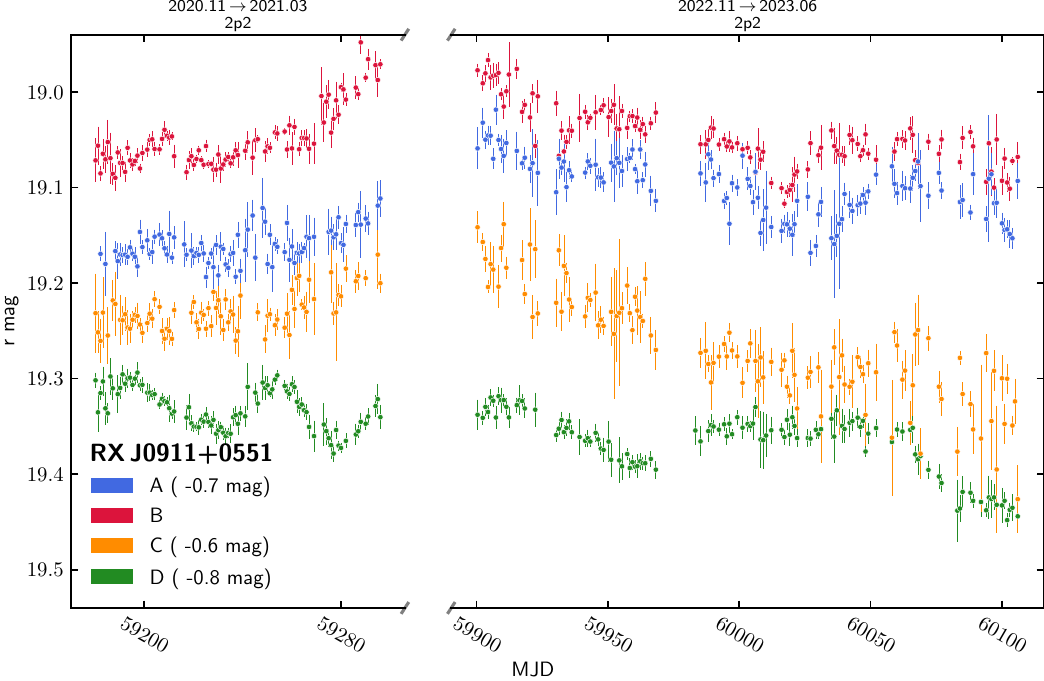}
    \caption{\appendixlccaption{RX$\,$J0911+0551}}
    \label{fig:lcsRXJ0911+0551}
\end{figure}

\paragraph{RX$\,$J0911$+$0551}
Discovered by~\citet{bade1997}, this is a quadruply imaged quasar with source redshift $z_{\rm s}\sim 2.763$ and lens redshift  $z_{\rm l}\sim0.769$~\citep{kneib2000}.
It was monitored at the 2p2 for two seasons, from October 2020 to March 2021 and from November 2021 to June 2022. The resulting curves are displayed in Fig.~\ref{fig:lcsRXJ0911+0551}.
The ABC$\;$D delay is known to be $150\pm6$ or 146$\pm4$ days~\citep{Eulaers2011}: a precision sufficient for TDC even with the potential multi-modality.
Two high cadence monitoring seasons should have been enough to much better constrain the D delay (due to a season of D overlapping both the first and second season), 
but the pandemic interruption of the first season leaves us with only one, short overlap.
This raises a similar problem to the case of SDSS$\,$J0832$+$0404, making the individual delays imprecise.
Nevertheless, we do have three delays involving D thanks to the A, B and C trio being reliably deblended by the forward-modelling.
Thus, the best precision on a single delay is 12\% for A$\,$D, and the combined precision estimate that will map to $H_0$ is 4\%, combining the delays relative to D.

\begin{table}[h!]
\tabcolsep=0.155cm
\centering
\caption{Same as Table \ref{tab:0029_vals} but for RX$\,$J0911+0551.}
\vspace{-0.1cm}\label{tab:j0911_cov}
\begin{tabular}{r|r|rrrrrr}
\toprule
 & \multirow{2}{*}{\textbf{Delays}} & \multicolumn{6}{c}{\textbf{Covariance Matrix}} \\
  &  & A$\,$B & A$\,$C & A$\,$D & B$\,$C & B$\,$D & C$\,$D \\
\midrule
A$\,$B & 4.8   & 1.3  &  &  &  &  &  \\
A$\,$C & -5.6  & 0.7  & 3.2 &  &  &  &  \\
A$\,$D & 154   & 0.3  & 0.5 & 77 &  &  &  \\
B$\,$C & -10.7 & -0.6 & 2.4 & 0.2 & 3.0 &  &  \\
B$\,$D & 150   & -1.0 & -0.2 & 75 & 0.8 & 79 &  \\
C$\,$D & 160   & -0.4 & -2.7 & 75 & -2.2 & 76 & 79 \\
\bottomrule
\end{tabular}
\end{table}

\begin{figure}
    \centering
    \includegraphics[width=0.49\textwidth]{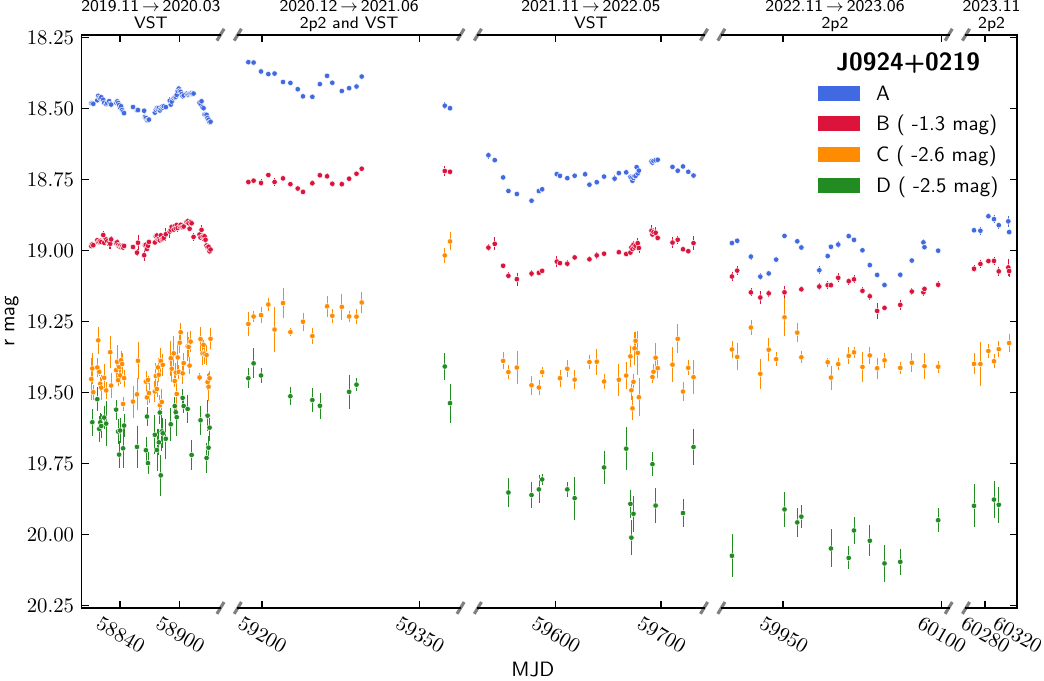}
    \caption{\appendixlccaption{J0924+0219}}
    \label{fig:lcsJ0924+0219}
\end{figure}

\paragraph{SDSS$\,$J0924$+$0219}
Discovered by~\citet{inada2003}, this is a quadruply lensed quasar with source redshift $z_{\rm s}\sim1.685$ and lens redshift $z_{\rm l}\sim0.393$.
Monitoring started with a daily cadence in October 2020 at the VST, until March 2021.
Then, three additional weekly cadenced seasons were acquired for microlensing studies. 
Two at the VST still, from December 2020 to April 2021 and from November 2021 to June 2022,
and the third one at the 2p2 from November 2022 to June 2023.
Shown in Fig.~\ref{fig:lcsJ0924+0219}, these weekly cadenced seasons propitiously coincided with fluctuations in the A and B images, providing a fast enough sampling to capture the variations appropriately.
The C curve is much noisier, but its first season shows a definite oscillation that can be matched to A and B.
The present A$\,$B estimation is more precise than, but also compatible with, the one presented in~\citet{Millon2020b}. 
A$\,$C on the other hand, was not given in ~\citet{Millon2020b} due to insufficient S/N, but 
was in the reanalysis by~\cite{donnan2021}. Their value (-30.9 days) is in slight (1.5$\sigma$) tension with the one presented herein.
The best precision on a single delay is 16\% for B$\,$C, and the combined precision estimate that will map to $H_0$ is 7\%, combining the delays relative to C.

\begin{table}[h!]
\tabcolsep=0.17cm
\centering
\caption{Same as Table \ref{tab:0029_vals} but for J0924+0219.}
\vspace{-0.1cm}\label{tab:j0924_cov}
\begin{tabular}{r|r|rrrrrr}
\toprule
 & \multirow{2}{*}{\textbf{Delays}} & \multicolumn{6}{c}{\textbf{Covariance Matrix}} \\
  &  & A$\,$B & A$\,$C & A$\,$D & B$\,$C & B$\,$D & C$\,$D \\
\midrule
A$\,$B & 3.7 & 0.9 &  &  &  &  &  \\
A$\,$C & -17.5 & 0.4 & 11.3 &  &  &  &  \\
A$\,$D & -1.6 & 0.5 & 0.9 & 13.1 &  &  &  \\
B$\,$C & -21.5 & -0.5 & 9.6 & 0.4 & 12.0 &  &  \\
B$\,$D & -5.2 & -0.4 & 0.5 & 12.4 & 1.0 & 13.8 &  \\
C$\,$D & 14.7 & 0.1 & -9.0 & 11.5 & -9.2 & 12.2 & 25.5 \\
\bottomrule
\end{tabular}
\end{table}

\begin{figure}
    \centering
    \includegraphics[width=0.49\textwidth]{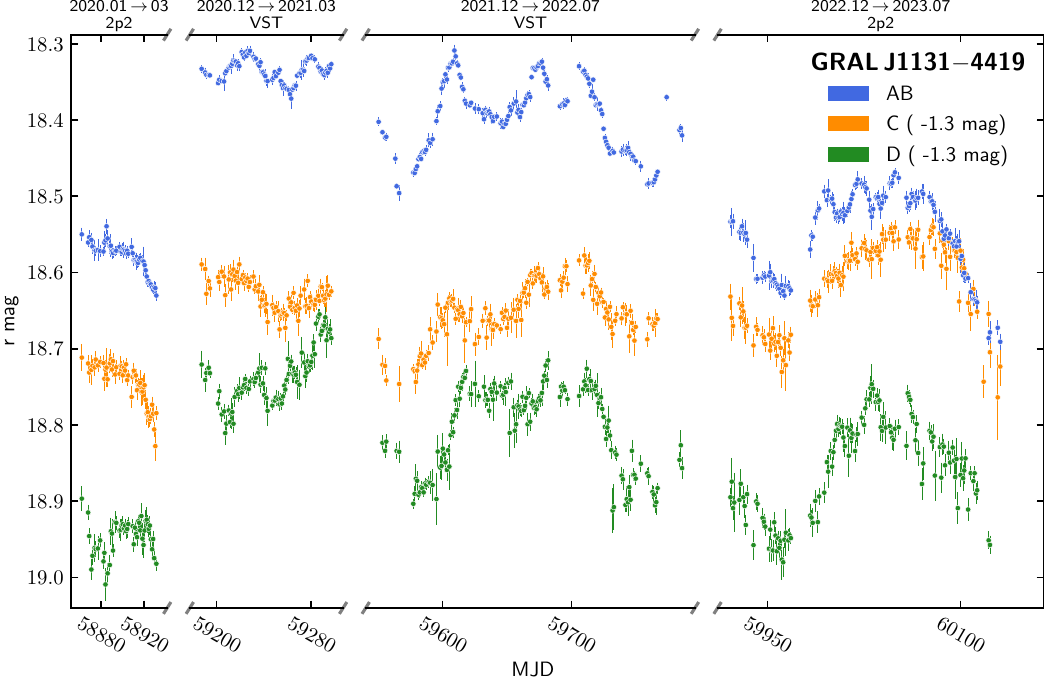}
    \caption{\appendixlccaption{GRAL$\,$J1131$-$4419}}
    \label{fig:lcsGRALJ1131-4419}
\end{figure}

\paragraph{GRAL$\,$J1131$-$4419}
Discovered by~\citet{kronemartins2018}, this is a quadruply imaged quasar with source redshift $z_{\rm s}=1.090$~\citep{wertz2019}.
It was first monitored at the 2p2 for a short season starting in December 2019, but quickly interrupted by the COVID-19 pandemic in March 2020.
Next, another short season was acquired at the VST starting in November 2020 and ending in April 2021.
Finally, two additional seasons could be fully completed: a first one at the VST starting in November 2021 and ending in July 2022, and the second at the 2p2 starting in November 2022 and ending in June 2023.
The narrow separation ($\sim$0$\farcs$45) of the A-B pair made it impossible to reliably deblend in most seeing conditions, so we only present the delays within the trio AB, C, and D, with AB the sum of the A and B fluxes.
The resulting curves, AB, C and D are provided in Fig.~\ref{fig:lcsGRALJ1131-4419}.
Thanks to rapid, high-amplitude intrinsic variations and the availability of four monitoring seasons, excellent precision could be achieved in the delays.
The best precision on a single delay is 7\% for C$\,$D, and the combined precision estimate that will map to $H_0$ is 4\%, combining the delays relative to D.

\begin{table}[h!]
\tabcolsep=0.17cm
\centering
\caption{Same as Table \ref{tab:0029_vals} but for GRAL$\,$J1131$-$4419.}
\vspace{-0.1cm}\label{tab:gral1131_cov}
\begin{tabular}{r|r|rrr}
\toprule
 & \multirow{2}{*}{\textbf{Delays}} & \multicolumn{3}{c}{\textbf{Covariance Matrix}} \\
  &  & AB$\,$C & AB$\,$D & C$\,$D \\
\midrule
AB$\,$C & 3.5 & 0.5 &  &  \\
AB$\,$D & -10.1 & 0.2 & 0.6 &  \\
C$\,$D & -13.6 & -0.3 & 0.4 & 0.7 \\
\bottomrule
\end{tabular}
\end{table}

\begin{figure}
    \centering
    \includegraphics[width=0.49\textwidth]{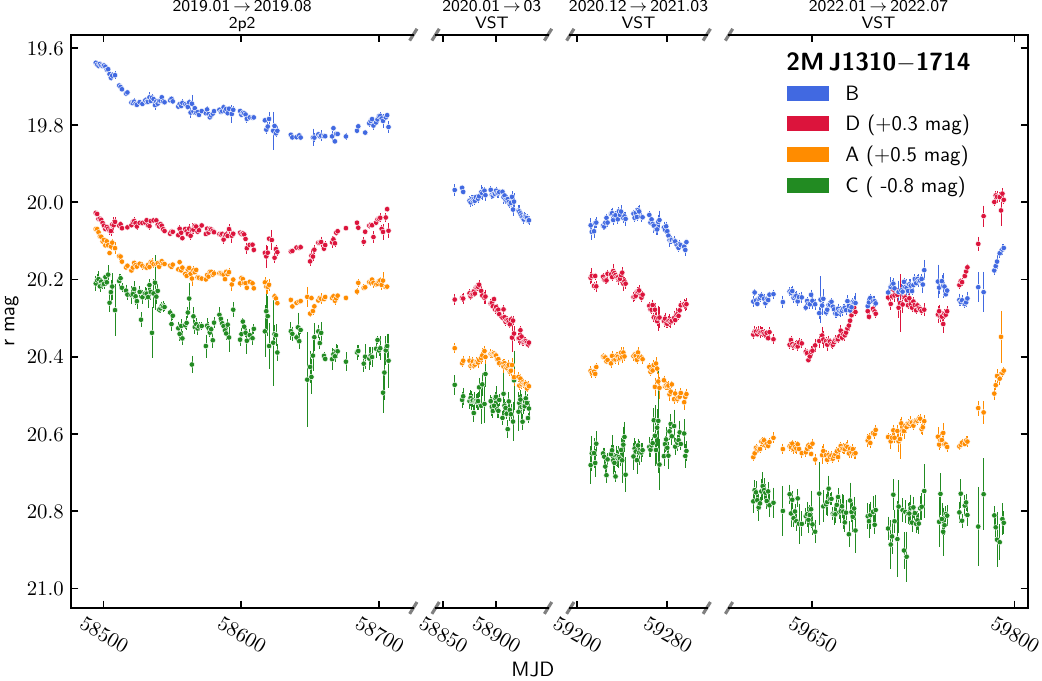}
    \caption{\appendixlccaption{2M$\,$J1310$-$1714}}
    \label{fig:lcs2M1310-1714}
\end{figure}

\paragraph{2M$\,$1310$-$1714}
Discovered by~\citet{Lucey2018}, this is a quadruply lensed quasar with source redshift $z_{\rm s}\sim1.975$ and double-galaxy lens redshift $z_{\rm l}\sim0.293$.
This object benefited from three monitoring seasons at the VST.
The first two, starting in December 2019 and 2020 respectively, both ended early in March of the following year because of the COVID-19 pandemic. Luckily, a third season, complete this time, was acquired starting in early January 2022 and ending in July 2022.
High-S/N light curves (Fig.~\ref{fig:lcs2M1310-1714}) were extracted from the three brightest images, A, B and C.
The light curve of the D image became noisier as monitoring progressed, as it got dimmer with time, reaching 21.6 mag by the end of the last VST season. Despite this, it still shows features that can be used to match it with the other three curves.
The reached precision in the time delays is excellent thanks to the sharp observed variations, availability of three seasons, and moderate extrinsic modulation needed to align the light curves.
The best precision on a single delay is 3\% for A$\,$B, and the combined precision estimate that will map to $H_0$ is 1\%, combining the delays relative to A.

\begin{table}[h!]
\tabcolsep=0.17cm
\centering
\caption{Same as Table \ref{tab:0029_vals} but for 2M$\,$J1310$-$1714.}
\vspace{-0.1cm}\label{tab:2m1310_cov}
\begin{tabular}{r|r|rrrrrr}
\toprule
 & \multirow{2}{*}{\textbf{Delays}} & \multicolumn{6}{c}{\textbf{Covariance Matrix}} \\
  &  & A$\,$B & A$\,$C & A$\,$D & B$\,$C & B$\,$D & C$\,$D \\
\midrule
A$\,$B & -55.9 & 2.3 &  &  &  &  &  \\
A$\,$C & -21.4 & 0.3 & 0.4 &  &  &  &  \\
A$\,$D & -14.9 & 0.2 & 0.2 & 0.3 &  &  &  \\
B$\,$C & 34.5 & -1.8 & 0.1 & 0.0 & 1.9 &  &  \\
B$\,$D & 41.0 & -1.9 & -0.1 & 0.1 & 1.8 & 2.0 &  \\
C$\,$D & 6.6 & 0.0 & -0.2 & 0.1 & -0.1 & 0.2 & 0.3 \\
\bottomrule
\end{tabular}
\end{table}

\paragraph{J1537$-$3010}
Discovered by~\citet{lemon2019a} and \citet{Delchambre2019}, this quadruply imaged quasar has source redshift $z_{\rm s}\sim1.721$.
This object was first monitored at the 2p2 before the COVID-19 pandemic, for a full season, from February to October 2019.
Next, two seasons were begun, at the 2p2 in February 2020 and at the VST in February 2021, but both were quickly interrupted due to the pandemic.
Finally, it was monitored for a final full season at the 2p2 again in 2022, starting in January and ending in September. The extracted light curves are those shown above, in Fig.~\ref{fig:lcj1537}.
The combined `two and a half' seasons provide the most stringent time-delay constraints obtained in this work, thanks to the source quasar's large and fast intrinsic variations which are easy to separate from the microlensing signal.
The best precision on a single delay is 2\% for B$\,$C, and the combined precision estimate that will map to $H_0$ is 1\%, combining the delays relative to B.

\begin{table}[h!]
\tabcolsep=0.17cm
\centering
\caption{Same as Table \ref{tab:0029_vals}, for J1537$-$3010.}
\vspace{-0.1cm}\label{tab:j1537_cov}
\begin{tabular}{r|r|rrrrrr}
\toprule
 & \multirow{2}{*}{\textbf{Delays}} & \multicolumn{6}{c}{\textbf{Covariance Matrix}} \\
  &  & A$\,$B & A$\,$C & A$\,$D & B$\,$C & B$\,$D & C$\,$D \\
\midrule
A$\,$B & -29.2 & 0.8 &  &  &  &  &  \\
A$\,$C & 8.5 & 0.1 & 0.5 &  &  &  &  \\
A$\,$D & -24.7 & 0.4 & 0.2 & 2.2 &  &  &  \\
B$\,$C & 37.7 & -0.5 & 0.3 & -0.3 & 0.7 &  &  \\
B$\,$D & 4.3 & -0.2 & 0.0 & 1.7 & 0.2 & 1.9 &  \\
C$\,$D & -33.3 & 0.3 & -0.2 & 1.9 & -0.5 & 1.6 & 2.1 \\
\bottomrule
\end{tabular}
\end{table}

\begin{figure}
    \centering
    \includegraphics[width=0.49\textwidth]{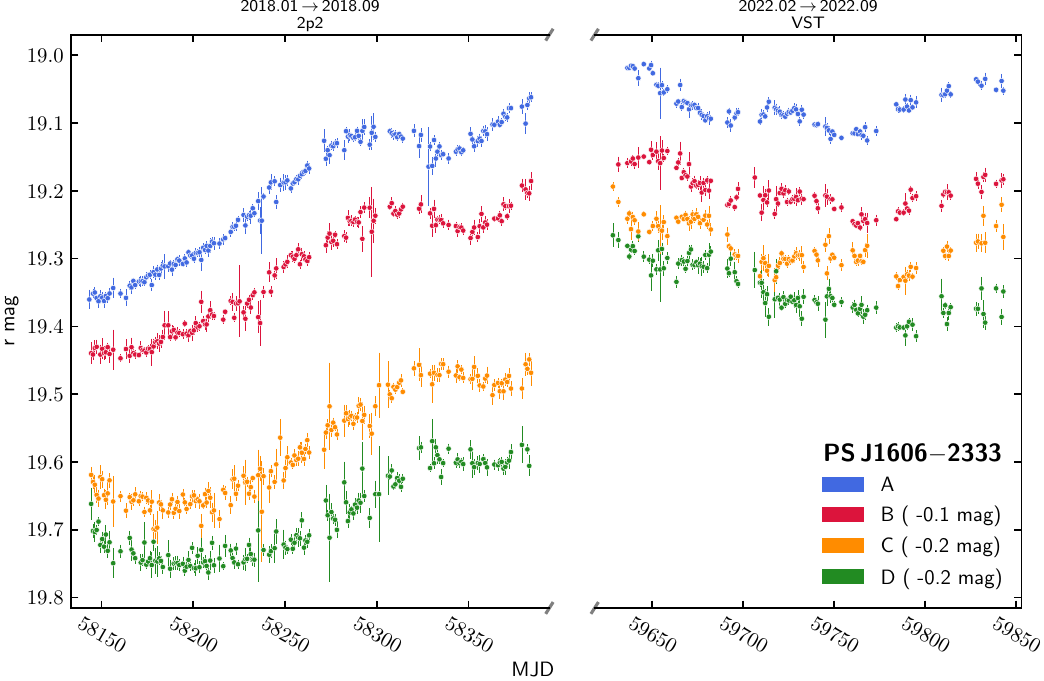}
    \caption{\appendixlccaption{PS$\,$J1606$-$2333}}
    \label{fig:lcsPSJ1606-2333}
\end{figure}

\paragraph{PS$\,$J1606$-$2333}
Discovered by~\citet{lemon2018b}, this quadruply lensed quasar has source redshift $z_{\rm s}\sim1.69$.
It was first monitored for a season at the 2p2 from February to September 2018. This first season was already presented in~\citet{Millon2020b}, but did not allow the reliable determination of the time delays involving the D lensed image. 
Thus, another season was planned at the VST, starting in late February 2022 until late September 2022.
Note that the curves were extracted from the 2p2 data again with the new method -- the combined result is displayed in Fig.~\ref{fig:lcsPSJ1606-2333}.
Together, the 2p2 and VST curves can this time constrain the D delay thanks to the higher frequency variations observed in the VST data. 
The delays within the A, B and C trio determined with the 2p2 data by~\citet{Millon2020b} are confirmed and made more precise thanks to this added season of data and re-analysis.
The best precision on a single delay is 5\% for A$\,$C, and the combined precision estimate that will map to $H_0$ is 2\%, combining the delays relative to A.

\begin{table}[h!]
\tabcolsep=0.17cm
\centering
\caption{Same as Table \ref{tab:0029_vals} but for PS$\,$J1606$-$2333.}
\vspace{-0.1cm}\label{tab:j1606_cov}
\begin{tabular}{r|r|rrrrrr}
\toprule
 & \multirow{2}{*}{\textbf{Delays}} & \multicolumn{6}{c}{\textbf{Covariance Matrix}} \\
  &  & A$\,$B & A$\,$C & A$\,$D & B$\,$C & B$\,$D & C$\,$D \\
\midrule
A$\,$B & -11.0 & 1.0 &  &  &  &  &  \\
A$\,$C & -28.9 & 0.5 & 2.0 &  &  &  &  \\
A$\,$D & -34.5 & 0.6 & 1.0 & 4.2 &  &  &  \\
B$\,$C & -17.9 & -0.5 & 1.3 & 0.5 & 1.9 &  &  \\
B$\,$D & -23.5 & -0.4 & 0.5 & 2.8 & 0.9 & 3.8 &  \\
C$\,$D & -5.5 & 0.1 & -0.8 & 2.4 & -0.8 & 2.3 & 3.4 \\
\bottomrule
\end{tabular}
\end{table}

\begin{figure}
    \centering
    \includegraphics[width=0.49\textwidth]{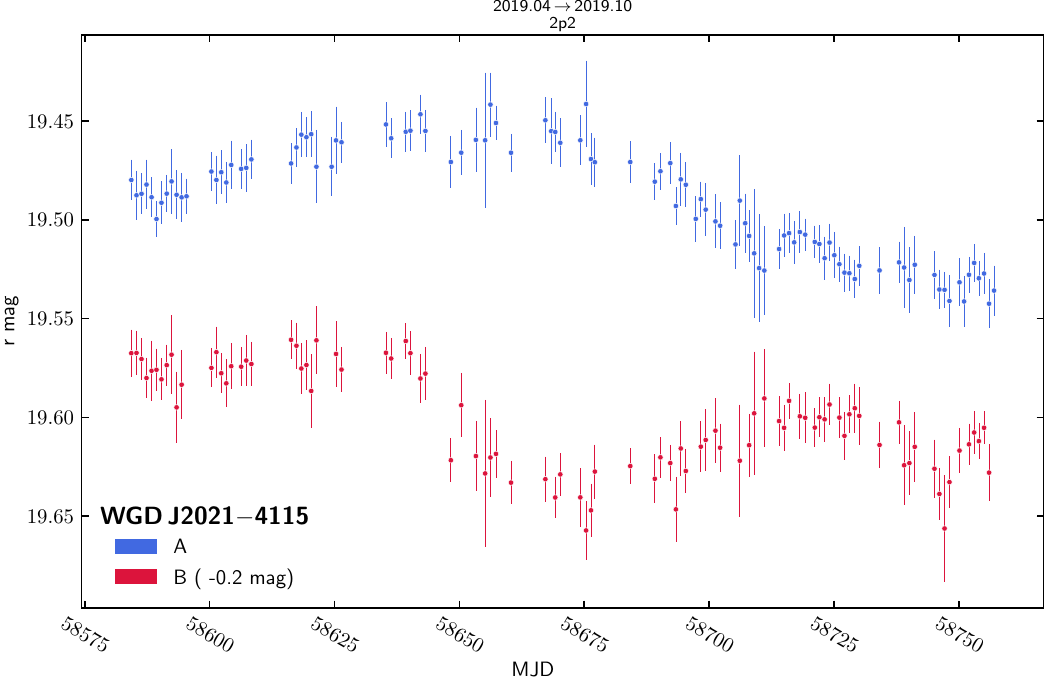}
    \caption{\appendixlccaption{WGD$\,$J2021$-$4115}}
    \label{fig:lcsWGD2021-4115}
\end{figure}

\paragraph{WGD$\,$J2021$-$4115}
This doubly imaged quasar, discovered by~\citet{agnello2018} has source redshift $z_{\rm s}\sim1.390$ and lens redshift $z_{\rm l}\sim0.335$.
Its monitoring at the 2p2 telescope from April to September 2019 resulted in two well-defined light curves (Fig.~\ref{fig:lcsWGD2021-4115}), albeit with slow variations somewhat degenerate with eventual microlensing, mainly due to the short overlaps between curves resulting from the relatively long -91 days delay.
The delay uncertainty would thereby dominate the error budget of an $H_0$ estimation with TDC.
However, the high-resolution fitted model does reveal signs of an Einstein ring, essential for constraining a precise mass model of the lens.
This could motivate the push for obtaining deep, sharp imaging of the object despite it being only doubly imaged.\footnote{This Einstein ring was later confirmed with sharp imaging from an HST gap program, Cameron Lemon, SNAP 17308.}
Furthermore, the long delay, the brightness of the two images (\textit{r}-mag 19.5 and 19.8) and their wide separation (2$\farcs$7), all constitute ingredients that will likely permit a time-delay determination with LSST.
Overall, we expect this object to become a compelling system for TDC after a few years of LSST operation, despite the relatively imprecise delay obtained herein.
The precision on the single delay, A$\,$B, is 11\%.

\begin{table}[h!]
\tabcolsep=0.17cm
\centering
\caption{Same as Table \ref{tab:0029_vals} but for WGD$\,$J2021$-$4115. }
\vspace{-0.1cm}\label{tab:j2021_cov}
\begin{tabular}{r|r|r}
\toprule
 & \multirow{2}{*}{\textbf{Delay}} & \multicolumn{1}{c}{\textbf{Variance}} \\
  &  & A$\,$B \\
\midrule
A$\,$B & -90.8 & 93.4 \\
\bottomrule
\end{tabular}
\tablefoot{The provided value is a variance, and the associated uncertainty is its square root. } 
\end{table}

\begin{figure}
    \centering
    \includegraphics[width=0.49\textwidth]{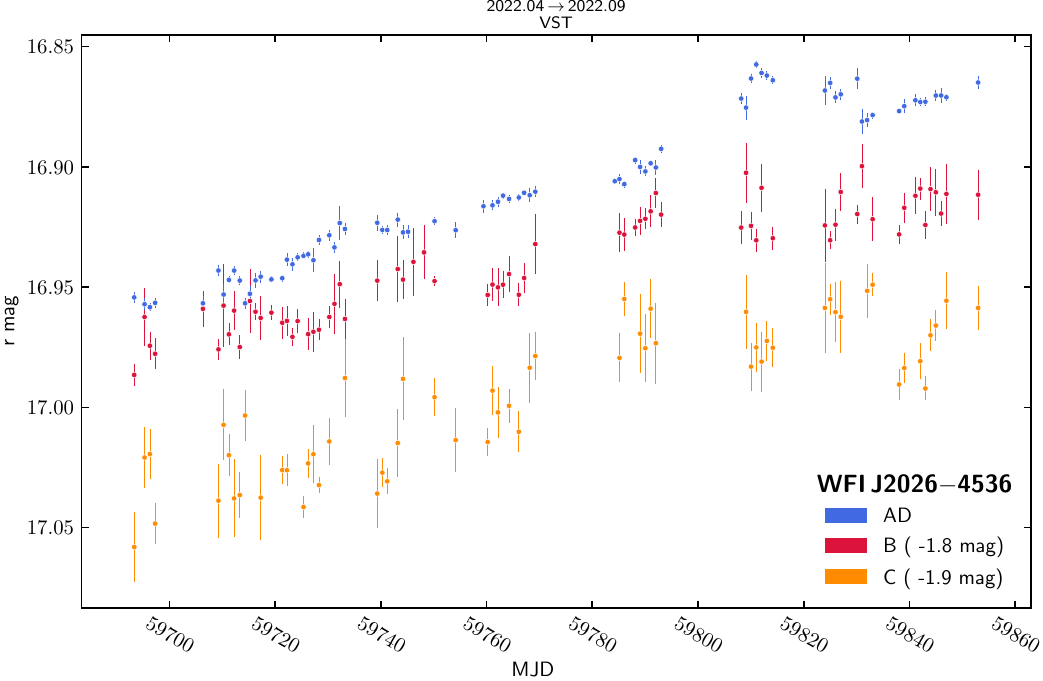}
    \caption{\appendixlccaption{WFI$\,$J2026$-$4536}}
    \label{fig:lcsWFIJ2026-4536}
\end{figure}

\paragraph{WFI$\,$J2026$-$4536}
Discovered by~\citet{morgan2004}, this quadruply imaged quasar has source redshift $z_{\rm s}\sim2.237$.
It was monitored at the VST for a short season between April and September 2022.
The two brightest images, A and D, are too narrowly separated (0$\farcs$3) to be reliably deblended. Their fluxes were thereby summed. 
Even though high-S/N features are present in the resulting set of light curves (Fig.~\ref{fig:lcsWFIJ2026-4536}), patches of bad weather made the time-delay estimation degenerate: possible solutions include the overlapping of the features of one curve with the empty regions of another. 
This is reflected in the high uncertainty in the delays, for example the AD$\,$C delay: determined with a relative precision of more than 50\%, it is nonetheless perfectly compatible with the value reported by~\citet{Millon2020}.
The best precision on a single delay is 19\% for AD$\,$B,
and the combined precision estimate that will map to $H_0$ is 12\%, combining the delays relative to B.

\begin{table}[h!]
\tabcolsep=0.17cm
\centering
\caption{Same as Table \ref{tab:0029_vals} but for WFI$\,$J2026$-$4536. }
\vspace{-0.1cm}\label{tab:j2026_cov}
\begin{tabular}{r|r|rrr}
\toprule
 & \multirow{2}{*}{\textbf{Delays}} & \multicolumn{3}{c}{\textbf{Covariance Matrix}} \\
  &  & AD$\,$B & AD$\,$C & B$\,$C \\
\midrule
AD$\,$B & 16.3 & 10.0 &  &  \\
AD$\,$C & -14.2 & 1.1 & 57.0 &  \\
B$\,$C & -30.6 & -7.0 & 50.6 & 100 \\
\bottomrule
\end{tabular}
\tablefoot{These do not include the estimation by~\cite{Millon2020b}, which can be taken as an extra, independent measurement in a TDC likelihood term. } 
\end{table}

\begin{figure}
    \centering
    \includegraphics[width=0.49\textwidth]{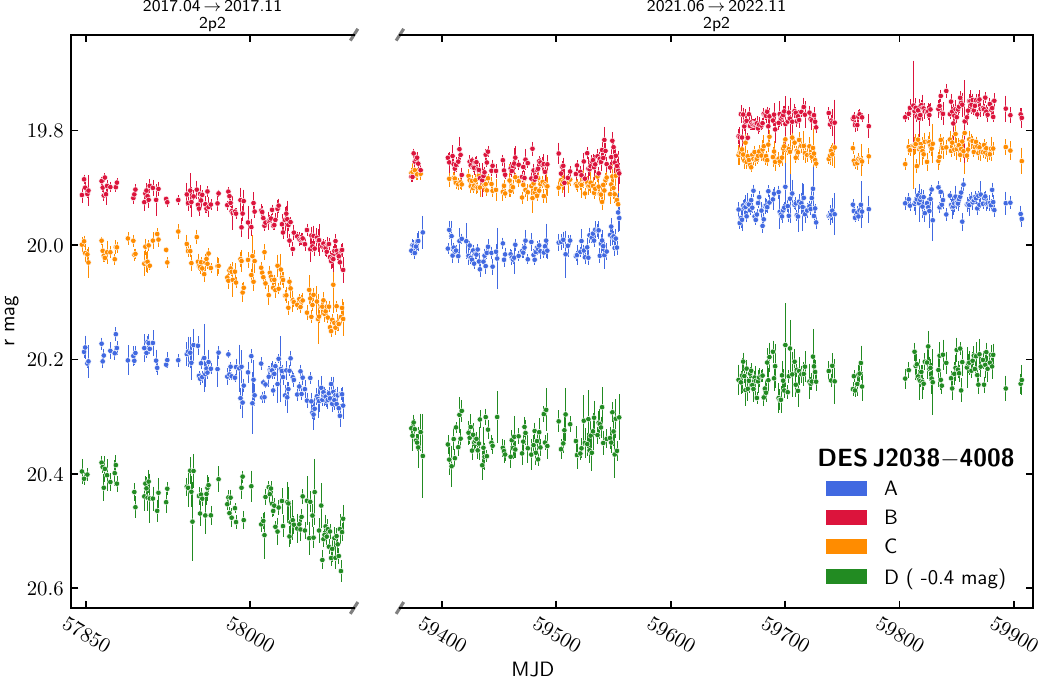}
    \caption{\appendixlccaption{WG$\,$J2038$-$4008}}
    \label{fig:lcsDES2038-4008}
\end{figure}

\paragraph{WG$\,$J2038$-$4008}
Discovered by~\citet{agnello2018}, this quadruply lensed quasar has source redshift $z_{\rm s}\sim0.777$ and lens redshift $z_{\rm l}\sim0.228$~\citep{stern2021}. 
It was monitored at the 2p2 telescope for three seasons: from April to December 2017, from June to December 2021, and from April to December 2022.
It was also monitored for five seasons at the Euler Swiss Telescope, a 1.2m facility at the same observatory, with a two-day cadence. 
When double coverage occurred, only the 2p2 data were used in the time-delay estimation due to its much higher quality and cadence.
Of note, the curves resulting from this dataset were already presented in~\cite{Wong2024}; a TDC analysis of the system.
Not only is this lens on the faint side of the targets monitored during this campaign, it also is the one whose source quasar varied the least on the monitoring baseline.
Fortunately, the combined datasets yield curves (Fig.~\ref{fig:lcsDES2038-4008}) that provide meaningful constraining power, thanks to the image D lagging behind the others by a longer time (about 30 days).
An additional problem was encountered with the A$\,$C delay however, for which two solutions were possible: one positive (7.9 days), and the other negative (-5.3 days). 
The negative one was favoured by the data, and, more importantly, switching solution changes the ordering of arrival of the images. 
The ordering of arrival is very well constrained by mass models of the lens, 
and allowed us to further prefer the negative solution.
Overall, the best precision on a single delay is 18\% for A$\,$D, and the combined precision estimate that would map to $H_0$ is 9\%, combining the delays relative to D.
However, the total error budget in the TDC analysis by~\cite{Wong2024} was roughly 25--30\%, showing how correlation between delays might have played a role in degrading the precision of the final estimate.

\begin{table}[h!]
\tabcolsep=0.17cm
\centering
\caption{Same as Table \ref{tab:0029_vals} but for WG$\,$J2038$-$4008.}
\vspace{-0.1cm}\label{tab:j2038_cov}
\begin{tabular}{r|r|rrrrrr}
\toprule
 & \multirow{2}{*}{\textbf{Delays}} & \multicolumn{6}{c}{\textbf{Covariance Matrix}} \\
  &  & A$\,$B & A$\,$C & A$\,$D & B$\,$C & B$\,$D & C$\,$D \\
\midrule
A$\,$B & -12.4 & 14.2 &       &   &   &  &  \\
A$\,$C & -5.3  & 6.1  & 14.8  &   &   &  &  \\
A$\,$D & -33.3 & 7.5  & 7.1   & 39.9  &  &  &  \\
B$\,$C & 7.1   & -6.9 & 8.6   & -0.4  & 16.1  &  &  \\
B$\,$D & -20.8 & -5.4 & 1.1   & 31.8  & 6.5   & 41.2 &  \\
C$\,$D & -28.1 & 1.4  & -7.6  & 32.1  & -9.0  & 30.7 & 40.9 \\
\bottomrule
\end{tabular}
\end{table}

\begin{figure}
    \centering
    \includegraphics[width=0.49\textwidth]{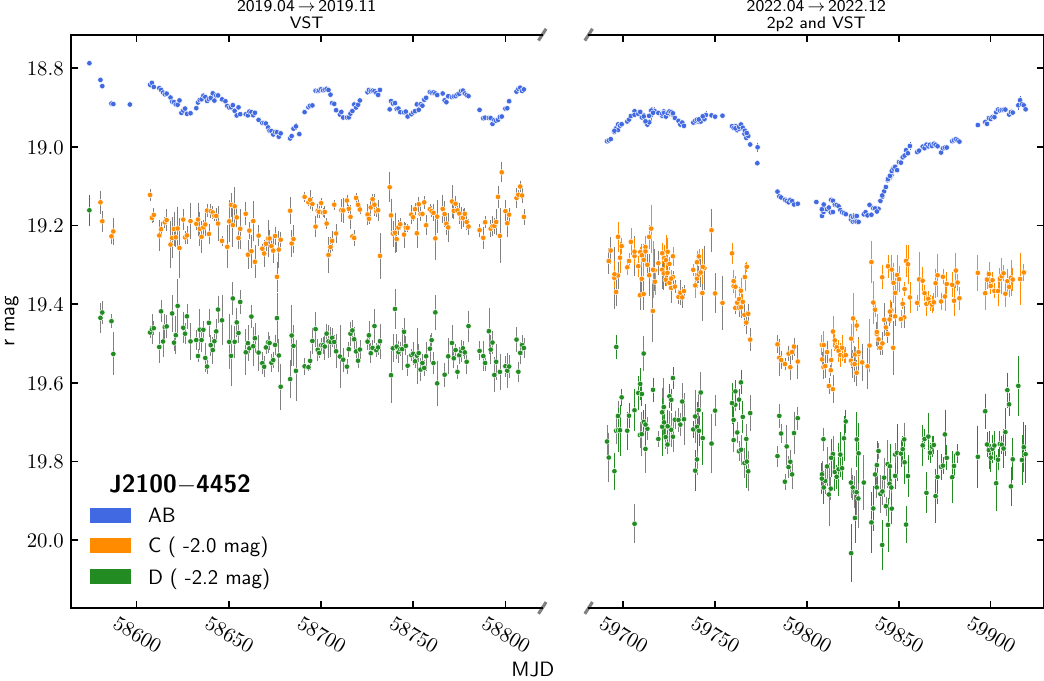}
    \caption{\appendixlccaption{WG$\,$J2100$-$4452}
        This lens is a special case, it was monitored at the same time both at the 2p2 and at the VST due to scheduling constraints. 
        The VST points are differentiated with gray error bars.
    }
    \label{fig:lcsJ2100-4452}
\end{figure}

\paragraph{WG$\,$J2100$-$4452}
Discovered by~\citet{agnello2019}, this is a quadruply lensed quasar with source redshift $z_{\rm s}\sim0.92$ and lens redshift $z_{\rm l}\sim0.203$~\citep{spiniello2019}.
It was first monitored at the VST for a season, between April and November 2019. Next, it was observed simultaneously by both the 2p2 and VST for an additional season, with combined coverage starting in April 2022 and ending in December of the same year. 
The double coverage occurred due to scheduling constraints and the uncertainty due to the COVID-19 pandemic, and proved beneficial given that the 2p2 monitoring was interrupted for almost two months due to a snowstorm at La Silla observatory. 
The extracted light curves are displayed in Fig.~\ref{fig:lcsJ2100-4452}. The VST and 2p2 points are differentiated by the colour of their error bars: gray for VST, matching colour for 2p2. 
This double coverage also offers a reassuring test of the photometric extraction, as curves produced by the two facilities match within their noise after a slight magnitude offset due to the mismatch in filter.
However, the two brightest and closest images A and B were too narrowly separated (0$\farcs$53) to allow for a reliable deblending.
Thus, their fluxes were summed, yielding the high-S/N AB light curve. C and D are much weaker, with light curves consequently much noisier. 
Fortunately, the oscillations starting at MJD 58700 are still well visible in both B and C as visible in Fig.~\ref{fig:lcsJ2100-4452}.
These have timescales way below those we allow for our microlensing modulation, as such the estimated delays are not overly sensitive to the freedom granted to the extrinsic modulation.
Thus, mainly due to the low S/N of the C and D curves, we obtain delays whose uncertainty budget is likely dominated by photon noise.
The best precision on a single delay is 12\% for C$\,$D.
Combining all delays relative to C, the best possible precision mapping to $H_0$ would be 8\%.

\begin{table}[h!]
\tabcolsep=0.17cm
\centering
\caption{Same as Table \ref{tab:0029_vals} but for WG$\,$J2100$-$4452.}
\vspace{-0.1cm}\label{tab:j2100_cov}
\begin{tabular}{r|r|rrr}
\toprule
 & \multirow{2}{*}{\textbf{Delays}} & \multicolumn{3}{c}{\textbf{Covariance Matrix}} \\
  &  & AB$\,$C & AB$\,$D & C$\,$D \\
\midrule
AB$\,$C & 4.6 & 1.0 &  &  \\
AB$\,$D & -7.3 & 0.1 & 1.5 &  \\
C$\,$D & -12.3 & -0.8 & 1.4 & 2.3 \\
\bottomrule
\end{tabular}
\end{table}

\begin{figure}
    \centering
    \includegraphics[width=0.49\textwidth]{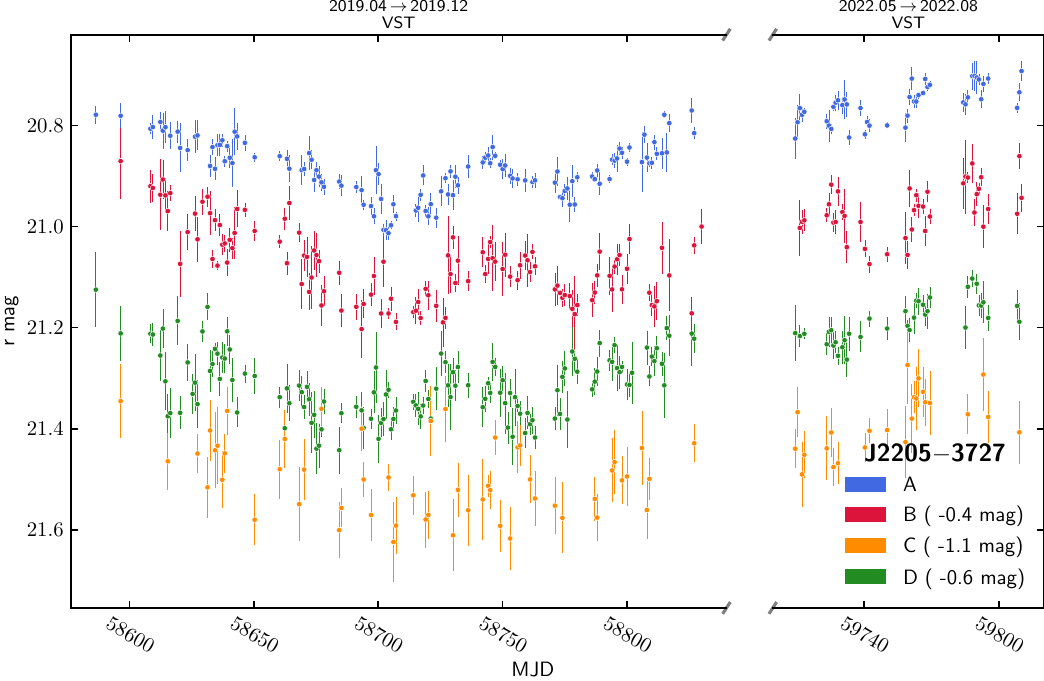}
    \caption{\appendixlccaption{J2205$-$3727}}
    \label{fig:lcsJ2205-3727}
\end{figure}

\paragraph{J2205$-$3727}
Discovered by~\citet{lemon2023}, this is a quadruply imaged quasar with source redshift $z_{\rm s}\sim1.848$.
This is the faintest lens monitored by this program: image C in particular, with an average \textit{r}-magnitude of 22.6, is the faintest quasar lensed image for which a time delay could be estimated.
J2205$-$3727 was monitored for two seasons at the VST, the first one started in April 2019 and ended in December 2019, and the second from May to August 2022. The resulting light curves are shown in Fig.~\ref{fig:lcsJ2205-3727}.
The curves of images A, B, and D all exhibit short-timescale features that can be used for alignment, but the flux of image C was often compatible with zero, especially on nights with poor seeing. Consequently, data points of the C curve from nights with seeing values exceeding 1.3 were discarded.
After this filtering was made, the C curve was fortunately still well sampled enough to reveal the patterns seen in the other three curves, but a part of its constraining power was lost.
This is well reflected in our estimated time-delay uncertainties, which are much larger in the delays involving it.
Overall, we find time delays well compatible with zero within the A, B, C trio, but longer delays in the pairs involving D, the image across from the lensing galaxy.
The error budget in the delays is mostly dominated by photon noise, as the estimated delays were insensitive to the complexity of the allowed extrinsic modulations.
The best precision on a single delay is 15\% for A$\,$D, and the combined precision estimate that will map to $H_0$ is 7\%, combining the delays relative to D.

\begin{table}[h!]
\tabcolsep=0.17cm
\centering
\caption{Same as Table \ref{tab:0029_vals}, but for J2205$-$3727.}
\vspace{-0.1cm}\label{tab:j2205_cov}
\begin{tabular}{r|r|rrrrrr}
\toprule
 & \multirow{2}{*}{\textbf{Delays}} & \multicolumn{6}{c}{\textbf{Covariance Matrix}} \\
  &  & A$\,$B & A$\,$C & A$\,$D & B$\,$C & B$\,$D & C$\,$D \\
\midrule
A$\,$B & -1.8 & 6.4 &  &  &  &  &  \\
A$\,$C & 0.2 & 1.4 & 18.2 &  &  &  &  \\
A$\,$D & 12.3 & 1.2 & 1.0 & 3.4 &  &  &  \\
B$\,$C & 1.7 & -5.0 & 16.7 & -0.2 & 22.2 &  &  \\
B$\,$D & 14.0 & -5.2 & -0.4 & 2.2 & 4.9 & 7.5 &  \\
C$\,$D & 11.9 & -0.2 & -17.1 & 2.3 & -17.0 & 2.6 & 19.7 \\
\bottomrule
\end{tabular}
\end{table}

\section{Light curves}
\label{appendix:lightcurves}
Here we list the light curves of the monitored targets.

\section{Observational statistics}
\newgeometry{margin=1cm} 
\begin{landscape}
\begin{table*}[]
\caption{Observational properties for each monitoring dataset of the lenses below R.A.$\,=9\,$h (see Table~\ref{tab:obsprops2} for the others), including per-curve noise statistics and quality of the monitoring data properties. 
\label{tab:obsprops}}
\def\arraystretch{1.08}\vspace{-0.2cm}
\begin{tabular*}{\linewidth}{@{\extracolsep{\fill}} lll|llllll|llllll}
Object                                 &   & Sep. $''$ & \multicolumn{6}{c}{VST}                                                                                                                                               & \multicolumn{6}{c}{2p2}                                                                                                                                               \\
                                       &   &           & mag range   & $\tilde{\sigma}_\mathrm{photon}$ & $\tilde{\Delta F}$ & ${\sigma}_{\rm rms}$ & seeing ($''$)                                 & cadence                  & mag range   & $\tilde{\sigma}_\mathrm{photon}$ & $\tilde{\Delta F}$ & ${\sigma}_{\rm rms}$ & seeing ($''$)                          & cadence                  \\\hline
\multirow{4}{*}{DES$\,$J0029$-$3814}   & A & 0.7       &             &                                  &                    &                            & \multirow{4}{*}{}                       & \multirow{4}{*}{}        & 20.68-20.81 & 34.6                             & 28.5               & 22.9                       & \multirow{4}{*}{$1.18_{-0.26}^{+0.40}$} & \multirow{4}{*}{1 (1.4)} \\
                                       & B & 0.7       &             &                                  &                    &                            &                                         &                          & 21.80-21.96 & 42.9                             & 70.8               & 40.1                       &                                         &                          \\
                                       & C & 0.7       &             &                                  &                    &                            &                                         &                          & 21.97-22.12 & 46.2                             & 82.8               & 45.5                       &                                         &                          \\
                                       & D & 1.7       &             &                                  &                    &                            &                                         &                          & 21.66-21.84 & 40                               & 31.4               & 17.5                       &                                         &                          \\\hline
\multirow{4}{*}{PS$\,$J0030$-$1525}    & A & 0.5       &             &                                  &                    &                            & \multirow{4}{*}{}                       & \multirow{4}{*}{}        & 19.35-19.39 & 15.8                             & 6.7                & 10.1                       & \multirow{4}{*}{$1.12_{-0.23}^{+0.38}$} & \multirow{4}{*}{1 (1.7)} \\
                                       & B & 1.5       &             &                                  &                    &                            &                                         &                          & 19.92-19.98 & 16.6                             & 7.3                & 7                          &                                         &                          \\
                                       & C & 1.8       &             &                                  &                    &                            &                                         &                          & 21.16-21.24 & 24.2                             & 20.3               & 23.1                       &                                         &                          \\
                                       & D & 0.5       &             &                                  &                    &                            &                                         &                          & 22.65-23.21 & 93.5                             & 219.5              & 250.9                      &                                         &                          \\\hline
\multirow{4}{*}{DES$\,$J0053$-$2012}   & A & 2.1       &             &                                  &                    &                            & \multirow{4}{*}{}                       & \multirow{4}{*}{}        & 19.48-19.52 & 17.3                             & 8.4                & 6                          & \multirow{4}{*}{$1.10_{-0.23}^{+0.38}$} & \multirow{4}{*}{1 (1.7)} \\
                                       & B & 1.0       &             &                                  &                    &                            &                                         &                          & 19.36-19.40 & 17.3                             & 8.4                & 5                          &                                         &                          \\
                                       & C & 1.0       &             &                                  &                    &                            &                                         &                          & 19.53-19.56 & 17.4                             & 9.8                & 5.8                        &                                         &                          \\
                                       & D & 2.1       &             &                                  &                    &                            &                                         &                          & 21.94-22.02 & 34.8                             & 52.1               & 32.8                       &                                         &                          \\\hline
\multirow{4}{*}{WG$\,$J0214$-$2105}    & A & 0.9       & 20.49-20.56 & 12.6                             & 18.7               & 8.8                        & \multirow{4}{*}{$0.94_{-0.18}^{+0.30}$} & \multirow{4}{*}{1 (2.0)} & 20.36-20.60 & 22.6                             & 19.1               & 10.3                       & \multirow{4}{*}{$1.10_{-0.22}^{+0.31}$} & \multirow{4}{*}{1 (1.4)} \\
                                       & B & 1.4       & 20.34-20.48 & 12.2                             & 15.6               & 9.7                        &                                         &                          & 20.24-20.60 & 21.8                             & 15.4               & 9.3                        &                                         &                          \\
                                       & C & 1.0       & 20.43-20.52 & 12.5                             & 18                 & 11.3                       &                                         &                          & 20.29-20.58 & 22                               & 17                 & 11.2                       &                                         &                          \\
                                       & D & 0.9       & 21.35-21.42 & 19.2                             & 46.9               & 22.2                       &                                         &                          & 21.20-21.36 & 34.4                             & 33.6               & 17.5                       &                                         &                          \\\hline
\multirow{4}{*}{HE$\,$J0230$-$2130}    & A & 0.7       & 18.94-19.09 & 4.2                              & 9.7                & 5.5                        & \multirow{4}{*}{$0.91_{-0.18}^{+0.23}$} & \multirow{4}{*}{1 (1.6)} & 19.27-19.59 & 39.6                             & 10.6               & 10.5                       & \multirow{4}{*}{$1.07_{-0.19}^{+0.29}$} & \multirow{4}{*}{1 (1.5)} \\
                                       & B & 0.7       & 19.07-19.16 & 4.5                              & 10.2               & 5.7                        &                                         &                          & 19.70-19.85 & 39.7                             & 15.7               & 12.7                       &                                         &                          \\
                                       & C & 1.5       & 19.78-19.87 & 6.2                              & 10                 & 4.8                        &                                         &                          & 20.47-20.62 & 40.7                             & 14.4               & 13                         &                                         &                          \\
                                       & D & 1.5       & 21.80-21.92 & 32.9                             & 63.9               & 30.6                       &                                         &                          & 22.29-22.46 & 59.6                             & 68.9               & 37.9                       &                                         &                          \\\hline
\multirow{4}{*}{WISE$\,$J0259$-$1635} & A & 0.8       & 19.73-19.78 & 9.7                              & 14                 & 7.3                        & \multirow{4}{*}{$0.97_{-0.22}^{+0.31}$} & \multirow{4}{*}{1 (1.9)} & 19.79-19.87 & 17.6                             & 15.1               & 9.5                        & \multirow{4}{*}{$1.09_{-0.19}^{+0.23}$} & \multirow{4}{*}{1 (1.7)} \\
                                       & B & 0.9       & 20.08-20.12 & 10.4                             & 16.1               & 10.3                       &                                         &                          & 19.99-20.06 & 17.8                             & 15.2               & 7.3                        &                                         &                          \\
                                       & C & 1.2       & 20.42-20.48 & 11.5                             & 17.9               & 11.1                       &                                         &                          & 20.28-20.37 & 18.3                             & 14.7               & 6.1                        &                                         &                          \\
                                       & D & 0.8       & 20.78-20.83 & 13.8                             & 29.7               & 15.2                       &                                         &                          & 20.69-20.73 & 19.7                             & 28.9               & 13.2                       &                                         &                          \\\hline
\multirow{4}{*}{DES$\,$J0420$-$4037}   & A & 1.0       & 18.93-18.99 & 11.1                             & 6.1                & 7.1                        & \multirow{4}{*}{$0.88_{-0.15}^{+0.21}$} & \multirow{4}{*}{1 (1.3)} &             &                                  &                    &                            & \multirow{4}{*}{}                       & \multirow{4}{*}{}        \\
                                       & B & 0.6       & 19.94-20.01 & 12                               & 14.8               & 8.3                        &                                         &                          &             &                                  &                    &                            &                                         &                          \\
                                       & C & 0.6       & 21.19-21.26 & 18.4                             & 45.8               & 23.2                       &                                         &                          &             &                                  &                    &                            &                                         &                          \\
                                       & D & 1.4       & 21.36-21.44 & 19.1                             & 40                 & 19.3                       &                                         &                          &             &                                  &                    &                            &                                         &                          \\\hline
\multirow{2}{*}{DES$\,$J0602$-$4335}   & A & 1.8       &             &                                  &                    &                            & \multirow{2}{*}{}                       & \multirow{2}{*}{}        & 19.48-19.51 & 8.8                              & 7.9                & 4.7                        & \multirow{2}{*}{$1.04_{-0.22}^{+0.26}$} & \multirow{2}{*}{1 (1.2)} \\
                                       & B & 1.8       &             &                                  &                    &                            &                                         &                          & 19.67-19.69 & 8.9                              & 8.7                & 3.8                        &                                         &                          \\\hline
\multirow{4}{*}{J0607$-$2152}          & A & 0.6       & 19.86-20.13 & 9.3                              & 15.6               & 7.7                        & \multirow{4}{*}{$0.84_{-0.19}^{+0.23}$} & \multirow{4}{*}{1 (1.6)} &             &                                  &                    &                            & \multirow{4}{*}{}                       & \multirow{4}{*}{}        \\
                                       & B & 0.6       & 21.88-22.09 & 43.7                             & 91.6               & 46.3                       &                                         &                          &             &                                  &                    &                            &                                         &                          \\
                                       & C & 1.1       & 21.15-21.61 & 25.1                             & 34.1               & 20.6                       &                                         &                          &             &                                  &                    &                            &                                         &                          \\
                                       & D & 1.2       & 21.55-21.78 & 31.7                             & 45.5               & 24.2                       &                                         &                          &             &                                  &                    &                            &                                         &                          \\\hline
\multirow{4}{*}{J0659+1629}            & A & 1.4       & 17.74-17.80 & 5.9                              & 3.2                & 1.5                        & \multirow{4}{*}{$0.90_{-0.17}^{+0.25}$} & \multirow{4}{*}{1 (1.6)} &             &                                  &                    &                            & \multirow{4}{*}{}                       & \multirow{4}{*}{}        \\
                                       & B & 1.9       & 19.30-19.34 & 8.2                              & 7.5                & 3.6                        &                                         &                          &             &                                  &                    &                            &                                         &                          \\
                                       & C & 1.4       & 19.29-19.40 & 8.3                              & 7.6                & 5.1                        &                                         &                          &             &                                  &                    &                            &                                         &                          \\
                                       & D & 4.7       & 19.12-19.15 & 7.4                              & 6.3                & 3.2                        &                                         &                          &             &                                  &                    &                            &                                         &                          \\\hline
\multirow{2}{*}{J0832+0404}            & A & 2.2       &             &                                  &                    &                            & \multirow{2}{*}{}                       & \multirow{2}{*}{}        & 19.73-19.77 & 19.3                             & 4.9                & 5.8                        & \multirow{2}{*}{$0.95_{-0.16}^{+0.24}$} & \multirow{2}{*}{1 (1.2)} \\
                                       & B & 2.2       &             &                                  &                    &                            &                                         &                          & 20.99-21.07 & 24.4                             & 15.4               & 15.3                       &                                         &                          \\
\bottomrule
\end{tabular*}
\textbf{Notes}: Seeing and noise quantities are measured with the median, and distance to the 20\textsuperscript{th} and 80\textsuperscript{th} percentiles for the lower and upper intervals respectively.
The third column denotes the separation of a given lensed image to its closest neighbour.
The magnitude range column gives the 1\textsuperscript{st} and 99\textsuperscript{th} percentiles of the magnitudes in the given lensed image.  
$\sigma_{\rm rms}$ is the root-mean-squared deviation from a smooth spline fitted to the curve at hand. 
$\tilde{\sigma}_\mathrm{photon}$,  $\tilde{\Delta F}$ and ${\sigma}_{\rm rms}$ are given in millimagnitudes. 
The cadence is given in days. Tildes denote a median over the dataset.
\end{table*}

\begin{table*}[]
\addtocounter{table}{-1}
\caption{Continued.
\label{tab:obsprops2}}
\def\arraystretch{1.08}
\begin{tabular*}{\linewidth}{@{\extracolsep{\fill}} lll|llllll|llllll}
Object                                 &   & Sep. $''$ & \multicolumn{6}{c}{VST}                                                                                                                                               & \multicolumn{6}{c}{2p2}                                                                                                                                               \\
                                       &   &           & mag range   & $\tilde{\sigma}_\mathrm{photon}$ & $\tilde{\Delta F}$ & ${\sigma}_{\rm rms}$ & seeing ($''$)                            & cadence                  & mag range   & $\tilde{\sigma}_\mathrm{photon}$ & $\tilde{\Delta F}$ & ${\sigma}_{\rm rms}$ & seeing ($''$)                           & cadence                  \\\hline
\multirow{4}{*}{RX$\,$J0911+0551}      & A & 0.6       &             &                                  &                    &                            & \multirow{4}{*}{}                       & \multirow{4}{*}{}        & 19.76-19.85 & 10.9                             & 22.3               & 8.9                        & \multirow{4}{*}{$1.03_{-0.17}^{+0.29}$} & \multirow{4}{*}{1 (1.4)} \\
                                       & B & 0.5       &             &                                  &                    &                            &                                         &                          & 19.02-19.07 & 8.5                              & 15.8               & 8.1                        &                                         &                          \\
                                       & C & 0.5       &             &                                  &                    &                            &                                         &                          & 19.84-19.92 & 11.4                             & 34.7               & 14.9                       &                                         &                          \\
                                       & D & 3.0       &             &                                  &                    &                            &                                         &                          & 20.12-20.17 & 12                               & 13.7               & 6.3                        &                                         &                          \\\hline
\multirow{4}{*}{J0924+0219}            & A & 0.7       & 18.45-18.72 & 11                               & 6.1                & 6.1                        & \multirow{4}{*}{$0.90_{-0.18}^{+0.29}$} & \multirow{4}{*}{1 (3.2)} & 18.93-19.05 & 19.5                             & 7                  & 4.5                        & \multirow{4}{*}{$0.93_{-0.12}^{+0.28}$} & \multirow{4}{*}{8 (8.2)} \\
                                       & B & 1.5       & 20.22-20.29 & 12.6                             & 12.5               & 5.9                        &                                         &                          & 20.37-20.45 & 19.9                             & 14.6               & 5.4                        &                                         &                          \\
                                       & C & 1.2       & 21.91-22.07 & 32.4                             & 66                 & 43.7                       &                                         &                          & 21.93-22.01 & 25.9                             & 53.4               & 17.6                       &                                         &                          \\
                                       & D & 0.7       & 22.09-22.41 & 45.2                             & 151.6              & 61.6                       &                                         &                          & 22.38-22.60 & 45                               & 148.7              & 50.1                       &                                         &                          \\\hline
\multirow{4}{*}{GRAL$\,$J1131$-$4419}  & A & 0.5       & 18.89-18.97 & 8.5                              & 11.2               & 11.1                       & \multirow{4}{*}{$0.95_{-0.20}^{+0.31}$} & \multirow{4}{*}{1 (1.4)} & 19.12-19.21 & 16.9                             & 16.7               & 11.1                       & \multirow{4}{*}{$0.98_{-0.18}^{+0.32}$} & \multirow{4}{*}{1 (1.6)} \\
                                       & B & 0.5       & 19.31-19.42 & 9.2                              & 16.4               & 14.2                       &                                         &                          & 19.43-19.52 & 17.3                             & 20.7               & 14.9                       &                                         &                          \\
                                       & C & 1.4       & 19.91-19.97 & 10.7                             & 14.3               & 7.7                        &                                         &                          & 19.87-20.03 & 18.2                             & 13.1               & 7.3                        &                                         &                          \\
                                       & D & 1.4       & 20.06-20.16 & 11.6                             & 14.4               & 10.1                       &                                         &                          & 20.13-20.27 & 19.4                             & 14.6               & 12.7                       &                                         &                          \\\hline
\multirow{4}{*}{2M$\,$J1310$-$1714}    & A & 2.9       & 20.02-20.26 & 19.2                             & 11.8               & 6.2                        & \multirow{4}{*}{$0.91_{-0.20}^{+0.25}$} & \multirow{4}{*}{1 (1.6)} & 19.74-19.81 & 10.9                             & 8.3                & 4.7                        & \multirow{4}{*}{$1.12_{-0.30}^{+0.51}$} & \multirow{4}{*}{1 (1.6)} \\
                                       & B & 3.8       & 19.93-20.05 & 18.4                             & 11.4               & 6                          &                                         &                          & 19.76-19.80 & 10.9                             & 9.3                & 7.5                        &                                         &                          \\
                                       & C & 2.9       & 19.91-20.13 & 18.7                             & 11.8               & 5.9                        &                                         &                          & 19.66-19.74 & 10.8                             & 8                  & 6.2                        &                                         &                          \\
                                       & D & 4.6       & 21.38-21.64 & 37                               & 38.1               & 17.8                       &                                         &                          & 21.07-21.22 & 24.6                             & 24.7               & 13.1                       &                                         &                          \\\hline
\multirow{4}{*}{J1537$-$3010}          & A & 1.9       & 20.23-20.30 & 11                               & 16.9               & 12.4                       & \multirow{4}{*}{$1.06_{-0.22}^{+0.36}$} & \multirow{4}{*}{1 (1.5)} & 20.15-20.31 & 18.7                             & 14.1               & 7.9                        & \multirow{4}{*}{$1.04_{-0.25}^{+0.44}$} & \multirow{4}{*}{1 (1.5)} \\
                                       & B & 2.1       & 20.44-20.48 & 11.2                             & 18                 & 8.9                        &                                         &                          & 20.27-20.45 & 19.6                             & 15.9               & 8.8                        &                                         &                          \\
                                       & C & 2.0       & 20.40-20.43 & 11.1                             & 15.5               & 10.8                       &                                         &                          & 20.33-20.49 & 19.9                             & 14.8               & 8.1                        &                                         &                          \\
                                       & D & 1.9       & 21.15-21.20 & 13.1                             & 31                 & 14.1                       &                                         &                          & 21.01-21.11 & 27.7                             & 22.8               & 12.3                       &                                         &                          \\\hline
\multirow{4}{*}{PS$\,$J1606$-$2333}    & A & 1.1       & 19.05-19.10 & 6.4                              & 8.2                & 3.9                        & \multirow{4}{*}{$0.99_{-0.23}^{+0.26}$} & \multirow{4}{*}{1 (1.9)} & 19.12-19.31 & 11                               & 8.3                & 3.5                        & \multirow{4}{*}{$0.96_{-0.19}^{+0.33}$} & \multirow{4}{*}{1 (1.4)} \\
                                       & B & 0.9       & 19.22-19.28 & 6.6                              & 11.9               & 6.4                        &                                         &                          & 19.29-19.47 & 11.4                             & 10.8               & 5.9                        &                                         &                          \\
                                       & C & 0.9       & 19.44-19.51 & 7.1                              & 13.3               & 5.6                        &                                         &                          & 19.69-19.86 & 13                               & 15.4               & 8.2                        &                                         &                          \\
                                       & D & 0.9       & 19.55-19.63 & 7.4                              & 14.2               & 6.3                        &                                         &                          & 19.85-19.99 & 13.5                             & 14.4               & 6.8                        &                                         &                          \\\hline
\multirow{2}{*}{WGD$\,$J2021$-$4115}   & A & 2.7       &             &                                  &                    &                            & \multirow{2}{*}{}                       & \multirow{2}{*}{}        & 19.46-19.53 & 20.9                             & 8.7                & 4.9                        & \multirow{2}{*}{$1.22_{-0.28}^{+0.38}$} & \multirow{2}{*}{1 (1.6)} \\
                                       & B & 2.7       &             &                                  &                    &                            &                                         &                          & 19.83-19.88 & 21.5                             & 12.3               & 7.6                        &                                         &                          \\\hline
\multirow{4}{*}{WFI$\,$J2026$-$4536}   & A & 0.3       & 17.57-17.61 & 7                                & 17                 & 9.8                        & \multirow{4}{*}{$1.10_{-0.27}^{+0.32}$} & \multirow{4}{*}{1 (1.9)} &             &                                  &                    &                            & \multirow{4}{*}{}                       & \multirow{4}{*}{}        \\
                                       & B & 1.2       & 18.67-18.71 & 7.2                              & 10.3               & 6.7                        &                                         &                          &             &                                  &                    &                            &                                         &                          \\
                                       & C & 0.8       & 18.88-18.95 & 7.4                              & 19.3               & 15.5                       &                                         &                          &             &                                  &                    &                            &                                         &                          \\
                                       & D & 0.3       & 17.69-17.78 & 7                                & 18.7               & 13.8                       &                                         &                          &             &                                  &                    &                            &                                         &                          \\\hline
\multirow{4}{*}{WG$\,$J2038$-$4008}   & A & 1.9       &             &                                  &                    &                            & \multirow{4}{*}{}                       & \multirow{4}{*}{}        & 19.93-20.22 & 24.4                             & 10.5               & 10.4                       & \multirow{4}{*}{$1.22_{-0.25}^{+0.36}$} & \multirow{4}{*}{1 (1.7)} \\
                                       & B & 1.5       &             &                                  &                    &                            &                                         &                          & 19.77-19.94 & 23.3                             & 10                 & 9.3                        &                                         &                          \\
                                       & C & 1.5       &             &                                  &                    &                            &                                         &                          & 19.84-20.04 & 23.7                             & 9.2                & 9                          &                                         &                          \\
                                       & D & 2.1       &             &                                  &                    &                            &                                         &                          & 20.64-20.85 & 33                               & 19.9               & 17.7                       &                                         &                          \\\hline
\multirow{4}{*}{WG$\,$J2100$-$4452}          & A & 0.5       & 19.21-19.32 & 9.3                              & 20.2               & 15.9                       & \multirow{4}{*}{$1.06_{-0.24}^{+0.42}$} & \multirow{4}{*}{1 (1.6)} & 19.24-19.39 & 19.9                             & 17.1               & 15.4                       & \multirow{4}{*}{$1.20_{-0.25}^{+0.34}$} & \multirow{4}{*}{1 (1.7)} \\
                                       & B & 0.5       & 20.26-20.45 & 14.7                             & 51.6               & 30.8                       &                                         &                          & 20.32-20.58 & 22.3                             & 54.7               & 27.7                       &                                         &                          \\
                                       & C & 2.1       & 21.16-21.38 & 26.3                             & 42.5               & 21.5                       &                                         &                          & 21.29-21.40 & 28.8                             & 50.7               & 22.1                       &                                         &                          \\
                                       & D & 2.1       & 21.46-21.72 & 35.5                             & 58.2               & 26                         &                                         &                          & 21.90-22.06 & 43.9                             & 86.4               & 42.8                       &                                         &                          \\\hline
\multirow{4}{*}{J2205$-$3727}          & A & 0.6       & 20.78-20.92 & 11.6                             & 31                 & 17.3                       & \multirow{4}{*}{$1.06_{-0.25}^{+0.43}$} & \multirow{4}{*}{1 (1.7)} &             &                                  &                    &                            & \multirow{4}{*}{}                       & \multirow{4}{*}{}        \\
                                       & B & 0.7       & 21.37-21.55 & 18.8                             & 56.5               & 28                         &                                         &                          &             &                                  &                    &                            &                                         &                          \\
                                       & C & 0.6       & 22.48-22.68 & 49.4                             & 158.7              & 56.2                       &                                         &                          &             &                                  &                    &                            &                                         &                          \\
                                       & D & 1.4       & 21.81-21.96 & 25.7                             & 48.8               & 26.9                       &                                         &                          &             &                                  &                    &                            &                                         &                         
\end{tabular*}
\end{table*}

\end{landscape}

\end{document}